\pdfoutput=1
\documentclass[aps,pre,twocolumn,superscriptaddress,floatfix]{revtex4}
\usepackage{graphicx,bm,amssymb,amsmath,dcolumn,hyperref}
\usepackage{graphicx}
\usepackage{verbatim}
\usepackage{color}
\usepackage{epstopdf}
\usepackage{subfigure}
\usepackage{mathrsfs}
\usepackage{amssymb}
\usepackage{float}
\usepackage{subfigure}
\usepackage{multirow}

\begin{document}

\definecolor{blue}{rgb}{0,0,1}
\definecolor{red}{rgb}{1,0,0}
\definecolor{green}{rgb}{0,1,0}
\newcommand{\blue}[1]{\textcolor{blue}{#1}}
\newcommand{\red}[1]{\textcolor{red}{#1}}
\newcommand{\green}[1]{\textcolor{green}{#1}}

\newcommand{\PB}{P_{\rm B}}

\newcommand{\calA}{ {\mathcal A}}
\newcommand{\calC}{ {\mathcal C}}
\newcommand{\calE}{ {\mathcal E}}
\newcommand{\calG}{ {\mathcal G}}
\newcommand{\calH}{ {\mathcal H}}
\newcommand{\calO}{ {\mathcal O}}
\newcommand{\calS}{ {\mathcal S}}
\newcommand{\calZ}{ {\mathcal Z}}

\title{Percolation thresholds of randomly rotating patchy particles on Archimedean lattices}

\author{Quancheng Wang}
\affiliation{School of Physics and Electronic Engineering, Anhui University, Hefei, Anhui 230601, China}

\author{Zhenfang He}
\affiliation{School of Physics and Electronic Engineering, Anhui University, Hefei, Anhui 230601, China}

\author{Junfeng Wang}
\affiliation{School of Physics, Hefei University of Technology, Hefei, Anhui 230009, China}

\author{Hao Hu}
\email[Corresponding author:\;]{huhao@ahu.edu.cn}
\affiliation{School of Physics and Electronic Engineering, Anhui University, Hefei, Anhui 230601, China}

\date{\today}

\begin{abstract}
	We study the percolation of randomly rotating patchy particles on $11$ Archimedean lattices in two dimensions. 
	Each vertex of the lattice is occupied by a particle, and in each model the patch size and number are monodisperse.
	When there are more than one patches on the surface of a particle, they are symmetrically decorated. 
	As the proportion $\chi$ of the particle surface covered by the patches increases, the clusters connected 
	by the patches grow and the system percolates at the threshold $\chi_c$.  
	We combine Monte Carlo simulations and the critical polynomial method to give precise estimates 
	of $\chi_c$ for disks with one to six patches and spheres with one to two patches on the $11$ lattices.
	For one-patch particles, we find that the order of $\chi_c$ values for particles on different lattices 
	is the same as that of threshold values $p_c$ for site percolation on same lattices,
	which implies that $\chi_c$ for one-patch particles mainly depends on the geometry of lattices. 
	For particles with more patches, symmetry become very important in determining $\chi_c$.
	With the estimates of $\chi_c$ for disks with one to six patches, by analyses related to symmetry, 
	we are able to give precise values of $\chi_c$ for disks with an arbitrary number of patches on all $11$ lattices.
	The following rules are found for patchy disks on each of these lattices: 
	(i) as the number of patches $n$ increases, values of $\chi_c$ repeat in a periodic way, 
	with the period $n_0$ determined by the symmetry of the lattice; 
	(ii) when $\mod(n,n_0)=0$, the minimum threshold value $\chi_{\rm min}$ appears, 
	and the model is equivalent to site percolation with $\chi_{\rm min}=p_c$;
	(iii) disks with $\mod(n,n_0)=m$ and $n_0-m$ ($m<n_0/2$) share the same $\chi_c$ value.
\end{abstract}
\maketitle 

\section{Introduction}
\label{sec1}

\begin{figure}
\begin{center}
\includegraphics[scale=0.165]{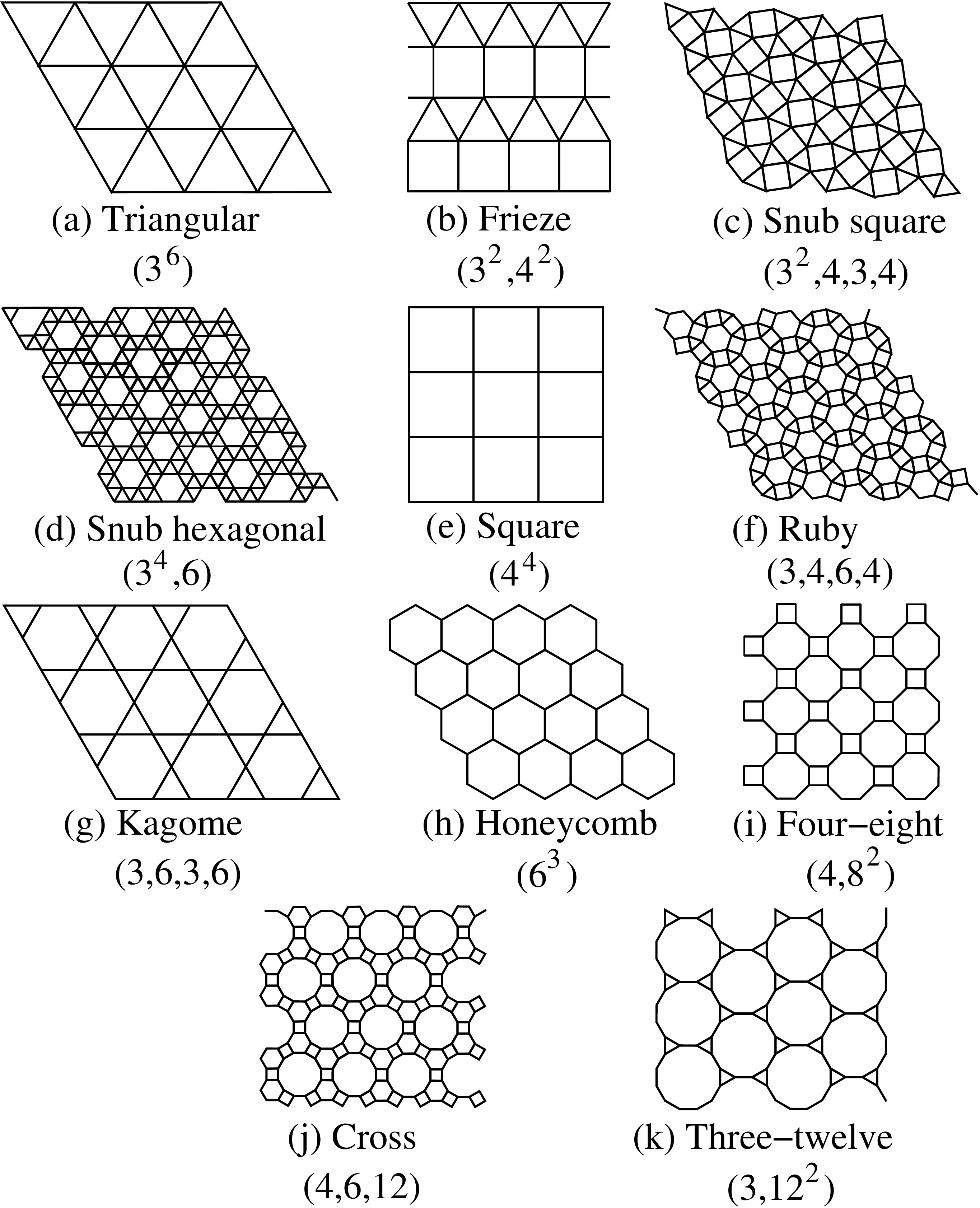}
	\caption{The $11$ Archimedean lattices in two dimensions. 
        For each lattice, all vertices are equivalent if the lattice size is infinite or the boundary conditions are periodic.
	Lattice names are used as in Ref.~\cite{Scullard2020}, 
	and under the names are symbols designated using the general notation proposed 
	by Gr\"unbaum and Shephard~\cite{Gruenbaum1987}. 
	A brief introduction of the history and nomenclature of the lattices are available in Ref.~\cite{Suding1999}.}
\label{Fig:archimedean}
\end{center}
\end{figure}

Patchy particles~\cite{Walther2013,Zhang2017} are created by modifying the surface of colloidal particles,
where each modified area is regarded as a patch. One-patch particles with two distinct surface areas 
are usually called Janus particles, and two-patch particles are called triblock Janus particles.
These particles can be designed in various shapes, e.g., spheres, dumbbells, disks, and rods.
And the patches can be decorated with different properties, e.g., chemical, optical, electrical, 
and magnetic properties. As model systems with anisotropic interactions, patchy particles
are used to study equilibrium gels and water~\cite{Sciortino2017,Smallenburg2014},
and they can self-assemble into open lattices~\cite{Chen2011,Mao2013,Li2020} 
such as the entropy stabilized kagome lattice~\cite{Chen2011,Mao2013}.
In two dimensions (2D), when patchy disks or spheres are compressed tightly, they form a triangular lattice. 
Putting Janus particles onto the densely packed triangular lattice, various continuous thermodynamic phase transitions
and critical phenomena have recently been observed~\cite{Mitsumoto2018,Patrykiejew2020,Patrykiejew2020-A}.

\begin{figure}
\begin{center}
\includegraphics[scale=0.34]{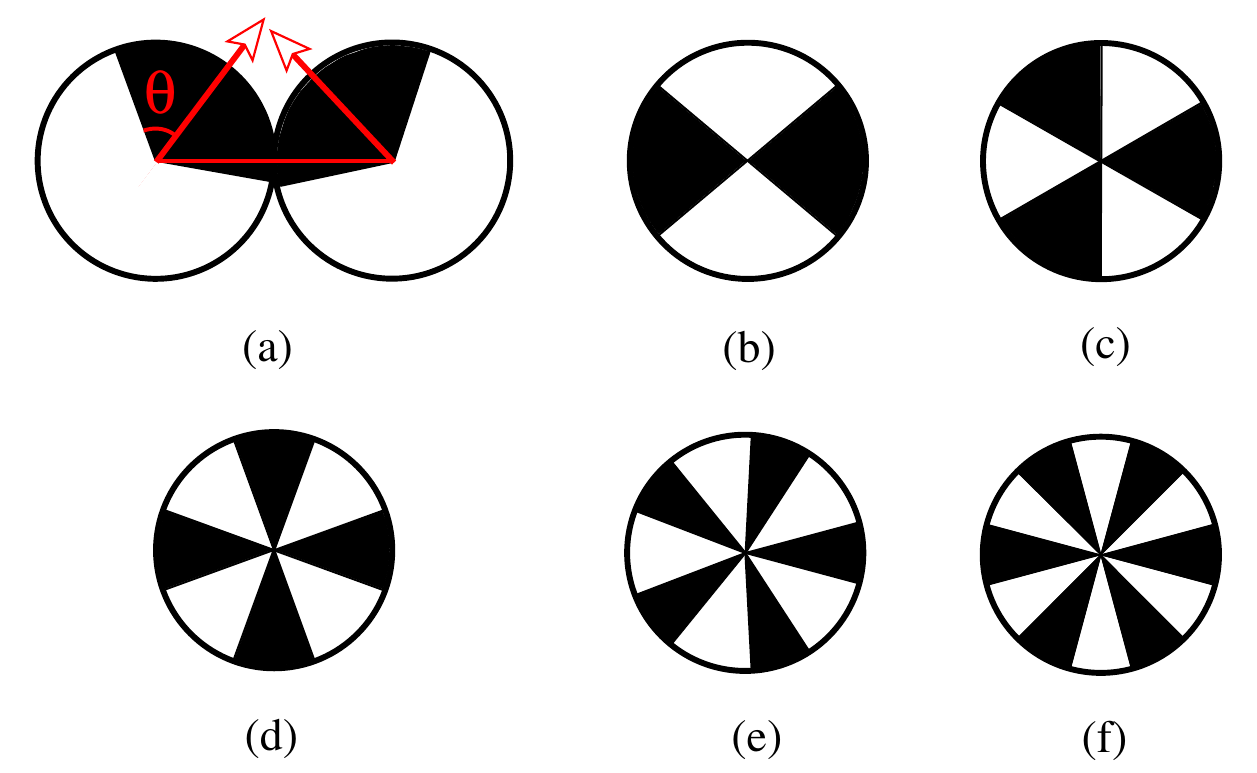}
	\caption{Schematic drawings of patchy disks. 
	On each disk the dark areas are patches, which share the same size 
	and locate symmetrically when their number is more than one. 
	The half-angle $\theta$ characterizes the size of a patch.
	(a)-(f) show particles containing one to six patches, respectively. 
	(a) also exemplifies the connection of two particles, i.e. they are considered as connected 
	if their patches touch each other. The red arrow indicates the direction of the particle, 
	which is drawn by connecting the center of the particle and that of one patch.  
	}
\label{Fig:patches}
\end{center}
\end{figure}

Percolation is extensively studied in stochastic processes, 
phase transitions and critical phenomena, and widely applied in various problems such as exploring gelation in polymers, 
transport behaviors in porous media, the spread of epidemics, the fractal structure of landscapes etc.~\cite{Stauffer1994, Saberi2015}
There exist many studies on percolation of patchy particles in the continuum space, for which recent examples 
include different percolated states in mixtures of patchy colloids~\cite{Seiferling2016}, reentrant percolation 
of inverse patchy colloids~\cite{Araujo2017} and of patchy colloids on patterned substrates~\cite{Treffenstadt2018},  
effects of surface heterogeneity on percolation thresholds of random patchy spheres~\cite{Wang2019},
the design of patchy particle gels with tunable percolation thresholds~\cite{Song2021}. 
However, except a study on directed percolation of patchy disks on the square lattice~\cite{Kartha2016},
we find that the percolation behavior of patchy particles on lattices remains largely unexplored, 
possibly due to that patchy particles are made in the continuum space~\cite{Walther2013,Zhang2017}.

In this work, we study percolation of patchy particles on different lattices, 
in order to get more understanding of collective behaviors of the particles in 2D. 
As model systems, we assume that each vertex of the lattice is occupied by a patchy particle, 
and that particles rotate randomly.
Two adjacent particles are considered as connected only when their patches are in contact with each other.
As the proportion $\chi$ of the particle surface occupied by the patch(es) increases, 
the clusters formed by connected particles will gradually become larger. At a threshold value $\chi_c$, 
a cluster that spans the entire lattice first forms, i.e. the percolation transition occurs. 
The threshold is an important parameter for percolation, and thus we focus on determining the $\chi_c$ 
values and exploring the dependence of these values on the symmetry and geometry of patchy particles and lattices.

For the purpose above, we first numerically study disks with one to six patches and spheres with one to two patches,
on all $11$ Archimedean lattices in 2D, which are shown in Fig.~\ref{Fig:archimedean}.
When there are more than one patches on a particle, the patches locate symmetrically and share the same size, 
as shown for patchy disks in Fig.~\ref{Fig:patches}. 
Monte Carlo (MC) simulations are conducted to produce independent configurations, 
and the recently developed critical polynomial method~\cite{Scullard2008, Scullard2010, Scullard2011, Scullard2012-A, Scullard2012, Jacobsen2014, Jacobsen2015, Scullard2016, Scullard2020, Xu2021} 
is combined with MC sampling to precisely estimate $\chi_c$ of these models.
For one-patch particles, it is found that the order of $\chi_c$ values on different Archimedean lattices is the same as 
that of threshold values $p_c$ for site percolation on same lattices. This suggests that, 
for one-patch particles, the lattice geometry is the most important factor which affects $\chi_c$ values. 
For particles with more patches, their $\chi_c$ values on different lattices do not follow the same order as 
those for one-patch particles, which reflects that symmetry significantly influence percolation thresholds 
of particles with more than one patches.

The role of symmetry in determining the $\chi_c$ values is further exhibited 
in analytic calculations of the probabilities of different patch-covering structures 
of a particle near the above numerically estimated values of $\chi_c$.
These calculations lead to expressions of the probabilities as a function of $\chi$,
which allow us to prove the equality of $\chi_c$ values for various models 
or explain the difference between close $\chi_c$ values for distinct models.
As results, we are able to give precise values of $\chi_c$ for disks with 
an arbitrary number of patches on all $11$ Archimedean lattices in 2D.
We find that $\chi_c$ values for patchy disks on each of the lattices are governed 
by the following rules:
(i) $\chi_c$ values repeat in a periodic way as the number of patches $n$ increases,
with the period $n_0$ determined by the symmetry of the lattice;
(ii) the minimum threshold value $\chi_{\rm min}$ appears when $\mod(n,n_0)=0$,
for which the model of patchy disks is equivalent to site percolation
with $\chi_{\rm min}=p_c$;
(iii) disks with $\mod(n,n_0)=m$ and $n_0-m$ ($m < n_0/2$) share the same value of $\chi_c$. 
These results could provide references for further studies, such as exploring
at finite temperatures the connectivity and phase behaviors of patchy particles on lattices in 2D~\cite{Mitsumoto2018,Patrykiejew2020,Patrykiejew2020-A}. 

The remainder of this paper is organized as follows. 
Section~\ref{sec-model} introduces the models and numerical methods. 
Section~\ref{sec-result} presents main results.
A brief conclusion and discussion is given in Section~\ref{sec-conclu}.
More details can be found in the supplemental material (SM)~\cite{sup-mat}.

\section{Models and Numerical Methods}
\label{sec-model}

\subsection{Models}
For the Archimedean lattices, all vertices are equivalent and the lengths of edges are equal,
as shown in Fig.~\ref{Fig:archimedean}. 
For the model of a given type of patchy particles on one of the Archimedean lattices,
each vertex of the lattice is occupied by a randomly rotating particle. The particle center locates 
right at the vertex and its diameter is set equal to the edge length of the lattice, 
thus two particles at the ends of an edge are in contact with each other.
Two neighboring particles are connected if one patch on one particle touch another patch 
on the other particle (the two patches both cover the same edge).
As shown in Fig.~\ref{Fig:patches}(a), the size of a patch is characterized by the half-angle $\theta$.
The one- and two-patch spheres can be drawn similar to Fig.~\ref{Fig:patches}(a-b),
except that a patch is a sphere cap with $\theta$ being the polar angle. 
For convenience of comparing thresholds of different types of particles,
we also define the size of patches by the proportion $\chi$ of the particle surface covered 
by the patches, which is related to $\theta$ as
\begin{equation}
\chi={2n \theta}/{2 \pi} = n \theta / \pi
\label{eq:chi-theta-1}
\end{equation}
for patchy disks, and as
\begin{equation}
	\chi=\frac{n}{4\pi} \int_0^{\theta} \sin \theta d \theta \int_0^{2\pi} d \phi ={n (1-\cos{\theta})}/{2}
\label{eq:chi-theta-2}
\end{equation}
for patchy spheres.
Here $n$ is the number of symmetrically distributed patches on a particle.
In experiments, the patch size can be designed through surface modification 
or compartmentalization~\cite{Walther2013,Zhang2017}.

\subsection{The critical polynomial}
The critical polynomial method is a powerful method proposed and developed in recent years
to calculate percolation thresholds in two dimensions~\cite{Scullard2008, Scullard2010, Scullard2011, Scullard2012-A, Scullard2012, Jacobsen2014, Jacobsen2015, Scullard2016, Scullard2020, Xu2021}.
It originated from the fact that every exactly solved percolation threshold in 2D can be expressed as 
the unique root of a polynomial.
For general (solved and unsolved) percolation models in 2D, the critical polynomial was first defined
using a linearity hypothesis and symmetry analyses~\cite{Scullard2008, Scullard2010, Scullard2011}.
Then the recursive deletion-contraction algorithm~\cite{Scullard2012-A} was proposed to find the polynomial.
Latest developments of the method are the alternative probabilistic, geometrical interpretation of 
the critical polynomial~\cite{Scullard2012} 
and transfer-matrix techniques for its calculation~\cite{Jacobsen2014, Jacobsen2015, Scullard2016, Scullard2020}.
Unprecedented estimates for thresholds of unsolved planar-lattice models have been obtained, e.g., 
for bond percolation on the kagome lattice, the precision of the estimate is in the order of $10^{-17}$~\cite{Scullard2020}. 
The critical polynomial has also been combined with MC sampling to provide high-precision estimates
of threshold values, e.g., for nonplanar and continuum percolation models~\cite{Xu2021},
for which transfer-matrix calculations are difficult. 

For a finite periodic lattice $B$ in 2D, the probabilistic, geometrical definition of 
the critical polynomial~\cite{Scullard2012} is
\begin{equation}
\PB \equiv R_2 - R_0.
\label{eq:PB}
\end{equation}
Here $R$ represents the wrapping (or crossing) probability~\cite{Langlands92, Pinson94},
where ``wrapping" means that, when putting the periodic lattice in 2D onto a torus, 
there exist a percolation cluster which wraps around the torus. 
The quantity $R_2$ is the probability of wrapping in two directions, and $R_0$ is the probability of
non-wrapping.
If filling the infinite space in 2D using copies of $B$ in some periodic way, wrapping in two directions
means that there is a open cluster which connects every copy, and non-wrapping means that 
no infinite copies of $B$ can be connected by open clusters~\cite{Scullard2012}.

Due to universality~\cite{Scullard2012}, the root $p(L)$ of $\PB(p,L)=0$ gives an estimate of 
the percolation threshold that becomes more accurate as the linear size $L$ of the lattice $B$ 
is increased. Here $p$ represents the control parameter for the percolation problem, 
e.g., the occupation probability ($p \in [0,1]$) for bond or site percolation.
From previous studies on the critical polynomial~\cite{Scullard2008, Scullard2010, Scullard2011, 
Scullard2012-A, Scullard2012, Jacobsen2014, Jacobsen2015, Scullard2016, Scullard2020}, 
it is known that $p(L)=p_c$ [$p_c \equiv p(L\rightarrow \infty)$] for exactly solved lattice 
models, even at the smallest $L$; and that, for unsolved percolation problems, $p(L)$ 
very quickly approaches $p_c$ as $L$ increases. For example, on Archimedean lattices, 
the unsolved bond percolation thresholds behave as 
$[p(L)-p_c] \simeq \sum_{k=1}^{\infty} A_k L^{-\Delta_k}$, 
with $\Delta_1=4$ or $6$, and $\Delta_i>\Delta_j$ when $i>j$~\cite{Scullard2020}. 
The finite-size correction of $\PB$ is much smaller than that for other quantities
such as wrapping probabilities~\cite{Xu2021}.

The critical polynomial $\PB$ is a dimensionless quantity, since $\PB(p_c, L \rightarrow \infty)=0$. 
Thus in the renormalization group formulation~\cite{Nightingale90}, $\PB$ has 
the finite-size scaling formula~\cite{Xu2021}
\begin{equation}
        \PB (t, u_1, u_2, L) = \PB (L^{y_t}t, L^{y_1} u_1, L^{y_2} u_2).
        \label{eq:PBscaling}
\end{equation}
Here $t \propto p-p_c$ is the relevant thermal renormalization scaling field,
and $y_t=1/\nu=3/4$ is the associated renomalization exponent.
And $u_1$ and $u_2$ represent two leading irrelevant scaling fields
with renormalization exponents $y_2<y_1<0$.
When assuming $y_2>2y_1$, by Taylor expansion to the first order, 
one gets
\begin{equation}
\PB (p,L) \simeq a_1 (p-p_c) L^{y_t} + b_1 L^{y_1}+ b_2 L^{y_2},
        \label{eq:PBfitting}
\end{equation}
where $a_1$, $b_1$ and $b_2$ are non-universal amplitudes.
The irrelevant exponents are related to $\Delta$ values as $y_1=y_t-\Delta_1$
and $y_2=y_t-\Delta_2$~\cite{Xu2021}.
Our MC data will be fitted by the above finite-size scaling formula,
with $p$ replaced by $\theta$,
though the irrelevant exponents may be different from those of bond percolation 
on Archimedean lattices~\cite{Scullard2020}.

\subsection{Monte Carlo simulation}
The MC method is used to sample independent configurations for $88$ models,
including six types of patchy disks and two types of patchy spheres, on all $11$ Archimedean lattices in 2D.
For a single configuration, a random direction is generated for each particle to
simulate the random rotation. 
In simulations, each lattice is encoded with the aid of a square or triangular array,
whose correspondence with the actual lattice is shown in Figs.S1-S11 of the SM~\cite{sup-mat},
and periodic boundary conditions are employed for the lattices.
The number of vertices (equivalent to the number of particles) on each lattice can 
be calculated from $L$, as shown in Table S1 of the SM~\cite{sup-mat}.
Hereafter $L$ is the linear size of the square or triangular array
for encoding the lattices, which is in proportion to the actual linear size.
We conduct simulations at different patch sizes and systems sizes,
and use sampled values of $\PB$ to determine the percolation threshold $\theta_c$ ($\chi_c$).

In our MC simulations, we simulated systems with up to $O(10^5)$ particles.
At least $10^8$ independent configurations were generated for 
each set of $(\theta,L)$ values, for all $88$ models (up to $10^{10}$ independent configurations for 
one-patch disks when $L<16$, as shown in Table S2 of the SM~\cite{sup-mat}). 
The total simulation time was about $14.6$ months if using a computer workstation with 
$2$ Intel Xeon Scalable Gold 6130 CPU and $8 \times 16$ GB DDR4 ECC Registered Shared Memory
(time for each model is shown in Table S3 of the SM~\cite{sup-mat}). 

\begin{figure}[hp]
\begin{center}
\includegraphics[scale=0.8]{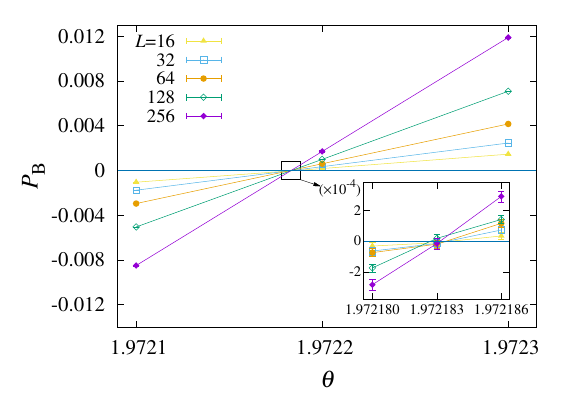}
	\caption{Plot of $\PB$ versus $\theta$ for one-patch disks on triangular lattices for different linear sizes $L$.
	The lines are added to guide the eyes.}
\label{Fig:tri-PB}
\end{center}
\end{figure}

\begin{table*}[ht]
\begin{center}
\caption{Fit results of the critical polynomial $\PB$ for one-patch disks on the triangular lattice.}
\label{Tab:PB-fit}
    \renewcommand\arraystretch{1.25}
    \setlength{\tabcolsep}{3.8mm}{
    \begin{tabular}{cccccccc}
    \hline
    \hline
        $L_{\rm min}$ & ${\chi}^2$/DOF & $\theta_c$ & $a_1$ & $b_1$ & $y_1$ & $b_2$ & $y_2$ \\
    \hline
    $6$ & $63.6/79$ & $1.972\,183\,20(7)$ & $-1.65(4)$ & $-2.1(6)$ & $-4.9(2)$ & \\
    $8$ & $62.6/76$ & $1.972\,183\,20(7)$ & $-1.65(4)$ & $-2.2(30)$ & $-4.9(6)$ & \\
    $4$ & $83.8/88$ & $1.972\,183\,20(7)$ & $-1.65(4)$ & $-2.2(25)$ & $-4.9(4)$ & $0.3(53)$ & $-6$ \\
    $2$ & $84.8/91$ & $1.972\,183\,21(7)$ & $-1.65(4)$ & $-3.7(2)$ & $-5.11(4)$ & $9.2(5)$ & $-7$  \\
    $4$ & $83.8/88$ & $1.972\,183\,20(7)$ & $-1.65(4)$ & $-2.2(13)$ & $-4.9(3)$ & $0.5(85)$ & $-7$ \\
    $2$ & $84.3/91$ & $1.972\,183\,21(7)$ & $-1.65(4)$ & $-2.7(1)$ & $-4.97(3)$ & $13.0(6)$ & $-8$ \\
    \hline
    \hline
    \end{tabular}}
\end{center}
\end{table*}

\section{Results}
\label{sec-result}
We first present numerical results for one-patch particles, then for particles with two to six patches. 
Precise estimates of $\chi_c$ for these particles on $11$ Archimedean lattices are obtained 
by combining MC simulations and the critical polynomial method. 
It is found that the lattice geometry mainly determines $\chi_c$ for one-patch particles,
and that the symmetry of patches and lattices significantly affects $\chi_c$ for particles with more patches.
Furthermore, with the above numerical estimates of $\chi_c$, by analyses related to symmetry,
we give values of $\chi_c$ for disks with an arbitrary number of patches on all $11$ Archimedean lattices.
We also present the rules governing $\chi_c$ values of patchy disks on these lattices.

\subsection{Numerical results for one-patch particles}
\label{sec-one-patch}
\subsubsection{One-patch disks}
\label{sec-one-patch-disk}

\begin{figure}
\includegraphics[width=0.8\linewidth]{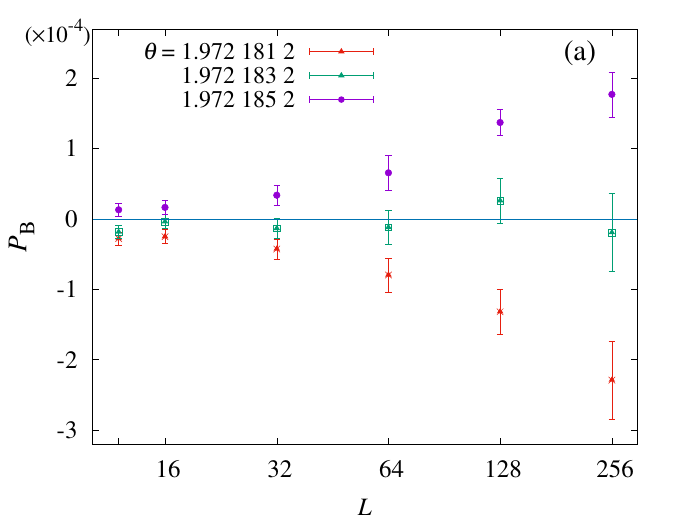} \\
\includegraphics[width=0.8\linewidth]{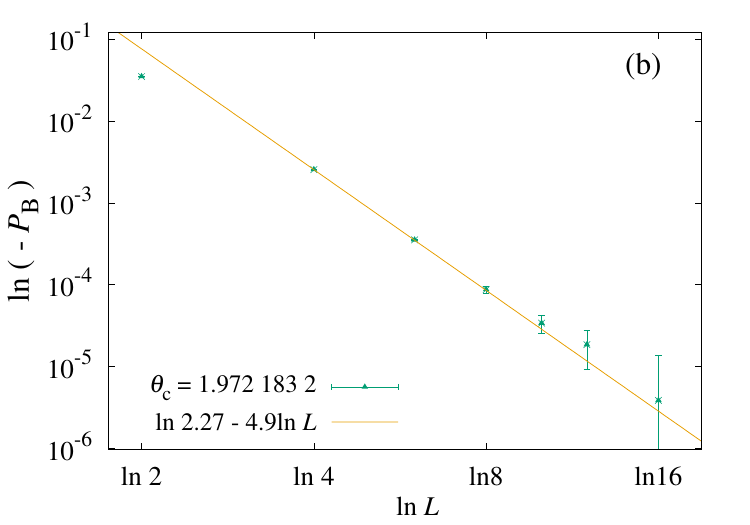}
	\caption{Plots for one-patch disks on the triangular lattice: 
		(a) $\PB$ versus $L$ at $\theta_c=1.972\,183\,2$ and two nearby values of $\theta$. As $\theta$ deviates from $\theta_c$, the curves bend upwards ($\theta>\theta_c$) or downwards ($\theta<\theta_c$). (b) $\ln(-\PB)$ versus $\ln L$ at $\theta_c$ for small sizes $L$. The slope of the straight line is given by the correction exponent $y_1\simeq -4.9$.}
\label{Fig:tri-PB-L}
\end{figure}

For one-patch disks on the triangular lattice,
the plot of $\PB$ versus $\theta$ is shown in Fig.~\ref{Fig:tri-PB}. 
It can be seen that curves for different sizes approximately cross 
near $\theta_c \simeq 1.972\,183$.
To more precisely estimate the percolation threshold $\theta_c$, 
we use Eq.~(\ref{eq:PBfitting}), with $p$ replaced by $\theta$, to fit the data of $\PB$ according to the least-square criterion.
When performing the fits, we gradually increase $L_{\rm min}$ and exclude data points for $L < L_{\rm min}$. 
In general, the fit results are acceptable, only when $\chi^2$ is less than or close to the degree of freedom (DOF), 
and the decrease of $\chi^2$ per DOF does not exceed one when increasing $L_{\rm min}$.
Fits are made with fixed $y_t=3/4$ and the results are shown in Table~\ref{Tab:PB-fit}. 
When setting $b_2=0$, the fit results show that the leading irrelevant exponent is $y_1 \simeq -4.9$. 
When $y_2$ is fixed at $-6$, $-7$ or $-8$, the fit results are also consistent with $y_1 \simeq -4.9$.
Thus we put our final estimate as $y_1 = -4.9(3)$. 
And from Table~\ref{Tab:PB-fit}, the percolation threshold of one-patch disks on the triangular lattice is 
estimated to be $\theta_c=1.972\,183\,20(8)$, i.e., $\chi_c=0.627\, 765\, 41(3)$. 
All error bars of quantities in this work can be regarded as one-sigma.

To demonstrate the above estimates of $\theta_c$, we plot in Fig.~\ref{Fig:tri-PB-L}(a) $\PB$ versus $L$
at $\theta_c$ and two nearby values. It can be seen that when $\theta$ deviates from $\theta_c$,
the curves bend upwards ($\theta>\theta_c$) or downwards ($\theta<\theta_c$) as the size $L$ increases.
Since $\PB \simeq b_1 L^{y_1}$ at $\theta_c$, the estimate of $y_1 \simeq -4.9$ is illustrated 
in Fig.~\ref{Fig:tri-PB-L}(b) as the slope of the straight line.

\begin{figure}[ht]
\begin{center}
\includegraphics[scale=0.37]{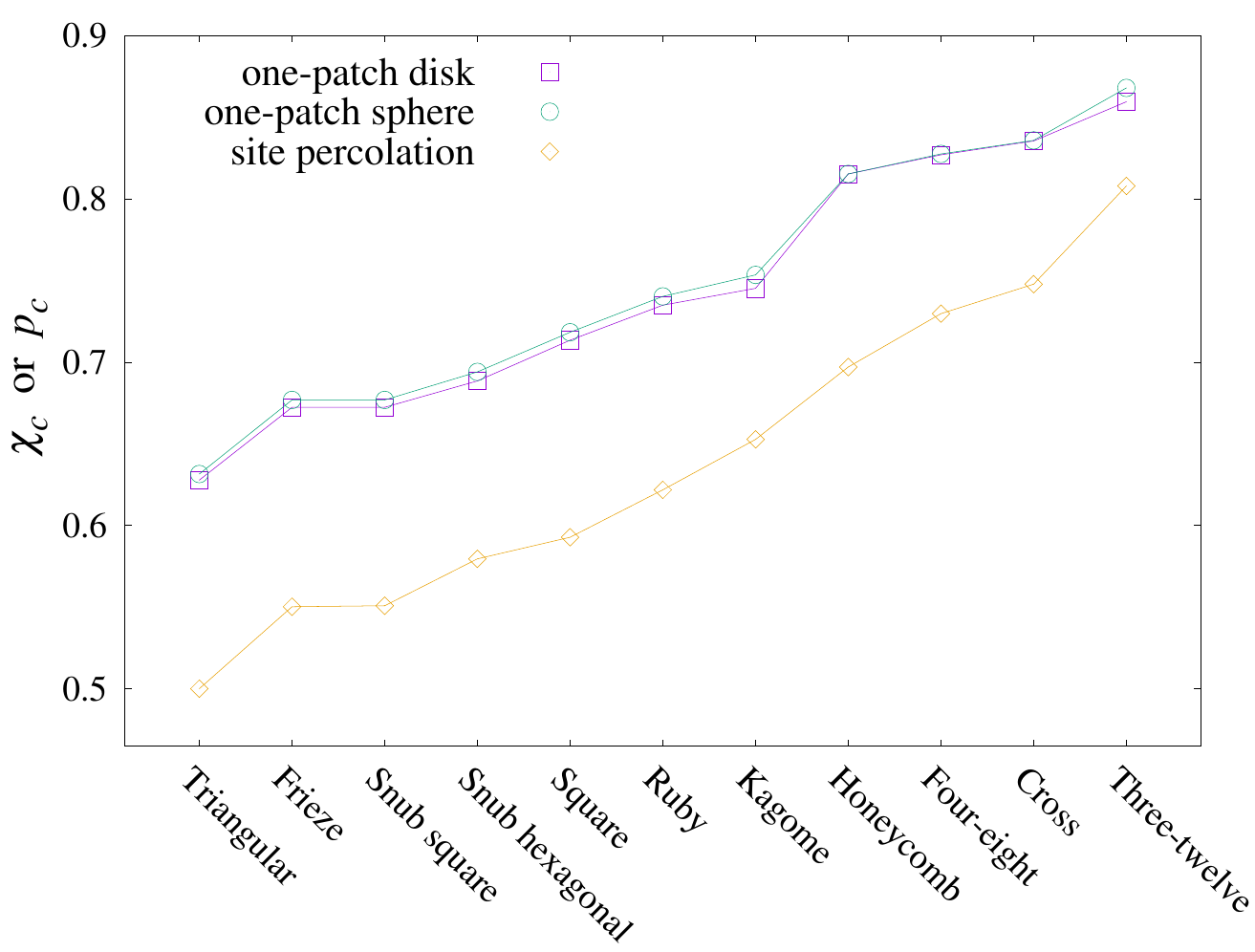}
	\caption{Plots of percolation thresholds $\chi_c$ for one-patch disks and spheres, and $p_c$ for site percolation. 
	The values of the thresholds are summarized in Table~\ref{Tab:chi-threshold}, where $p_c$ values come from 
	the cited references.  
	The lines are added to guide the eyes.}
\label{Fig:one-patch}
\end{center}
\end{figure}

For one-patch disks on other $10$ Archimedean lattices, we also conduct similar analyses for the simulation data.
Plots of $\PB$ versus $\theta$ are shown in Fig.~S12 of the SM~\cite{sup-mat}.
Fits of the data also yield precise values of the percolation threshold $\chi_c$ for these lattices, 
as summarized in Table~\ref{Tab:chi-threshold}.
By observing the above $\chi_c$ values for one-patch disks on all $11$ Archimedean lattices 
and $p_c$ values for site percolation on same lattices, we find that the two sets of values follow 
the same order, as plotted in Fig.~\ref{Fig:one-patch}.
The order can be regarded as from lattices with large coordination numbers to those with small coordination numbers,
and when the coordination numbers are the same, it is from lattices with small variances of angles (around a vertex) 
to those with large variances. 
Thus, similar to $p_c$ for site percolation, $\chi_c$ for one-patch disks on Archimedean lattices 
are mainly influenced by the lattice geometry.

\begin{table*}
\begin{center}
\caption{Percolation thresholds ${\chi}_c$ of patchy particles on Archimedean lattices. 
	The site-percolation thresholds $p_c$ for these lattices are also included for comparison.}
\label{Tab:chi-threshold}
    \renewcommand\arraystretch{1.18}
    \setlength{\tabcolsep}{1.8mm}{
    \begin{tabular}[t]{llllll}
    \hline
    \hline
        & Lattice & Triangular & Frieze & Snub square & Snub hexagonal   \\
    \hline
        & One-patch & $0.627\,765\,41(3)$ & $0.672\,338\,8(1)$ & $0.672\,346\,35(4)$ & $0.688\,526\,01(4)$ \\
        & Two-patch & $0.554\,469\,9(4)$ & $0.624\,383\,7(6)$ & $0.622\,832\,9(4)$ & $0.625\,385\,2(6)$  \\
   \multirow{2}*{Disk} & Three-patch & $ 0.558\,806\,6(7)$ & $0.623\,505\,(1)$ & $0.620\,411\,9(5)$ & $0.617\,753\,2(7)$ \\
        & Four-patch & $0.554\,469\,2(4)$ & $0.645\,671\,6(5)$ & $0.670\,484\,3(5)$ & $0.625\,384\,5(6)$ \\
        & Five-patch & $0.627\,765\,6(4)$ & $0.658\,762\,8(6)$ & $0.627\,557\,4(8)$ & $0.688\,525\,6(5)$ \\
        & Six-patch & $0.500\,000\,1(5)$ & $0.692\,899\,0(9)$ & $0.756\,361\,(1)$ & $0.579\,497\,9(9)$ \\
        \hline
   \multirow{2}*{Sphere} & One-patch & $0.631\,475\,6(3)$ & $0.676\,869\,4(4)$ & $0.676\,877\,1(3)$ & $0.694\,042\,7(3)$ \\
              & Two-patch & $0.532\,379\,5(7)$ & $0.606\,503\,6(4)$ & $0.605\,554\,2(5)$ & $0.611\,814\,2(6)$ \\
        \hline
   \multicolumn{2}{l}{\;\;\;\;\;\;\;\;$p_c$} & $1/2$ & $0.550\,213\,(3)$~\cite{Suding1999} & $0.550\,806\,(3)$~\cite{Suding1999} & $0.579\,498\,(3)$~\cite{Suding1999} \\
    \hline
    \hline
	    & Lattice & Square & Ruby & Kagome & Honeycomb  \\
    \hline
        & One-patch & $0.713\,444\,50(3)$ & $0.734\,894\,0(1)$ & $0.745\,229\,66(5)$ & $0.815\,301\,86(3)$ \\
        & Two-patch & $0.676\,345\,5(3)$ & $0.717\,490\,7(3)$ & $0.687\,495\,0(2)$ & $0.815\,301\,6(3)$ \\
   \multirow{2}*{Disk} & Three-patch & $0.713\,444\,6(3)$ & $0.712\,619\,8(6)$ & $0.725\,743\,3(6)$ & $0.697\,040\,4(9)$ \\
        & Four-patch & $0.592\,746\,5(4)$ & $0.775\,605\,0(8)$ & $0.687\,494\,9(4)$ & $0.815\,301\,9(3)$ \\
        & Five-patch & $0.713\,444\,4(8)$ & $0.726\,257\,4(7)$ & $0.745\,229\,4(8)$ & $0.815\,301\,8(5)$ \\
        & Six-patch & $0.676\,345\,4(3)$ & $0.764\,013\,5(7)$ & $0.652\,701\,(2)$ & $0.697\,040\,3(8)$ \\
     \hline
   \multirow{2}*{Sphere} & One-patch & $0.718\,297\,8(6)$ & $0.740\,310\,8(3)$ & $0.753\,481\,7(3)$ & $0.815\,301\,9(3)$ \\
        & Two-patch & $0.657\,338\,0(3)$ & $0.706\,489\,1(8)$ & $0.670\,797\,2(7)$ & $0.806\,134\,8(4)$ \\
     \hline
   \multicolumn{1}{l}{ } \multirow{2}*{$p_c$}  & & $0.592\,746\,050$ & $0.621\,812\,07(7)$~\cite{Jacobsen2014} & $1-2\sin(\pi/18)$ & $0.697\,040\,230 (5)$~\cite{Jacobsen2014} \\
     \multicolumn{2}{c}{}  & $792\,10(2)$~\cite{Jacobsen2015}  &   & $=0.652\,703\,644...$~\cite{Sykes1964}  &   \\
     \hline
    \hline
        &Lattice & Four-eight & Cross & Three-twelve & \\
    \hline	
        & One-patch & $0.827\,011\,22(3)$ & $0.835\,468\,95(7)$ & $0.859\,494\,83(5)$ & \\
        & Two-patch & $0.827\,010\,8(2)$ & $0.835\,469\,0(4)$ & $0.859\,495\,2(4)$ & \\
   \multirow{2}*{Disk} & Three-patch & $0.815\,649\,0(4)$ & $0.839\,888\,2(5)$ & $0.859\,495\,3(5)$ & \\
        & Four-patch & $0.856\,560\,1(4)$ & $0.865\,225\,2(7)$ & $0.859\,494\,7(5)$ & \\
        & Five-patch & $0.815\,649\,6(4)$ & $0.821\,517\,4(7)$ & $0.843\,143\,7(8)$ & \\
        & Six-patch & $0.827\,011\,3(3)$ & $0.839\,888\,4(5)$ & $0.903\,950\,3(5)$ & \\
    \hline
 \multirow{2}*{Sphere} & One-patch & $0.827\,533\,2(2)$ & $0.835\,822\,4(4)$ & $0.867\,995\,5(3)$ & \\
        & Two-patch & $0.818\,019\,5(4)$ & $0.826\,182\,5(5)$ & $0.852\,055\,5(5)$ & \\
    \hline
	    \multicolumn{1}{c}{ } \multirow{2}*{$p_c$} & & $0.729\,723\,2(5)$~\cite{Jacobsen2014} & $0.747\,800\,2(2)$~\cite{Jacobsen2014} & ${[1-2\sin{(\pi/18)}]}^{1/2}$ & \\	
     \multicolumn{1}{c}{ } &   &   &   & $=0.807\,900\,764...$~\cite{Suding1999}  & \\
    \hline
    \hline
    \end{tabular}}
\end{center}
\end{table*}

The threshold values of one-patch disks on the frieze and snub square lattices are very close 
since the two lattices share the same coordination number and variance of angles. 
We find that their order can also be understood from geometry as follows.
The patch of a disk can cover several neighboring edges connecting to the disk center,
and various patch-covering structures occur with different probabilities, 
which can be obtained by analyzing the configurations as exemplified by the calculations for the frieze lattice in the SM~\cite{sup-mat}.
In Table~\ref{Tab:proba-expression}, we show these probabilities for patch sizes $\chi$ 
near $\chi_c$ of the two lattices.
From the table, it is observed that $3$-edge patch-covering occurs with same probabilities
on the two lattices, as well as $4$-edge patch-covering.
However, for $3$-edge patch-covering, the frieze lattice has a structure with two adjacent $\pi/2$-angles;
and for $4$-edge covering, the frieze lattice has two structures (both occurring with probability $\chi-2/3$) 
with two adjacent $\pi/2$-angles, while the snub square lattice only has one structure 
(occurring with probability $\chi-2/3$) with two $\pi/2$-angles. Since structures with two adjacent $\pi/2$-angles 
are more open than other structures, the frieze lattice has a slightly lower percolation threshold than
the snub square lattice, which is confirmed by our numerical estimates in Table~\ref{Tab:chi-threshold}.

\begin{table*}[htbp]
\begin{center}
\caption{Probabilities of different patch-covering structures of a particle, for one-patch disks on the frieze and snub square lattices, 
	as a function of $\chi$ near estimated $\chi_c$ of the two lattices. 
        Bold lines indicate that corresponding edges are covered by the patch of a disk placed on the central vertex.	
	Some details for calculating these probabilities are given in the SM~\cite{sup-mat}.
	}
\label{Tab:proba-expression}
    \renewcommand\arraystretch{1.45}
    \setlength{\tabcolsep}{1.8mm}{
    \begin{tabular}[t]{lllll}
    \hline
    \hline
       \multirow{1}*{Lattice} & \multirow{1}*{Type} & \multirow{1}*{Structure} & \multirow{1}*{Probability} & Probability Sum \\
   \hline
	       & \multirow{2}*{$3$-edge} & \includegraphics[scale=0.2]{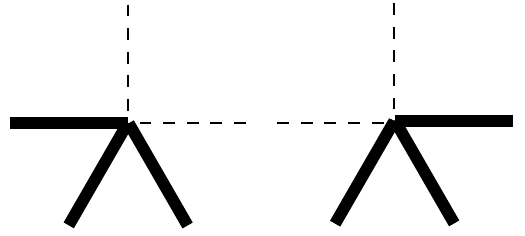} & \multirow{1}*{$2(3/4-\chi)$} & \multirow{2}*{\vspace{-4mm}$4-5{\chi}$} \\
	   Frieze    &   & \includegraphics[scale=0.2]{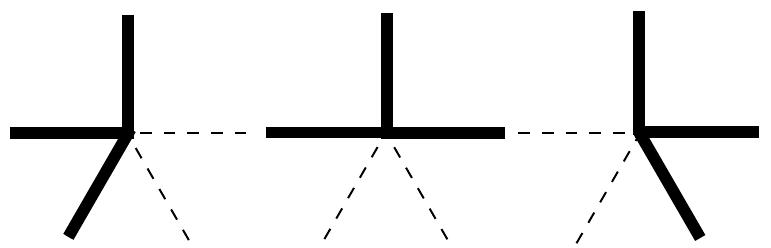} & \multirow{1}*{$3(5/6-\chi)$} &  \\  \cline{2-5}
	    & \multirow{4}*{$4$-edge}  & \includegraphics[scale=0.2]{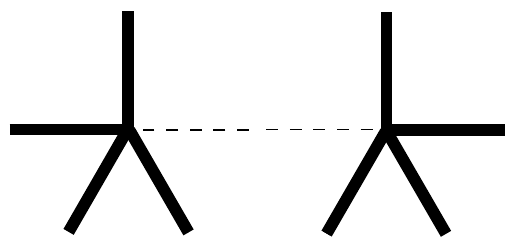} & \multirow{1}*{$2(\chi-7/12)$} & \multirow{4}*{\vspace{-3.5mm}$5{\chi}-3$} \\
               &  & \includegraphics[scale=0.2]{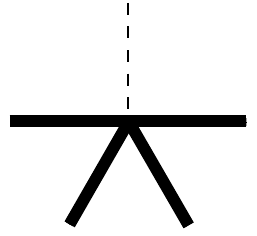}  & \multirow{1}*{$\chi-1/2$}  &  \\ 
	       &   & \includegraphics[scale=0.2]{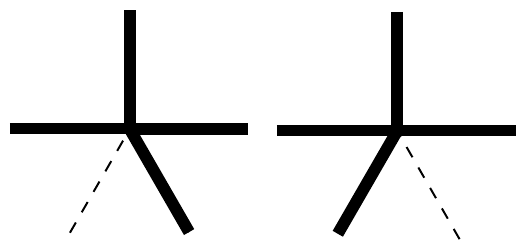}  & \multirow{1}*{$2(\chi-2/3)$}  &  \\
    \hline
	    & \multirow{2}*{$3$-edge} & \includegraphics[scale=0.2]{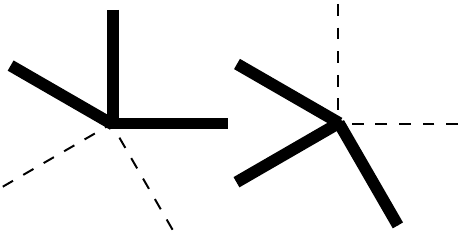} & \multirow{1}*{$2(3/4-\chi)$} & \multirow{2}*{\vspace{-4mm}$4-5{\chi}$} \\
	    Snub square  &   & \includegraphics[scale=0.2]{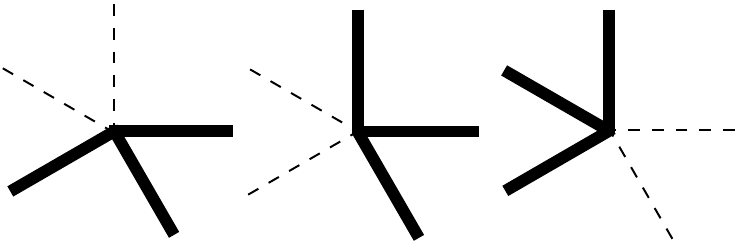} & \multirow{1}*{$3(5/6-\chi)$} &  \\  \cline{2-5}
	    & \multirow{4}*{$4$-edge}  & \includegraphics[scale=0.2]{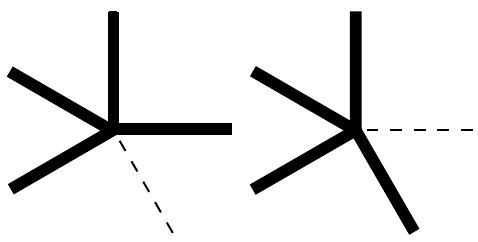} & \multirow{1}*{$2(\chi-7/12)$} & \multirow{4}*{\vspace{-3.5mm}$5{\chi}-3$} \\
	       &  & \includegraphics[scale=0.2]{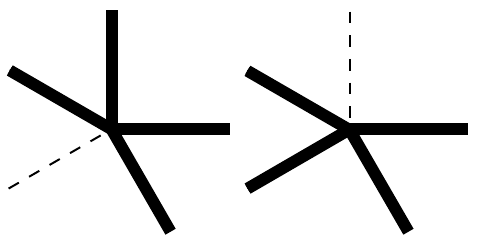}  & \multirow{1}*{$2(\chi-7/12)$}  &  \\ 
               &   & \includegraphics[scale=0.2]{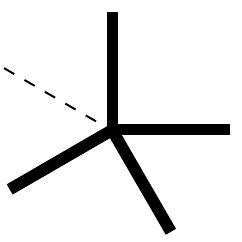}  & \multirow{1}*{$\chi-2/3$}  &  \\    
    \hline
    \hline
    \end{tabular}}
\end{center}
\end{table*}

\begin{table}[htbp]
\begin{center}
\caption{Results of $y_1$ from fitting the data of the critical polynomial $\PB$, 
	for one-patch disks or spheres on different Archimedean lattices.
	Dashed entries indicate that the values cannot be determined with
	the current data.}
\label{Tab:y1}
    \renewcommand\arraystretch{1.16}
    \setlength{\tabcolsep}{3.8mm}{
    \begin{tabular}[t]{lll}
    \hline
    \hline
	    Lattice & $y_1$(disk) & $y_1$(sphere) \\
    \hline
       Triangular & $-4.9(3)$ & $-3.1(1)$ \\
       Frieze & $-3.3(1)$ & $-3.3(1)$ \\
       Snub square & $-3.7(1)$ & $-3.9(1)$ \\
       Snub hexagonal & ~~~--- & ~~~--- \\
       Square & $-3.3(4)$ & $-3.5(2)$ \\
       Ruby & ~~~--- & ~~~--- \\
       Kagome & $-5.2(4)$ & $-5.4(1)$ \\
       Honeycomb & $-3.7(1)$ & $-4.2(1)$ \\
       Four-eight & $-4.7(4)$ & $-3.9(1)$ \\
       Cross & ~~~--- & ~~~--- \\
       Three-twelve & $-3.1(2)$ & ~~~--- \\
    \hline
    \hline
    \end{tabular}}
\end{center}
\end{table}

Besides $\chi_c$, we also get estimates of the leading irrelevant exponent $y_1$ for one-patch disks 
on other Archimedean lattices. The results of $y_1$ are summarized in Table~\ref{Tab:y1}. 
For lattices whose values of $y_1$ are absent in the table, the finite corrections are also small,
but the current data are not sufficient for determining $y_1$.
The presented estimates of $y_1$ are close to $y_1=y_t-\Delta_1=-3.25$ or $-5.25$ 
for unsolved bond percolation models on Archimedean lattices~\cite{Scullard2020}.
Besides the triangular lattice, we also make plots of $\PB$ versus $L$ at the estimated thresholds 
for one-patch disks on other lattices, as shown in Fig. S20 of the SM~\cite{sup-mat}, 
which demonstrate the scaling $\PB(\theta_c,L) \simeq b_1 L^{y_1}$ with the estimated 
values of $y_1$.

\subsubsection{One-patch spheres}
To observe the effect of particle shape on the percolation threshold, we also conduct simulations for 
one-patch spheres on Archimedean lattices. Similar analyses are performed for the simulation data.
The resulting threshold values $\chi_c$ are also summarized in Table~\ref{Tab:chi-threshold}
and plotted in Fig.~\ref{Fig:one-patch}. 
It can be seen that, except on the honeycomb lattice, $\chi_c$ values for one-patch spheres 
are slightly larger than those for one-patch disks,
and the calculated differences [$(\chi_c^{\rm sphere}-\chi_c^{\rm disk})/\chi_c^{\rm disk}$] 
are at most $1.1\%$. On the honeycomb lattice, within error bars, one-patch disks and spheres 
are found to share the same threshold. 
The thresholds $\chi_c$ for one-patch spheres on different lattices also follow the same order as
that for one-patch disks, implying that for one-patch spheres $\chi_c$ are also mainly influenced 
by the lattice geometry. 

We also get the estimates of $y_1$ for one-patch spheres on the lattices,
as summarized in Table~\ref{Tab:y1}. The values of $y_1$ for one-patch spheres and disks 
are very close for most lattices. 
Plots of $\PB$ versus $\theta$ or $L$ for one-patch spheres are presented in Fig.S13 and S21 of the SM~\cite{sup-mat}, 
which demonstrate our estimates of $\chi_c$ and $y_1$.

\begin{figure}[hb]
\begin{center}
\includegraphics[scale=1.2]{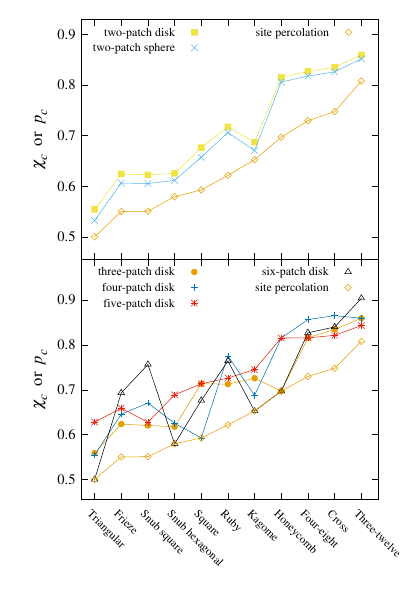}
	\caption{Plots of percolation thresholds $\chi_c$ for particles with two to six patches, and $p_c$ for site percolation, 
	on $11$ Archimedean lattices. The values of the thresholds are summarized in Table~\ref{Tab:chi-threshold}, 
	where $p_c$ values come from the cited references.
	The lines are added to guide the eyes.}
\label{Fig:Janus}
\end{center}
\end{figure}

\subsection{Numerical results for particles with two to six patches}
\label{sec:2-6}

We also conduct MC simulations 
for disks with two to six patches and spheres with two patches, on all $11$ Archimedean lattices.
Plots of $\PB$ versus $\theta$ are shown in Fig.~S14 to S19 of the SM~\cite{sup-mat}, 
in which approximate intersections of lines for different sizes  can be observed. 
By fitting the data of $\PB$ near the intersection points, we also obtain precise percolation thresholds 
for these models, as summarized in Table~\ref{Tab:chi-threshold}. 
These threshold values are plotted together with those for site percolation, as in Fig.~\ref{Fig:Janus}. 

\subsubsection{Two-patch particles} 
From plots for two-patches particles (disks or spheres) in Fig.~\ref{Fig:Janus}, 
comparing with plots for one-patches particles in Fig.~\ref{Fig:one-patch},
an obvious feature is that the curves become non-monotonic.
The local minimum appears at the kagome lattice, for which the symmetry of 
the two triangles connected to a vertex matches the symmetry of the two patches
on a particle. Thus symmetry become very important in determining $\chi_c$ values 
of two-patches particles. 

From Table~\ref{Tab:chi-threshold}, if one compares the $\chi_c$ values for two-patch particles with those for one-patch particles in more detail,
it can be found that, while models of one-patch particles have slightly lower $\chi_c$ values on the frieze lattice
than on the snub square lattice, models of two-patches particle have slightly higher $\chi_c$ values 
on the frieze lattice than on the snub square lattice. This can also be understood by calculating 
probabilities of different patch-covering structures of a particle, similar to calculations leading to 
Table~\ref{Tab:proba-expression}. For two-patch particles on the snub square lattice, the symmetry of two patches 
on the particle approximately matches the symmetry of two squares (or two triangles) connected to a vertex,
which causes the $\chi_c$ value on the snub square lattice to be slightly lower than that on the frieze lattice.

In Sec.~\ref{sec-one-patch}, for one-patch particles, it is found that differences 
of $\chi_c$ between disks and spheres are very small.  However, for two-patch particles, 
in Fig.~\ref{Fig:Janus}, it is seen that on most Archimedean lattices the differences 
of $\chi_c$ between disks and spheres are significantly larger than those for one-patch particles.
And on a given lattice, while one-patch spheres has slightly higher $\chi_c$ than one-patch disks
(except that on the honeycomb lattice one-patch disks and spheres share the same $\chi_c$), 
two-patch spheres have lower $\chi_c$ than two-patch disks. 
Therefore, though not affecting the order of $\chi_c$ values for different lattices,
the particle shape is still important in determining the values of $\chi_c$.

\begin{figure}
\begin{center}
\includegraphics[scale=0.68]{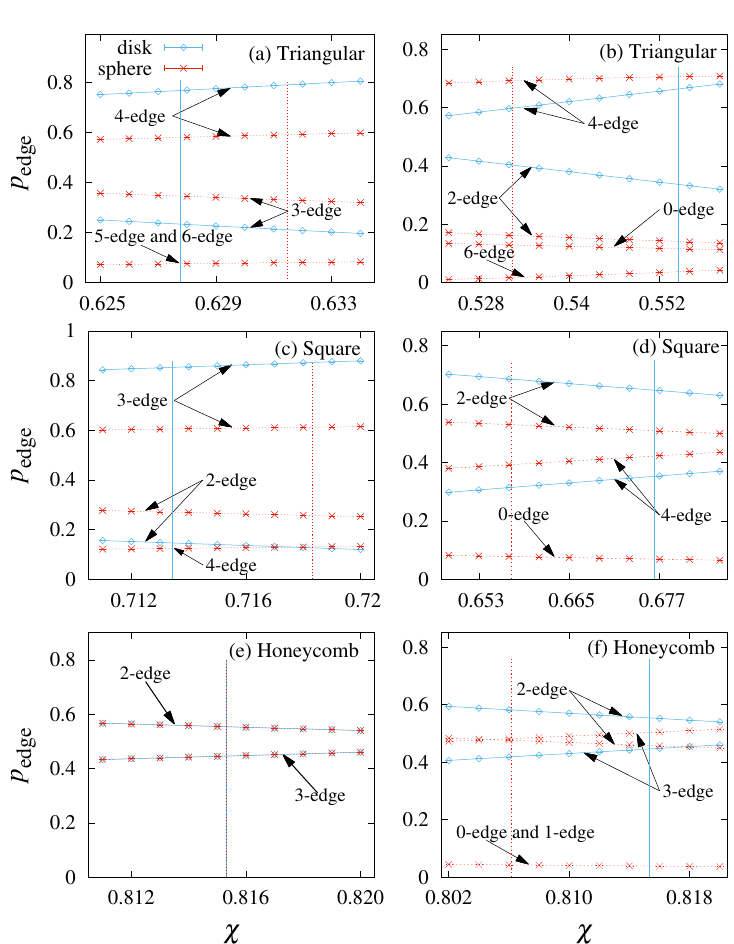}
	\caption{Probabilities of different patch-covering structures of a particle for $\chi$ near $\chi_c$,
	on the triangular, square and honeycomb lattices. 
	Plots (a), (c) and (e) are for one-patch particles,
	and plots (b), (d) and (f) are for two-patch particles.
	Vertical solid and dashed lines indicate values of $\chi_c$ for the disks and spheres, respectively.
	For one-patch disks and spheres on the honeycomb lattice, the probabilities are same, which leads 
	to the equality of $\chi_c$ for spheres and disks.}
\label{Fig:edge}
\end{center}
\end{figure}

To understand the change of $\chi_c$ values between one-patch and two-patch particles,
for three regular lattices (triangular, square and honeycomb), 
we plot the probabilities of different patch-covering structures of a particle for $\chi$ near $\chi_c$, 
as in Fig.~\ref{Fig:edge}.  From Fig.~\ref{Fig:edge}(a) and (c), it can be seen that, for one-patch particles, 
the lower value of $\chi_c$ for disks (comparing with that for spheres) is correlated with 
higher probabilities of large-edge patch-covering  structures, e.g., $4$-edge 
and $3$-edge for the triangular and square lattices, respectively.
From Fig.~\ref{Fig:edge}(b), (d) and (f), one also see that, for two-patch particles, 
the lower value of $\chi_c$ for spheres (comparing with that for disks) is also correlated with 
higher probabilities of large-edge patch-covering structures, e.g., $4$ and $6$-edge 
for the triangular lattice, $4$-edge for the square lattice, and $3$-edge for the honeycomb lattice.

\subsubsection{Three- to six-patch disks} 
For disks with three to six patches, the role of symmetry is more explicitly exhibited. 
From Fig.~\ref{Fig:Janus}, it can be seen that curves for these patchy particles are all non-monotonic.
Similar to that for two-patch particles at the kagome lattice, the local minimum of $\chi_c$ for four-patch disks
at the kagome lattice and for five-patch disks at the snub square lattice can also be understood by the approximate matching 
of the symmetry of patches on a particle and the symmetry of the lattice.
From Fig.~\ref{Fig:Janus} and Table~\ref{Tab:chi-threshold},
it is interesting to find that $\chi_c$ values of some models are numerically equal to $p_c$ values of site percolation
on same lattices, such as three-patch disks on the honeycomb lattice, four-patch disks on the square lattice,
six-patch disks on the triangular, snub hexagonal, kagome, and honeycomb lattices.
For each of these models, the symmetry of the patches on a particle perfectly matches the symmetry of the lattice,
which leads to the fact that patches of a particle either cover all edges connected to the particle center
with probability $\chi$ or cover no edge with probability $1-\chi$. Thus the models are exactly equivalent to
the corresponding site percolation models with $\chi_c=p_c$.
For other models considered, at $\chi=p_c$, the probability of all-edge patch-covering structure is smaller than $p_c$
and other patch-covering structures lead to lower connection. Therefore, their $\chi_c$ values should be larger
than $p_c$ for site percolation on same lattices, which is supported by results in Table~\ref{Tab:chi-threshold}
and can be easily seen in Figs.~\ref{Fig:one-patch} and ~\ref{Fig:Janus}.

From Table~\ref{Tab:chi-threshold}, it is also interesting to see that, 
within error bars, $\chi_c$ of some different patchy particles on a given lattice share the same value.
These include: 
\begin{itemize}
\item on the triangular (or snub hexagonal, or kagome) lattice, 
$\chi_c$ of one- and five-patch disks, $\chi_c$ of two- and four-patch disks;
\item on the square lattice, $\chi_c$ of one-, three- and five-patch disks, 
and $\chi_c$ of two- and six-patch disks;
\item on the honeycomb lattice, $\chi_c$ of one-, two-, four-, five-patch disks, and of one-patch spheres, 
and $\chi_c$ of three- and six-patch disks;
\item on the four-eight lattice, $\chi_c$ of one-, two- and six-patch disks,
and $\chi_c$ of three- and five-patch disks;
\item on the cross lattice, $\chi_c$ of one- and two-patch disks,
and $\chi_c$ of three- and six-patch disks;
\item on the three-twelve lattice, $\chi_c$ of one- to four-patch disks.
\end{itemize}
The above equalities of $\chi_c$ values can be understood 
by calculating probabilities of different patch-covering structures of a particle for $\chi$ near $\chi_c$. 
For example, on the honeycomb lattice, for one-patch spheres,
it can be verified numerically that, the patch-covering probabilities are equal to those 
for one-patch disks [e.g., see Fig.~\ref{Fig:edge}(e)]. 
For patchy disks, similar to results in Table~\ref{Tab:proba-expression}, by analytical calculations
we can obtain the probabilities of different patch-covering structures of a particle as a function 
of $\chi$ near $\chi_c$, as presented in the following section.
Since percolation of patchy particles can be understood as connection of patch-covering structures
of particles at the vertices, and there is only a single percolation threshold for a given model, same expressions (as functions of $\chi$) 
for these probabilities near $\chi_c$ can prove that these equalities of $\chi_c$ values hold exactly. 

\begin{table*}[htbp]
\begin{center}
\caption{Probabilities of different patch-covering structures as a function of $\chi$ near $\chi_c$, for $n$-patch disks on the honeycomb, square, triangular, snub hexagonal, kagome and four-eight lattices.
	 For each lattice, the last row shows estimates of percolation thresholds $\chi_c$ by combining numerical estimates in Table~\ref{Tab:chi-threshold}.
         *Regarding $2$-edge structures on the kagome lattice, for $n$-patch disks with$\mod(n,6)=3$, the two edges have a $2\pi/3$ angle, while for disks with$\mod(n,6)=2$ or $4$, the two edges are on the same straight line.}
\label{Tab:contact-n}
    \footnotesize
    \renewcommand\arraystretch{1.55}
    \setlength{\tabcolsep}{1.0mm}{
    \begin{tabular}[t]{llllllll}
    \hline
    \hline
	       & \multirow{2}*{Type} & $\mod(n,3)$ & & & &  \\
               & & $1$ & $2$ &  & &  \\
    \cline{2-8}
      \multirow{1}*{Honeycomb} & {$0$}-edge &   &   &  & & & \\
                               & {$2$}-edge & $3-3{\chi}$ & $3-3{\chi}$ &  & & \\
                               & {$3$}-edge    & $3{\chi}-2$ & $3{\chi}-2$ &  & & & \\
    \cline{2-8}
            &   $\chi_c$ & $0.815\,301\,8(3)$ & $0.815\,301\,8(3)$ &   \\
    \hline
	       & \multirow{2}*{Type} & $\mod(n,4)$ & & & & &  \\
               & & $1$ & $2$ & $3$ &  & &  \\
    \cline{2-8}
    \multirow{2}*{Square}  & {$0$}-edge &      &       &     &  &    &  \\
               & {$2$}-edge & $3-4{\chi}$ & $2-2{\chi}$ & $3-4{\chi}$ & & &    \\
               & {$3$}-edge & $4{\chi}-2$ &  &  $4{\chi}-2$  &    &  &  \\
               & {$4$}-edge &     & $2{\chi}-1$ &     &  &    &  \\
    \cline{2-8}
             &  $\chi_c$ & $0.713\,444\,5(4)$ & $0.676\,345\,5(4)$ & $0.713\,444\,5(4)$ &       \\
    \hline
               & \multirow{2}*{Type} & $\mod(n,6)$ & & & & &  \\
               & & $1$ & $2$ & $3$ & $4$ & $5$ &   \\
    \cline{2-8}

    Triangular           & {$2$}-edge &   & $2-3{\chi}$  &   &  $2-3{\chi}$ &   &     \\
               & {$3$}-edge & $4-6{\chi}$ &  & $2-2{\chi}$  &  & $4-6{\chi}$ &   \\
               & {$4$}-edge & $6{\chi}-3$ & $3{\chi}-1$ &   & $3{\chi}-1$ & $6{\chi}-3$ &   \\
               & {$6$-edge} &   &   & $2{\chi}-1$  &   &   &    \\
    \cline{2-8}
            &   $\chi_c$ & $0.627\,765\,5(2)$ & $0.554\,469\,6(4)$ & $0.558\,806\,6(7)$ & $0.554\,469\,6(4)$ & $0.627\,765\,5(2)$ &  \\
    \hline
               & \multirow{2}*{Type} & $\mod(n,6)$ & & & & &  \\
               & & $1$ & $2$ & $3$ & $4$ & $5$ &   \\
    \cline{2-8}

         & {$1$}-edge &   & $2/3-{\chi}$  &   & $2/3-{\chi}$  &   &   \\
         Snub hexagonal      & {$2$}-edge &   & $4/3-2{\chi}$  & $1-{\chi}$  &  $4/3-2{\chi}$ &   & \\
               & {$3$}-edge & $10/3-4{\chi}$ & $2{\chi}-2/3$ & $1-{\chi}$  & $2{\chi}-2/3$ & $10/3-4{\chi}$ &  \\
               & {$4$}-edge & $3{\chi}-5/3$ & ${\chi}-1/3$ &   & ${\chi}-1/3$ & $3{\chi}-5/3$ &   \\
               & {$5$}-edge & ${\chi}-2/3$  &   & $2{\chi}-1$  &   & ${\chi}-2/3$  &    \\
    \cline{2-8}
               & $\chi_c$ & $0.688\,525\,8(3)$ & $0.625\,384\,9(7)$ & $0.617\,753\,2(7)$ & $0.625\,384\,9(7)$  & $0.688\,525\,8(3)$ &  \\
    \hline
               & \multirow{2}*{Type}& $\mod(n,6)$ & & & & &  \\
               & & $1$ & $2$ & $3$ & $4$ & $5$ &   \\
    \cline{2-8}
Kagome
               & {$2$}-edge & $5/3-2{\chi}$  & $2-2{\chi}$  & $(2-2{\chi})^*$  &  $2-2{\chi}$ & $5/3-2{\chi}$  &  \\
               & {$3$}-edge & $2/3$ &  &   &  & $2/3$ &   \\
               & {$4$}-edge & $2{\chi}-4/3$ & $2{\chi}-1$ & $2{\chi}-1$  & $2{\chi}-1$ & $2{\chi}-4/3$ &  \\
    \cline{2-8}
               & $\chi_c$ & $0.745\,229\,5(5)$ & $0.687\,494\,9(4)$ & $0.725\,743\,3(6)$ & $0.687\,494\,9(4)$ & $0.745\,229\,5(5)$ &  \\
    \hline
               & \multirow{2}*{Type} & $\mod(n,8)$ & & & &  \\
               & & $1$ or $7$  & $2$ or $6$ & $3$ or $5$ & $4$ &  \\
    \cline{2-8}
      \multirow{2}*{Four-eight} & {$1$}-edge &   &   & $7/4-2{\chi}$ & $1-{\chi}$ & & \\
                               & {$2$}-edge & $3-3{\chi}$ & $3-3{\chi}$ & ${\chi}-1/2$ & $1-{\chi}$ & \\
                               & {$3$}-edge    & $3{\chi}-2$ & $3{\chi}-2$ & ${\chi}-1/4$ & $2{\chi}-1$ & & \\
    \cline{2-8}
            &   $\chi_c$ & $0.827\,011\,0(1)$ & $0.827\,011\,0(1)$ & $0.815\,649\,3(4)$ & $0.856\,560\,1(4)$ \\
    \hline
    \hline
    \end{tabular}}
\end{center}
\end{table*}

\begin{table*}[htbp]
\begin{center}
\caption{Probabilities of different patch-covering structures as a function of $\chi$ near $\chi_c$, for $n$-patch disks on the frieze, snub square, ruby, cross and three-twelve lattices.
	 For each lattice, the last row shows estimates of percolation thresholds $\chi_c$ by combining numerical estimates in Table~\ref{Tab:chi-threshold}.
	 *On the cross lattice, the $1$ and $2$-edge structures for $n$-patch disks with $\mod(n,12)=4$ are different from those for disks with $\mod(n,12)=3$ or $6$.}
\label{Tab:contact-n-b}
    \footnotesize
    \renewcommand\arraystretch{1.55}
    \setlength{\tabcolsep}{1.0mm}{
    \begin{tabular}[t]{llllllll}
    \hline
    \hline
               & \multirow{2}*{Type}& $\mod(n,12)$ & & & & &  \\
               & & $1$ or $11$ & $2$ or $10$ & $3$ or $9$ & $4$ or $8$ & $5$ or $7$ & $6$  \\
    \cline{2-8}
           & {$1$}-edge &   &   &   & $4/3-2{\chi}$  & $4/3-2{\chi}$  & $1-{\chi}$  \\
 Frieze    & {$2$}-edge &   & $2-3{\chi}$  & $3/2-2{\chi}$  &  ${\chi}-1/3$ & ${\chi}-1/3$  & \\
               & {$3$}-edge & $4-5{\chi}$ & ${\chi}$ & $1/2$  & $2/3-{\chi}$ & $1/6$ &  \\
               & {$4$}-edge & $5{\chi}-3$ & $2{\chi}-1$ & ${\chi}-1/2$  & $2{\chi}-2/3$ & $1/3$ & $1-{\chi}$  \\
               & {$5$}-edge &   &   & ${\chi}-1/2$  &   & ${\chi}-1/2$  &   $2{\chi}-1$ \\
    \cline{2-8}
               & $\chi_c$ & $0.672\,338\,8(1)$ & $0.624\,383\,7(6)$ & $0.623\,505\,(1)$ & $0.645\,671\,6(5)$  & $0.658\,762\,8(6)$ & $0.692\,899\,0(9)$ \\
    \hline

           & \multirow{2}*{Type} & $\mod(n,12)$ & & & & &  \\
               & & $1$ or $11$ & $2$ or $10$ & $3$ or $9$ & $4$ or $8$ & $5$ or $7$ & $6$  \\
    \cline{2-8}
           & {$0$}-edge &    &    &   &    & $2/3-{\chi}$  &  \\
           & {$1$}-edge &    &    &   &    & $1/6$  &  \\
               & {$2$}-edge &    & $2-3{\chi}$  & $3/2-2{\chi}$  &   & $1/6$  & $1-{\chi}$ \\
  Snub square  & {$3$}-edge & $4-5{\chi}$ & ${\chi}$ & $1-{\chi}$  & $2-2{\chi}$  & $1/6$ & $1-{\chi}$  \\
           & {$4$}-edge & $5{\chi}-3$ & $2{\chi}-1$ & $3{\chi}-3/2$  & $1/3$ & $1/6$ &  \\
           & {$5$}-edge &    &    &   & $2{\chi}-4/3$ & ${\chi}-1/3$ & $2{\chi}-1$ \\
    \cline{2-8}
               & $\chi_c$ & $0.672\,346\,35(4)$ & $0.622\,832\,9(4)$ & $0.620\,411\,9(5)$ & $0.670\,484\,3(5)$ & $0.627\,557\,4(8)$ & $0.756\,361\,(1)$ \\
    \hline
               & \multirow{2}*{Type} & $\mod(n,12)$ & & & & &  \\
               & & $1$ or $11$ & $2$ or $10$ & $3$ or $9$ & $4$ or $8$ & $5$ or $7$ & $6$  \\
    \cline{2-8}
           & {$1$}-edge &    &    & $3/2-2{\chi}$  &    & $3/2-2{\chi}$  &  \\
         Ruby  & {$2$}-edge & $7/3-3{\chi}$  & $5/3-2{\chi}$  & ${\chi}-1/2$  &  $2-2{\chi}$ & ${\chi}-1/3$  & $2-2{\chi}$ \\
           & {$3$}-edge & $2{\chi}-2/3$ & $2/3$ & $1/2$  &  & $1/6$ &   \\
           & {$4$}-edge & ${\chi}-2/3$ & $2{\chi}-4/3$ & ${\chi}-1/2$  & $2{\chi}-1$ & ${\chi}-1/3$ & $2{\chi}-1$ \\
    \cline{2-8}
               & $\chi_c$ & $0.734\,894\,0(1)$ & $0.717\,490\,7(3)$ & $0.712\,619\,8(6)$ & $0.775\,605\,0(8)$ & $0.726\,257\,4(7)$ & $0.764\,013\,5(7)$ \\
    \hline

               & \multirow{2}*{Type} & $\mod(n,12)$ & & & & &  \\
               & & $1$ or $11$ & $2$ or $10$ & $3$ or $9$ & $4$ or $8$ & $5$ or $7$ & $6$ \\
    \cline{2-8}
    \multirow{2}*{Cross}  & {$1$}-edge &      &       &  $1-{\chi}$   & $(1-{\chi})^*$ &  $11/12-{\chi}$  & $1-{\chi}$ \\
                 & {$2$}-edge & $3-3{\chi}$ & $3-3{\chi}$ & $1-{\chi}$ & $(1-{\chi})^*$ & $7/6-{\chi}$ & $1-{\chi}$   \\
                 & {$3$}-edge & $3{\chi}-2$ & $3{\chi}-2$ &  $2{\chi}-1$  &  $2{\chi}-1$  & $2{\chi}-13/12$ & $2{\chi}-1$ \\

    \cline{2-8}
             &  $\chi_c$ & $0.835\,469\,0(2)$ & $0.835\,469\,0(2)$ & $0.839\,888\,3(5)$ & $0.865\,225\,2(7)$  & $0.821\,517\,4(7)$ & $0.839\,888\,3(5)$ \\
    \hline
               & \multirow{2}*{Type} & $\mod(n,12)$ & & & & &  \\
               & & $1$ or $11$ & $2$ or $10$ & $3$ or $9$ & $4$ or $8$ & $5$ or $7$ & $6$ \\
    \cline{2-8}
               & {$1$}-edge &   &   &   &   & $11/6-2{\chi}$  & $1-{\chi}$ \\
   Three-twelve  & {$2$}-edge & $3-3{\chi}$  & $3-3{\chi}$  &  $3-3{\chi}$ &  $3-3{\chi}$ & $1/6$  & $1-{\chi}$    \\
               & {$3$}-edge & $3{\chi}-2$ & $3{\chi}-2$ & $3{\chi}-2$  & $3{\chi}-2$ & $2{\chi}-1$ & $2{\chi}-1$ \\

    \cline{2-8}
            &   $\chi_c$ & $0.859\,495\,0(4)$ & $0.859\,495\,0(4)$ & $0.859\,495\,0(4)$ & $0.859\,495\,0(4)$ & $0.843\,143\,7(8)$ & $0.903\,950\,3(5)$ \\
    \hline
    \hline

    \end{tabular}}
\end{center}
\end{table*}

\subsection{Results for disks with an arbitrary number of patches}
In the previous subsection, it is found that several models of patchy disks are equivalent with site percolation on same lattices.
This result can be generalized to disks with more patches: $n$-patch ($n>0$) disks with $\mod(n,3)=0$ on the honeycomb lattice, 
with $\mod(n,4)=0$ on the square lattice, with $\mod(n,6)=0$ on the triangular, snub hexagonal and kagome lattices,
with $\mod(n,8)=0$ on the four-eight lattice, with $\mod(n,12)=0$ on the frieze, snub square, ruby, cross and three-twelve lattices.
These models are equivalent to site percolation since the patches on a disk either cover all edges connecting to the disk center
with probability $\chi$ or cover no edge with probability $1-\chi$. 

It is also found in previous subsections that several models of patchy disks can share the same $\chi_c$ value.
Considering the above equivalences with site percolation, and observing these equalities of $\chi_c$ values 
[e.g. on the square lattice the equality of $\chi_c$ for disks with one (two) and five (six) patches],
we wonder if $\chi_c$ values on a given lattice appear in a periodic way as the number of patches $n$ increases,
and if there is any other rule governing the $\chi_c$ values. 

To explore possible periodic behaviors, we first try to calculate probabilities of different patch-covering structures of a disk 
as a function of $\chi$ for three regular lattices (honeycomb, square and triangular), near previously estimated values of $\chi_c$
in Table~\ref{Tab:chi-threshold}. Exemplary calculations for the triangular lattice and some calculation details for 
other two lattices are included in the SM~\cite{sup-mat}. It is found that indeed there are periodic behaviors for 
these probabilities of different patch-covering structures, as summarized in Table~\ref{Tab:contact-n}. 
As mentioned in the previous subsection, same expressions as functions of $\chi$ for these probabilities near estimated $\chi_c$
prove that the equality of $\chi_c$ values hold exactly.
Thus, from results in Table~\ref{Tab:chi-threshold}, as the number of patches on a disk $n$ increases, $\chi_c$ values appear with a period $n_0=3$,$4$ and $6$ 
for the honeycomb, square and triangular lattices, respectively. Since the snub hexagonal and kagome lattices can be regarded 
as sub lattices of the triangular lattice, we then calculate probabilities of different patch-covering structures of a disk 
on these two lattices  by making use of the triangular lattice, as shown in Table S5 and S6 of the SM~\cite{sup-mat}.
The results for the snub hexagonal and kagome are also summarized in Table~\ref{Tab:contact-n}, which confirm that
on these two lattices $\chi_c$ values appear with a period $n_0=6$.

For calculating probabilities of different patch-covering structures of a patchy disk at a vertex of 
the four-eight lattice, we can first place the disk at the center of a regular octagon 
and consider the patch-covering of edges connecting the center and the vertices of the octagon,
then use these intermediate results to get the final results. Some details are shown in Tables S7 and S8 of the SM~\cite{sup-mat},
and the final results are also summarized in Table~\ref{Tab:contact-n}, which shows that 
on the four-eight lattice $\chi_c$ values appear with a period $n_0=8$.
Similarly, we can make use of the regular dodecagon to get probabilities of 
different patch-covering structures of a patchy disk at a vertex of other five Archimedean lattice in 2D.
Some details for these calculations are included in Tables S9 to S23 of the SM~\cite{sup-mat}.
Final results for these five lattices are summarized in Table~\ref{Tab:contact-n-b}, 
which prove that $\chi_c$ values on these lattices appear with a period $n_0=12$.

It should be noted that, for the lattices with periods $n_0=12$ and $8$, 
when performing the calculations for probabilities of different patch-covering structures of a disk,
we have made use of an assumed symmetry from observing expressions of other lattices. 
Namely, for the honeycomb, square, triangular, snub hexagonal and kagome lattices, 
it is found that there is a symmetry between models with $\mod(n,n_0)=m$ ($0<m<n_0/2$) and $\mod(n,n_0)=n_0-m$.
Our results confirm that indeed this symmetry also holds for the lattices with $n_0=8$ and $12$.
This symmetry allows us to give the $\chi_c$ values of patchy disks with $\mod(n,n_0)>6$,
without performing additional numerical simulations.
We have tried to understand this symmetry by observing the patch-covering structures,
and found that this symmetry is associated with some symmetries of the structures, 
as shown in Fig.~S25 of the SM~\cite{sup-mat}, but we have not got a simple reasoning why this symmetry exists.

We also note that, for a fixed value of $j$, when probabilities of $j$-edge patch-covering are the same for
different models, the detailed structures with $j$-edge patch-covering still can be different. This can be clearly seen 
from Table~\ref{Tab:proba-expression}. Here, for Table~\ref{Tab:contact-n}, on the kagome lattice,
though probabilities of $j$-edge patch-covering are the same for $n$-patch disks with $\mod(n,6)=2$, $3$ and $4$,
the detailed structures of $2$-edge patch-covering for disks with $\mod(n,6)=3$ are different from those with $\mod(n,6)=2$ and $4$.
Since $2$-edge structures for disks with $\mod(n,6)=2$ and $4$ are more open than those for disks with $\mod(n,6)=3$,
the $\chi_c$ value of the latter is larger than that of the other two. 
Similarly, on the cross lattice, since the $2$-edge structures for disks with $\mod(n,12)=3$ and $6$ are more open
than that for disks with $\mod(n,12)=4$, the $\chi_c$ value of the latter is larger than that of the former two. 

\section{Conclusion and discussion}
\label{sec-conclu}

\begin{figure}
\begin{center}
\includegraphics[scale=1.2]{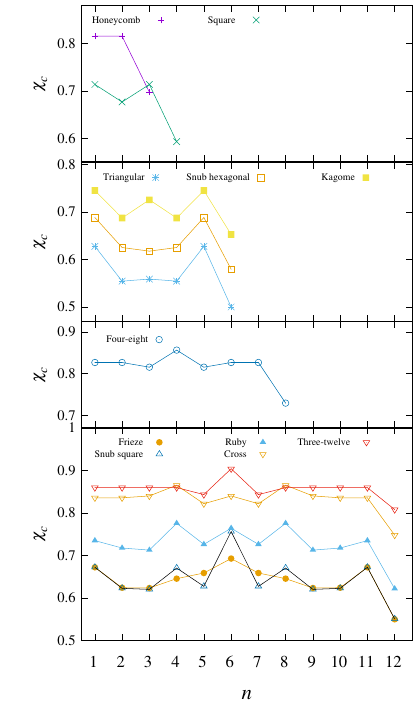}
	\caption{The percolation threshold $\chi_c$ versus the number of patches on a disk $n$. 
	As $n$ increases, $\chi_c$ values appear in periodic ways, with the period being $n_0=3$ for the honeycomb lattice, 
	$n_0=4$ for the square lattice, $n_0=6$ for the triangular, snub hexagonal and kagome lattices, 
	$n_0=8$ for the four-eight lattice, and $n_0=12$ for the remaining five lattices. 
	For each lattice, the minimum threshold value $\chi_{\rm min}$ appears when $\mod(n,n_0)=0$, 
	and $\chi_{\rm min}$ is equal to the threshold $p_c$ of site percolation on the same lattice.
	Moreover, $\chi_c$ for $\mod(n,n_0)=m$ ($0<m<n_0/2$) is the same as $\chi_c$ for $\mod(n,n_0)=n_0-m$.
	Detailed values of $\chi_c$ (or $p_c$) can be read from Table~\ref{Tab:chi-threshold} for $\mod(n,n_0)=0$,
	and from Tables~\ref{Tab:contact-n} and \ref{Tab:contact-n-b} for $\mod(n,n_0) \ne 0$.}
\label{Fig:kai-n}
\end{center}
\end{figure}

To summarize, we study in this work the percolation of patchy particles which are randomly rotating 
on $11$ Archimedean lattices in 2D. We combine MC simulations with the recently developed 
critical polynomial method to give precise estimates of the threshold values $\chi_c$ for $88$ models,
including disks with one to six patches and spheres with one to two patches, on the $11$ lattices.
These estimates are summarized in Table~\ref{Tab:chi-threshold}.
For one-patch particles, it is found that $\chi_c$ values on different lattices 
follow the same order as $p_c$ values for site percolation on same lattices, which implies that
in this case $\chi_c$ is mainly influenced by the geometry of the lattices. 
When there are more symmetrically distributed patches on a particle, the symmetry of 
the patches play an important role, in addition to the geometry of the lattices. 
The $\chi_c$ values on different lattices do not follow the same order as those for one-patch particles.  

Furthermore, to explore the role of symmetry, we consider $\chi_c$ of disks with an arbitrary number ($n>0$)
of symmetrically distributed patches. By analyzing probabilities of different patch-covering structures
of a patchy disk at a vertex as functions of $\chi$ near the above estimates of $\chi_c$, we give
$\chi_c$ values for these $n$-patch disks on all $11$ Archimedean lattices in 2D. 
The $\chi_c$ values are plotted in Fig.~\ref{Fig:kai-n}, in which the following rules are summarized:
(i) for a given lattice, $\chi_c$ values appear in a periodic way, with the period $n_0$ determined 
by the symmetry of the lattice. We find that $n_0=3$ for the honeycomb lattice, $n_0=4$ for the square lattice, 
$n_0=6$ for the triangular, snub hexagonal and kagome lattices, $n_0=8$ for the four-eight lattice, 
and $n_0=12$ for the remaining five lattices. (ii) the minimum threshold value $\chi_{\rm min}$ of a lattice presents when $\mod(n,n_0)=0$. 
Actually, at this condition, the model is equivalent to site percolation on the same lattice,
with $\chi_{\rm min}$ being equal to the site-percolation threshold $p_c$, whose value is given in Table~\ref{Tab:chi-threshold}. 
(iii) for each lattice, $\chi_c$ for $\mod(n,n_0)=m$ ($0<m<n_0/2$) is the same as that for $\mod(n,n_0)=n_0-m$.
In addition, we find that there exist other equalities between $\chi_c$ values, 
such as $\chi_c$ values of disks with $\mod(n,8)=1$ and $2$ on the four-eight lattice,
those of disks with $\mod(n,12)=1$ and $2$, and of disks with $\mod(n,12)=3$ and $6$, both on the cross lattice,
and those of disks with $\mod(n,12)=1$,$2$,$3$ and $4$ on the three-twelve lattice.
The precise values of $\chi_c$ for $n$-patch disks with $\mod(n,n_0) \ne 0$ are summarized 
in Tables~\ref{Tab:contact-n} and \ref{Tab:contact-n-b}.

Precise values of $\chi_c$ for one- and two-patch spheres are also available in Table~\ref{Tab:chi-threshold}.
Comparing with one- and two-patch disks, our numerical results for the spheres suggest that changing the shape 
from disk to sphere dose not affect the order of $\chi_c$ values for different lattices, but does affect 
the values of $\chi_c$. It may be interesting to investigate $\chi_c$ of spheres with more patches,
since more symmetries of patches are available on the sphere surface. 
Besides, the current work can also be extended to other lattices.
One example is the Lieb lattice in 2D. Percolation of four-patch disks on this lattice is equivalent to site percolation 
on the same lattice.  Similar to the results in Sec.~\ref{sec:2-6}, this can be proved as the patches on a disk
either cover all neighboring edges with probability $\chi$ or cover no edge with probability $1-\chi$.
By MC simulations of site percolation and the critical polynomial method, we determine the threshold value as 
$\chi_c = p_c = 0.739\,706\,0(6)$~\cite{sup-mat}, which is much more precise than a recent estimate $0.739\,6(5)$~\cite{Oliveira21}.
It should be noted that site percolation on the Lieb lattice in 2D is equivalent to site-bond percolation
on the square lattice with same site and bond occupation probabilities, for which empirical formulas
combined with latest threshold values of both pure bond and pure site percolation lead to 
$p_c \simeq 0.742\,2$~\cite{Yanuka1990} or $p_c \simeq 0.737\,6$~\cite{Tarasevich1999}. 
These values are quite close to our estimate above.

In the above models, we assume that each particle is randomly rotating with its center being fixed 
at a vertex of the lattice.  
For a system in which there are interactions between patchy particles, this corresponds to  
the high-temperature limit of the system at full occupancy of the lattice.
Since there is no correlation between different particles, 
the universality class of the percolation transitions is the same as ordinary percolation in 2D. 
It will be interesting to investigate the interplay of percolation 
and different thermodynamic phases for patchy particles at finite temperatures on lattices in 2D~\cite{Mitsumoto2018,Patrykiejew2020,Patrykiejew2020-A}, 
where universal properties might be different.
Percolation and phase transitions of patchy particles in continuum space in 2D also demand more investigations~\cite{Huang2020, Liang2021}.

Finally, we note that our calculation of the probabilities of different patch-covering structures of a particle 
implies a new way to define the models: the ``patch-covering" structures of a particle can be regarded as discrete states
of a vertex, and various states of a vertex occur with different probabilities. This transforms 
the continuously rotating particle model to a discrete ``spin" model on the lattices, for which 
more efficient numerical simulations may be performed. New models are obtained if one allows the states 
to be different from those of the patchy particles.  

\acknowledgments
J. Wang and H. Hu thank W. Xu and Y. Deng for a previous collaboration on combining MC simulations and the critical polynomial method to study nonplanar and continuum percolation models~\cite{Xu2021}.
This work was supported by the National Natural Science Foundation of China under Grant No.~11905001, and by the Anhui Provincial Natural Science Foundation under Grant No.~1908085QA23. We also acknowledge the High-Performance Computing Platform of Anhui University for providing computing resources.

\end{document}


\definecolor{blue}{rgb}{0,0,1}
\definecolor{red}{rgb}{1,0,0}
\definecolor{green}{rgb}{0,1,0}
\newcommand{\blue}[1]{\textcolor{blue}{#1}}
\newcommand{\red}[1]{\textcolor{red}{#1}}
\newcommand{\green}[1]{\textcolor{green}{#1}}

\newcommand{\PB}{P_{\rm B}}

\newcommand{\calA}{ {\mathcal A}}
\newcommand{\calC}{ {\mathcal C}}
\newcommand{\calE}{ {\mathcal E}}
\newcommand{\calG}{ {\mathcal G}}
\newcommand{\calH}{ {\mathcal H}}
\newcommand{\calO}{ {\mathcal O}}
\newcommand{\calS}{ {\mathcal S}}
\newcommand{\calZ}{ {\mathcal Z}}

\title{Supplemental material for: ``Percolation thresholds of randomly rotating patchy particles on Archimedean lattices"}

\author{Quancheng Wang}
\affiliation{School of Physics and Electronic Engineering, Anhui University, Hefei, Anhui 230601, China}

\author{Zhenfang He}
\affiliation{School of Physics and Electronic Engineering, Anhui University, Hefei, Anhui 230601, China}

\author{Junfeng Wang}
\affiliation{School of Physics, Hefei University of Technology, Hefei, Anhui 230009, China}

\author{Hao Hu}
\email[Corresponding author:\;]{huhao@ahu.edu.cn}
\affiliation{School of Physics and Electronic Engineering, Anhui University, Hefei, Anhui 230601, China}

\maketitle

\setcounter{equation}{0}
\renewcommand\theequation{S\arabic{equation}}

{\bf Further simulation details.} In our simulations, the vertices and edges of lattices are encoded with the aid of square or triangular arrays, 
whose correspondences with the actual lattices are shown below in Figs.~\ref{Fig:triangle} to \ref{Fig:three-twelve}.
For each figure (except the figure for the square or triangular lattice), the plot on the left side shows 
the array for encoding, and the one on the right side shows the actual lattice.
On the square or triangular array, a region of size $L_x \times L_y$ is shown, whose correspondence
in the actual lattice is also drawn, with the length of each edge being one.
The linear size in the main text is defined as $L \equiv L_x$.
The number of vertices in each lattice can be calculated from $L$ as shown in Table~\ref{Tab:lattice-size}.
The number of independent samples taken for one-patch disks on the lattices are summarized in Table~\ref{Tab:sample-size}. 
The simulation time consumed for each model is presented in Table~\ref{Tab:time}. 

{\bf More plots for estimating the thresholds values.} In the main text we only show plots for
estimating the threshold value of one-patch disks on the triangular lattice. Below Figs.~\ref{Fig:one-patch-disk} to 
\ref{Fig:six-patch-disk} show the intersection of $\PB$ near the percolation thresholds for other $87$ patchy particle models. 
For one-patch particles, to demonstrate our estimates of the correction exponent $y_1$, we also make plots of $\PB$ 
for small sizes $L$ at the percolation thresholds, as in Figs.~\ref{Fig:one-patch-disk-small-size} 
and \ref{Fig:one-patch-sphere-small-size}. 
For site percolation on the Lieb lattice in 2D, Fig.~\ref{Fig:Lieb} 
demonstrates our estimates of $p_c=0.739\,706\,0(6)$ and $y_1 \simeq -3.6$.

{\bf Calculations for one-patch disks on the frieze lattice.}
Here we present details for calculating probabilities of different patch-covering structures, 
for one-patch disks on the frieze lattice. The final results have already been given in Table III of the main text.
Calculations for other patchy particles on the frieze or other lattices can be performed in a similar way.

When a one-patch disk rotates randomly at a vertex of the frieze lattice, for patch sizes $\chi$ near $\chi_c=0.672\,338\,8(1)$,
there exist only $3$-edge and $4$-edge patch-covering structures. 
There are two types of $3$-edge patch-covering structures: one type has the blank sector covering 
two edges at an angle of $\pi/2$, as shown in Fig.~\ref{Fig:frieze-edge-covering} (a) and (b); 
and the other type has the blank sector covering two edges at an angle of $\pi/6$,
as shown in Fig.~\ref{Fig:frieze-edge-covering} (c), (d) and (e).
For the first type, if not changing the patch-covering states of edges, the disk can rotate within an angle $\chi_{3,1}$
(values of $\chi$, $\chi_i$ or $\chi_{i,j}$ are angles in units of $2\pi$), whose value can be calculated as
\begin{equation}
\chi_{3,1}=1-{\chi}-\dfrac{1}{4} = \dfrac{3}{4}-{\chi} \;.
\label{eq:frieze-3-1}
\end{equation}
For the other type,  if not changing the patch-covering states, the disk can rotate within another angle $\chi_{3,2}$ as
\begin{equation}
\chi_{3,2}=1-{\chi}-\dfrac{1}{6} = \dfrac{5}{6}-{\chi} \;.
\label{eq:frieze-3-2}
\end{equation}
Thus the probability of $3$-edge patch-covering structures is the summation of $\chi_{i,j}$ corresponding to
Fig.~\ref{Fig:frieze-edge-covering} (a-e)
\begin{equation}
\chi_{3}= 2 \chi_{3,1} + 3 \chi_{3,2} = 4-5{\chi}.
\label{eq:frieze-3}
\end{equation}

There are three types of $4$-edge patch-covering structures: the first type has the blank sector covering 
edge $3$ or $1$, as shown in Fig.~\ref{Fig:frieze-edge-covering} (f) and (g), respectively;
the second type has the blank sector covering edge $2$, as shown in Fig.~\ref{Fig:frieze-edge-covering} (h); 
and the last type has the blank sector covering edge $5$ or $4$, as shown 
in Fig.~\ref{Fig:frieze-edge-covering} (i) and (j), respectively.
The angles the disk can rotate without changing the patch-covering sates of edges are correspondingly given for the three types as
\begin{equation}
	\chi_{4,1}=\dfrac{1}{4}+\dfrac{1}{6}-(1-{\chi})= {\chi}-\dfrac{7}{12} \;,
\label{eq:frieze-4-1}
\end{equation}
\begin{equation}
\chi_{4,2}=\dfrac{1}{4} \times 2 -(1-{\chi}) = {\chi}-\dfrac{1}{2} \;,
\label{eq:frieze-4-2}
\end{equation}
\begin{equation}
\chi_{4,3}=\dfrac{1}{6} \times 2 -(1-{\chi}) = {\chi}-\dfrac{2}{3} \;.
\label{eq:frieze-4-3}
\end{equation}
Thus the total probability of $4$-edge structures is
\begin{equation}
\chi_{4}= 2 \chi_{4,1} + \chi_{4,2} + 2 \chi_{4,3}= 5{\chi}-3 \;.
\label{eq:frieze-4}
\end{equation}

{\bf Calculations for disks with an arbitrary number of patches.}
Here we present details for obtaining probabilities of different patch-covering structures 
in Tables V and VI of the main text. 
We take the model of $n$-patch ($n$ being positive integers) disks 
on the triangular lattice as an example to illustrate the calculations, 
and perform calculations on other lattices in a similar way.

For $n$-patch disks on the triangular lattice, we classify the particles by their values of $\mod(n,6)$
and conduct calculation for each value of $\mod(n,6)$.
For the case of $\mod(n,6)=1$ ($n=6m+1$, with $m$ being non-negative integers),
the assumed value of $\chi_c$ is $0.627\,765\,5(2)$, as taken from the main text. 
For $\chi$ near $\chi_c$, there are two types of patch-covering structures, i.e. $3$-edge and $4$-edge structures, 
and it can be proved that (the $3$ or $4$) edges are covered by the patches consecutively. 
For convenience, we label patches and edges in a counterclockwise way, as exemplified for seven-patch disks 
in Fig.~\ref{Fig:seven-patch-rotate}. Initially we set the right (assuming the observer stands at the disk center 
and faces outwards) boundary of patch $1$ at edge $1$ of the lattice.
When the disk rotates counterclockwise within an angle $\chi_1$, 
the structure always belongs to the $3$-edge type;
and when the disk rotates clockwise within an angle $\chi_2$, the structure always belongs to the $4$-edge type.
The values of $\chi_1$ and $\chi_2$ are determined as follows.

When the disk rotates counterclockwise by a small angle, for calculating $\chi_1$, 
since initially edges $2$, $3$ and $4$ are being covered by the patches, 
the structure only changes its type if these three edges keep being covered and a new edge ($5$ or $6$)
becomes covered. The angle between edge $5$ and the nearest patch to its right side is 
\begin{equation}
	\chi_{5,{\rm right}} = 2m \cdot \dfrac{1}{n}+\dfrac{1-\chi}{n}-\dfrac{1}{3} \;\;.
\label{eq:tri-7}
\end{equation}
Here $1/3$ is the fraction of the disk between edge $5$ and edge $1$;
$1/n$ is the fraction the disk occupied by a patch and a neighboring blank sector, 
and there are $2m+1/3$ such units between edge $5$ and edge $1$; 
and $(1-\chi)/n$ is the fraction of the disk occupied by a blank sector. 
We also derive the angle between edge $6$ and the nearest patch to its right side as 
\begin{equation}
	\chi_{6,{\rm right}} = m \cdot \dfrac{1}{n}+\dfrac{1-\chi}{n}-\dfrac{1}{6} \;\;.
\label{eq:tri-5}
\end{equation}
Since $\chi_{6,right}-\chi_{5,right}=1/[6(6m+1)]>0$, when the disk rotates counterclockwise,
the patches cover edge $5$ earlier than edge $6$. 
If representing $\chi_{k,right}$ as the angle between edge $k$ ($k=2,3,4$) and the right boundary 
of the patch covering the edge, in a similar way, we get $\chi_{5,right} < \chi_{k,right}$, 
which means that, when the disk rotates counterclockwise, the patches encounter edge $5$ before
leaving edge $k$.
Thus we have $\chi_1=\chi_{5,right}$, and by substituting $n=6m+1$ to Eq.~(\ref{eq:tri-7}) we get 
\begin{equation}
	\chi_1 = \dfrac{1-\chi}{n}+\dfrac{n-1}{3} \cdot \dfrac{1}{n}-\dfrac{1}{3} \;\;.
\label{eq:tri-theta1}
\end{equation}

When the disk rotates clockwise by a small angle, for calculating $\chi_2$, 
we need consider when one of edges $(2,3,4)$ becomes not covered, and ensures that
this happens before any patch covers edge $5$ or $6$. 
Using calculations similar to the above, we find that edge $4$ is the first one to 
become uncovered, without covering of edge $5$ or $6$, and the value of $\chi_2$
is given by
\begin{equation}
	\chi_2 = \dfrac{\chi}{n}+\dfrac{n-1}{2} \cdot \dfrac{1}{n}-\dfrac{1}{2} \;\;.
\label{eq:tri-theta2}
\end{equation}

Starting from the initial condition above, if the disk rotates persistently in counterclockwise
direction, one shall observe changes of patch-covering structures in a periodic way:
within an angle $\chi_1$ it belongs to the $3$-edge type, within another angle $\chi_2$ 
it belongs to the $4$-edge type, and this repeats six times within a total interval of $1/n$, 
where $1/n$ is the angle of a patch and a blank sector.
Each time the patch-covering structure changes its type, the boundary of one patch is at an edge,
and the patch is to the left or right side of the edge.
These are exemplified using the seven-patch disk in Fig.~\ref{Fig:seven-patch-rotate}.
After a rotation of total angle $1/n$, the condition is the same as the initial condition,
except that the label of the patch is different. Thus we get the probability of $3$-edge structure 
as 
\begin{equation}
 \chi_1 \cdot 6 \cdot n = \left(\dfrac{1-{\chi}}{n}+\dfrac{n-1}{3} \cdot \dfrac{1}{n}-\dfrac{1}{3}\right) \cdot 6 \cdot n = 4-6{\chi} \;\;,
\label{eq:tri-1}
\end{equation}
and the probability of $4$-edge structure as
\begin{equation}
 \chi_2 \cdot 6 \cdot n = \left(\dfrac{{\chi}}{n}+\dfrac{n-1}{2} \cdot \dfrac{1}{n}-\dfrac{1}{2}\right) \cdot 6 \cdot n = 6{\chi}-3 \;\;.
\label{eq:tri-2}
\end{equation}

The calculations for other models are performed in a similar way.
For cases of other $\mod(n,6)$ on the triangular lattice, of $\mod(n,4)$ on the square lattice,  
and of $\mod(n,3)$ on the honeycomb lattice,  
the results are summarized in Table~\ref{Tab:contact-proba}.
For the snub hexagonal and kagome lattices, the results are obtained by using intermediate results on the triangular lattice,
as shown in Tables~\ref{Tab:contact-proba-snub-hex} and \ref{Tab:contact-proba-kagome}, respectively.
For the four-eight lattice, the results are derived by considering patch-covering of a disk at the center of a regular octagon, 
as shown in Tables~\ref{Tab:contact-proba-four-eight-a} and \ref{Tab:contact-proba-four-eight-b}.
For the remaining five lattices, results are derived by considering patch-covering of a disk at the center of a regular dodecagon,
as shown in Tables~\ref{Tab:contact-proba-frieze-a} to \ref{Tab:contact-proba-three-twelve-c}.

{\bf Demonstration for symmetries between $n$-patch disks with $\mod(n,n_0)=m$ and $n_0-m$.}
Through the calculations of probabilities of patch-covering structures, we have proved the equality 
of $\chi_c$ values between $n$-patch disks with $\mod(n,n_0)=m$ ($0<m<n_0/2$) and $n_0-m$ on a given lattice.
To understand this equality, we find that there exist some symmetries between the patch-covering structures, 
as demonstrated for three regular lattices (honeycomb, square and triangular) in Fig.~\ref{Fig:symmetry-m}.

\setcounter{figure}{0}
\renewcommand\thefigure{S\arabic{figure}}

\vspace{2.8cm}
\begin{figure*}[htp]
\begin{center}
\includegraphics[scale=0.54]{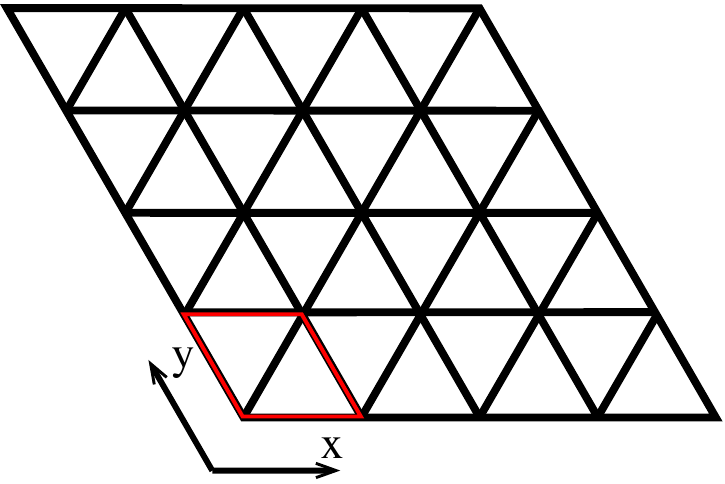}

  \hspace{10cm}    $1 \times 1$    \hspace{\stretch{1}}
	\caption{Triangular lattice. }
    \label{Fig:triangle}
\end{center}
\end{figure*}

\begin{figure*}[htbp]
\begin{center}
\includegraphics[scale=0.27]{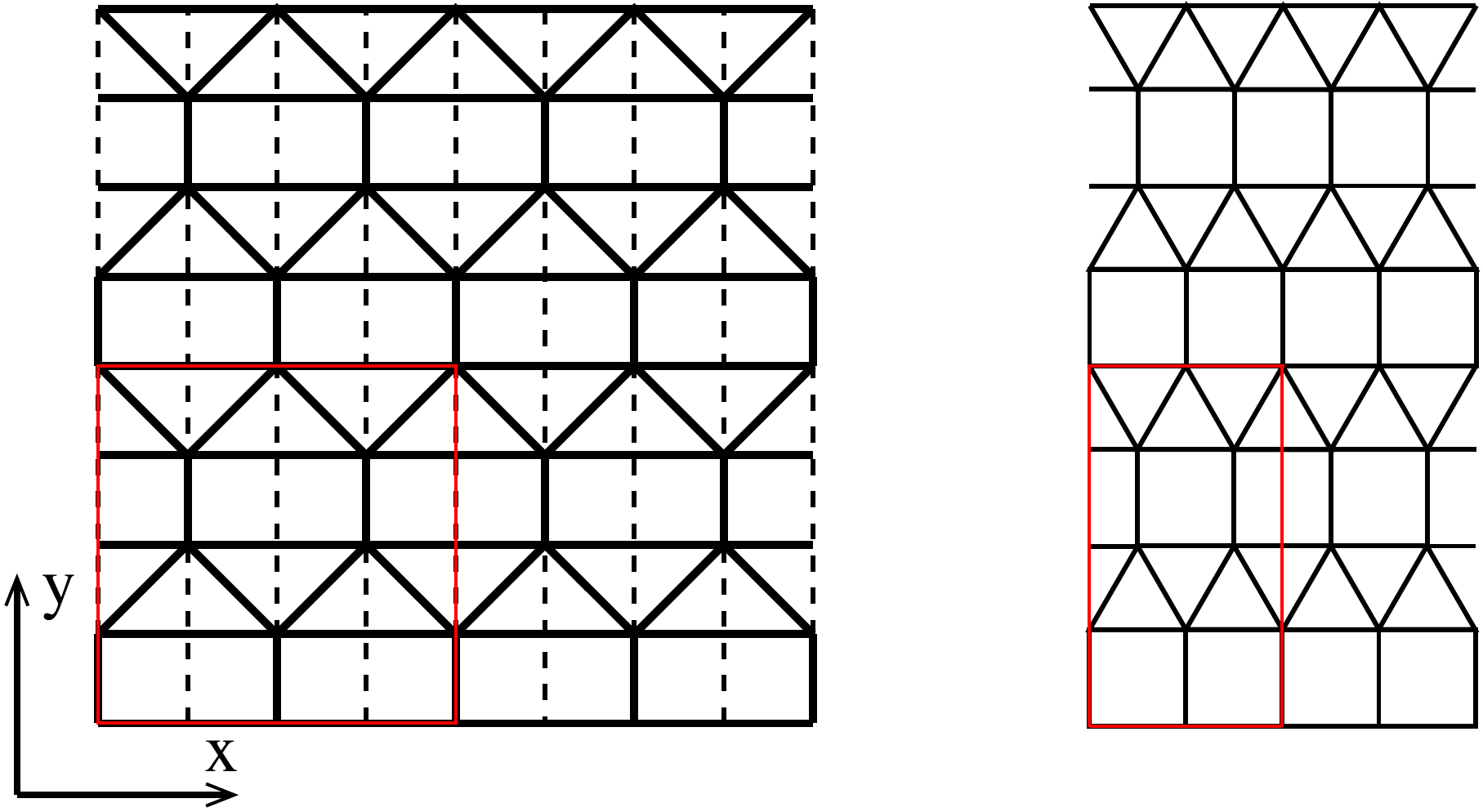}

  \hspace{1.4cm}    $4 \times 4$    \hspace{3.4cm}  $2 \times \left(2+\sqrt{3}\right)$
	\caption{Frieze lattice. }
    \label{Fig:frieze}
\end{center}
\end{figure*}

\newpage

\vspace{1.4cm}
\begin{figure*}[ht]
\begin{center}
\includegraphics[scale=0.26]{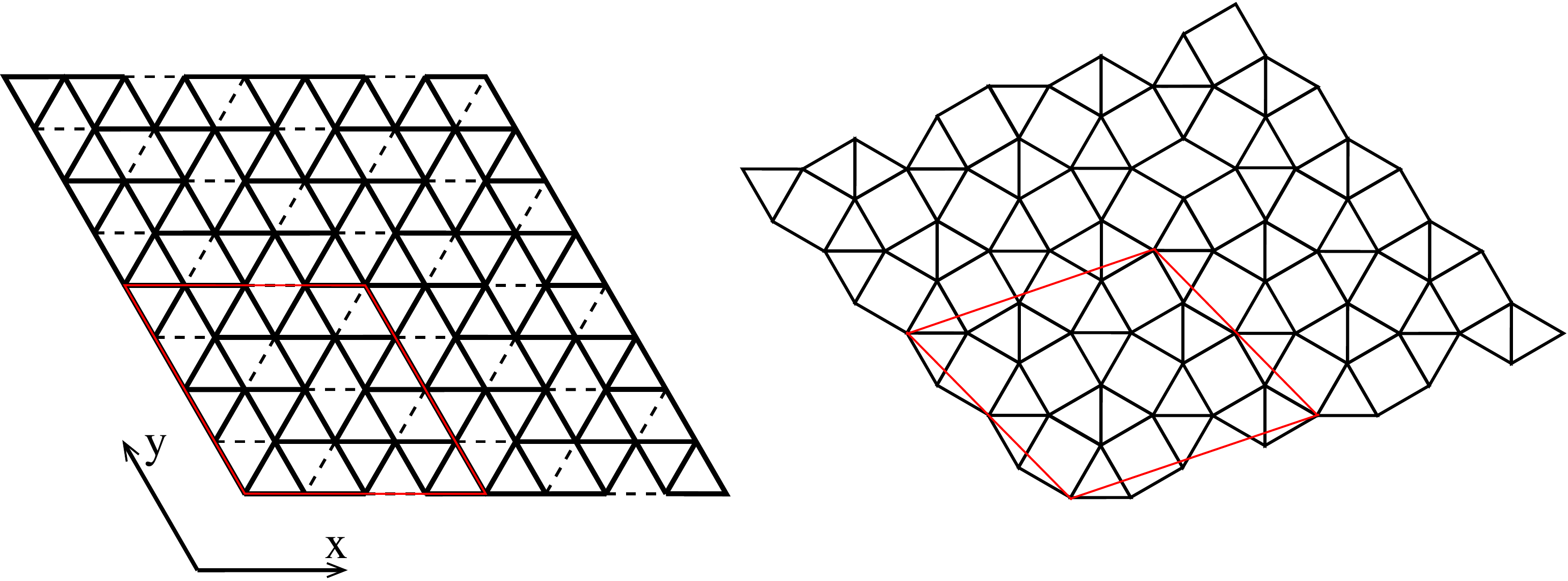}

	\hspace{2.8cm}    $4 \times 4$    \hspace{2.8cm}  $\sqrt{10+5\sqrt{3}} \times 2\sqrt{2+\sqrt{3}}$
	\caption{Snub square lattice. }
    \label{Fig:snub-square}
\end{center}
\end{figure*}

\vspace{1.3cm}
\begin{figure*}[htbp]
\begin{center}
\includegraphics[scale=0.15]{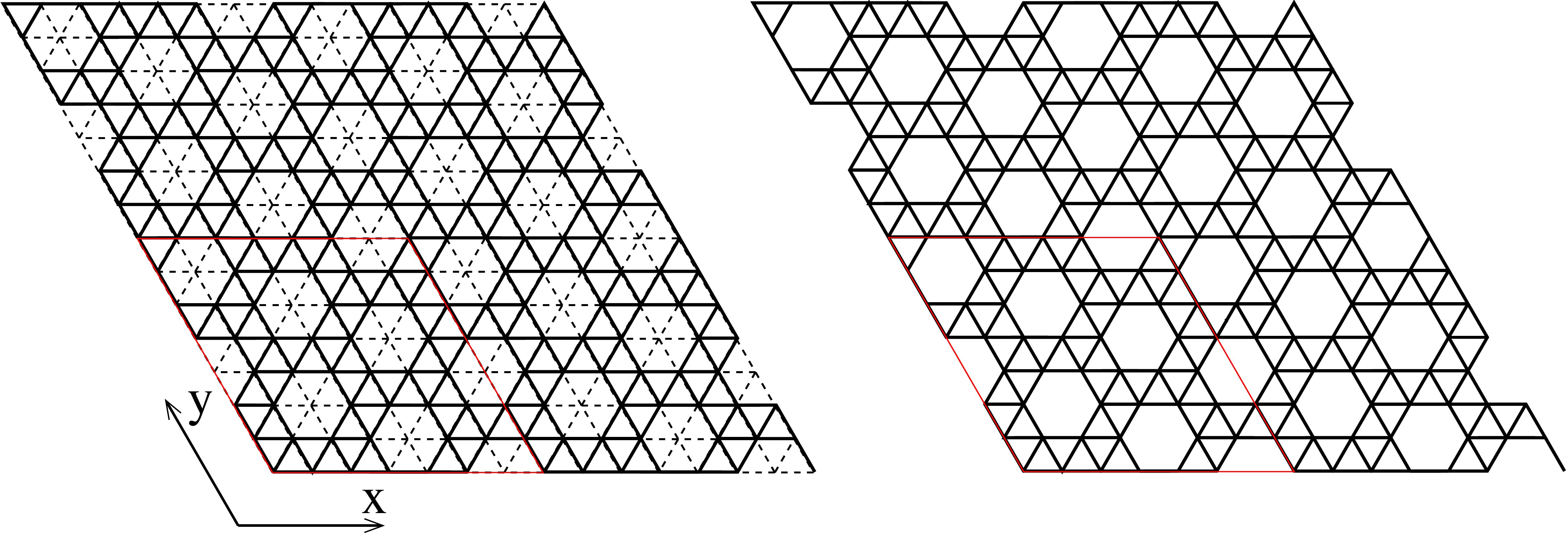}

  \hspace{1.8cm}    $7 \times 7$    \hspace{4.8cm}  $7 \times 7$
	\caption{Snub hexagonal lattice. }
    \label{Fig:snub-hexagonal}
\end{center}
\end{figure*}

\begin{figure*}[!htbp]
\begin{center}
\includegraphics[scale=0.48]{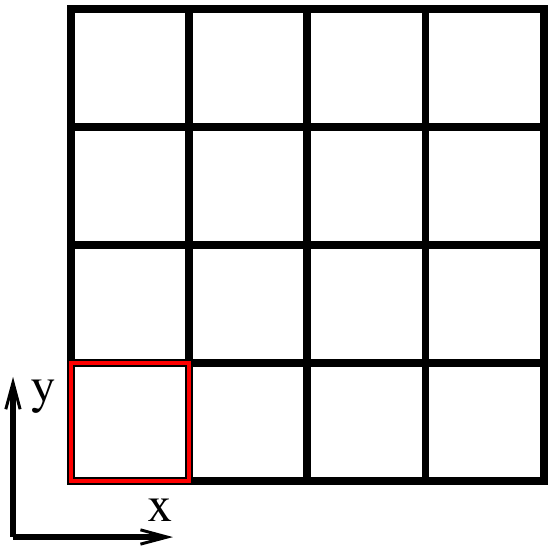}

  \hspace{9cm}    $1 \times 1$    \hspace{\stretch{1}}
	\caption{Square lattice. }
    \label{Fig:square}
\end{center}
\end{figure*}

\newpage

\vspace{0.6cm}
\begin{figure*}[htbp]
\begin{center}
\includegraphics[scale=0.155]{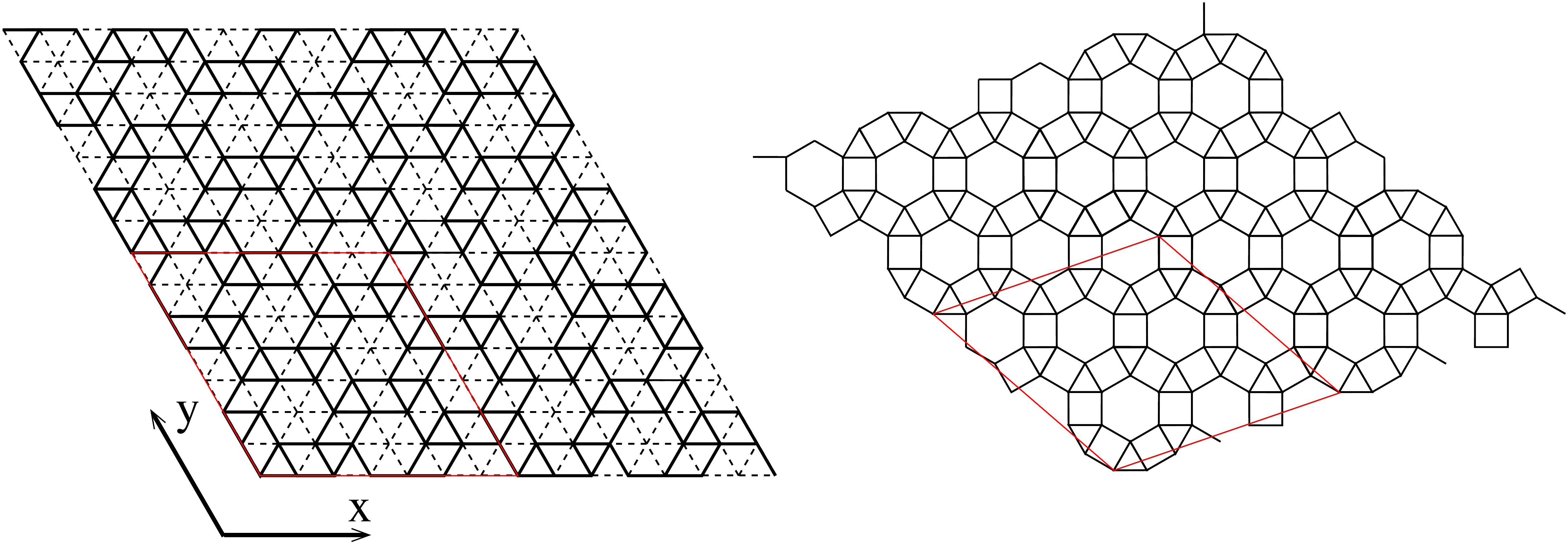}

	\hspace{3.2cm} $7 \times 7$ \hspace{2.6cm} $\sqrt{28+14\sqrt{3}} \times \sqrt{28+14\sqrt{3}}$
	\caption{Ruby lattice. }
    \label{Fig:ruby}
\end{center}
\end{figure*}

\vspace{1.5cm}
\begin{figure*}[!htbp]
\begin{center}
\includegraphics[scale=0.36]{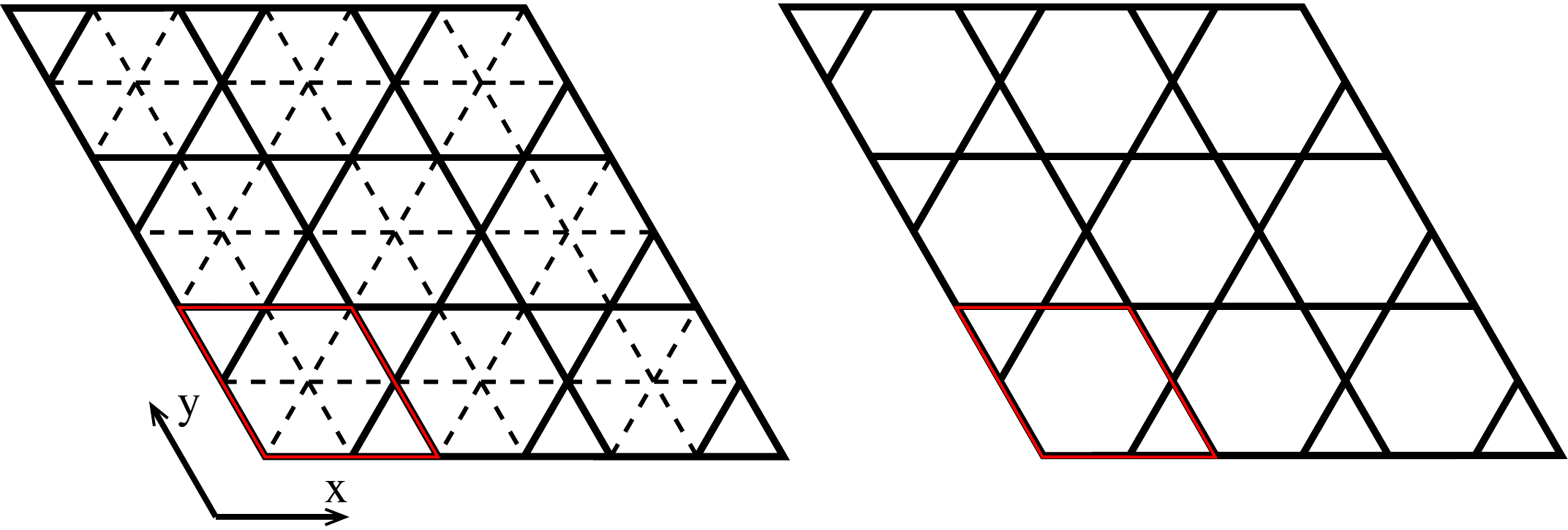}

  \hspace{2cm}    $2 \times 2$    \hspace{5.6cm}  $2 \times 2$
	\caption{Kagome lattice. }
    \label{Fig:kagome}
\end{center}
\end{figure*}

\begin{figure*}[htbp]
\begin{center}
\includegraphics[scale=0.29]{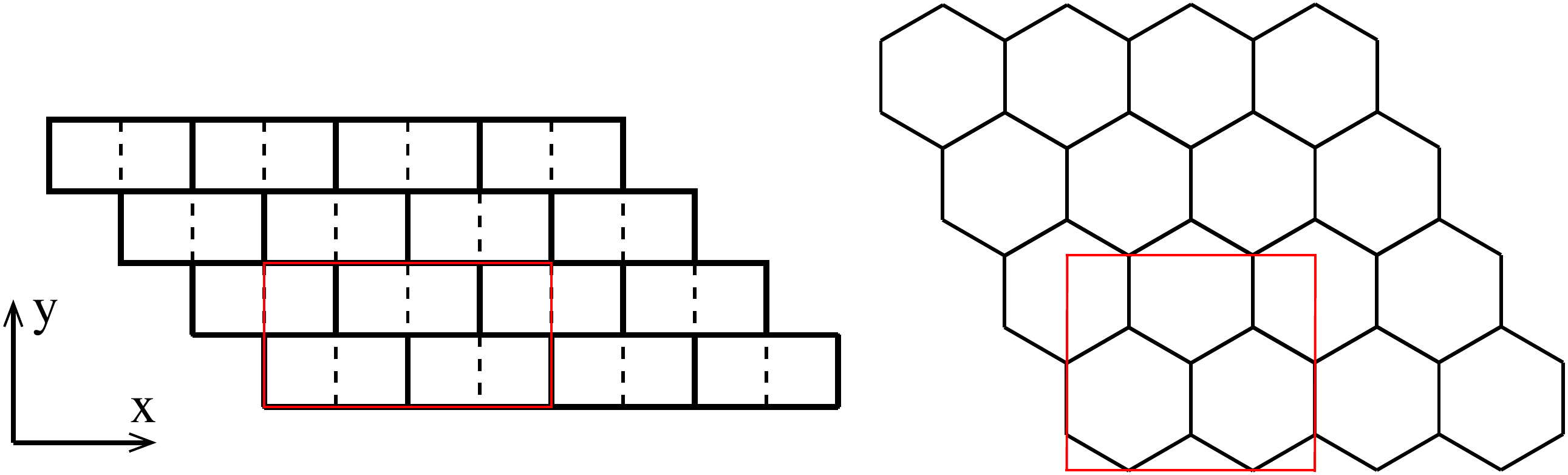}

  \hspace{2.0cm}    $4 \times 2$    \hspace{5cm}  $2\sqrt{3} \times 3$
	\caption{Honeycomb lattice. }
    \label{Fig:honeycomb}
\end{center}
\end{figure*}

\newpage

\vspace{1.0cm}
\begin{figure*}[htbp]
\begin{center}
\includegraphics[scale=0.20]{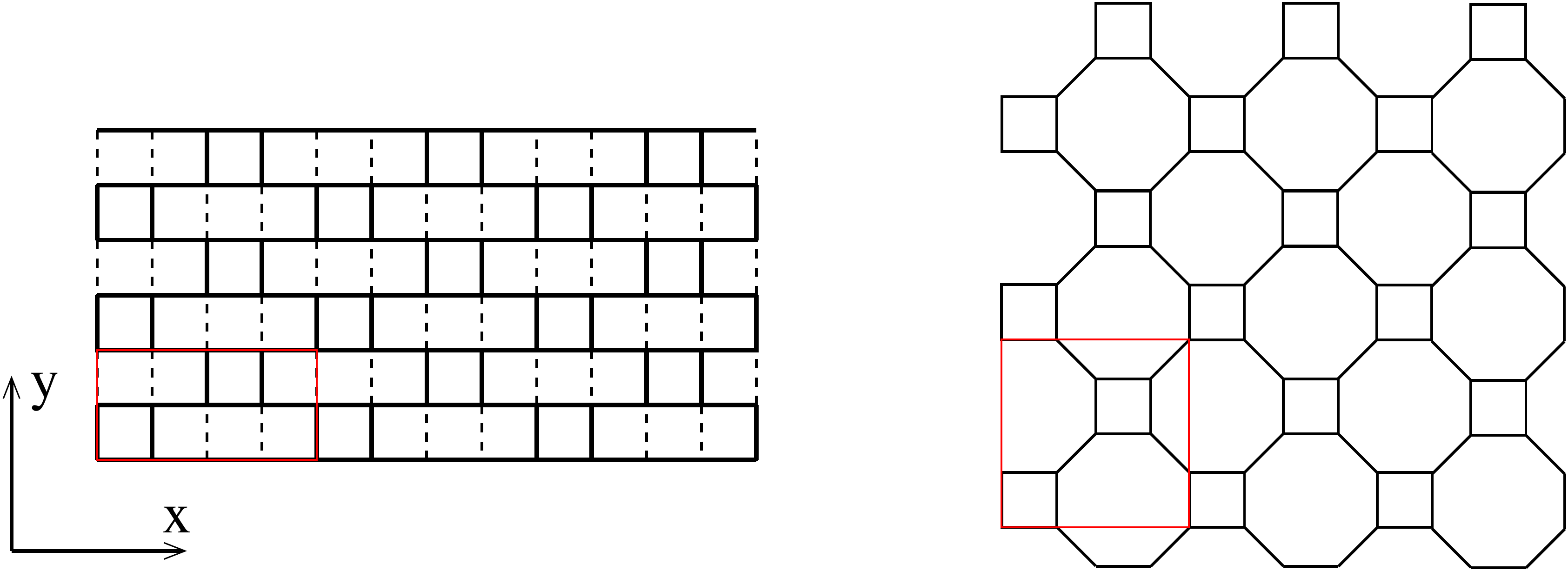}

  \hspace{2.5cm}    $4 \times 2$    \hspace{4cm}  $\left(2+\sqrt{2}\right) \times \left(2+\sqrt{2}\right)$
	\caption{Four-eight lattice. }
    \label{Fig:four-eight}
\end{center}
\end{figure*}

\vspace{1.1cm}
\begin{figure*}[h]
\begin{center}
\includegraphics[scale=0.12]{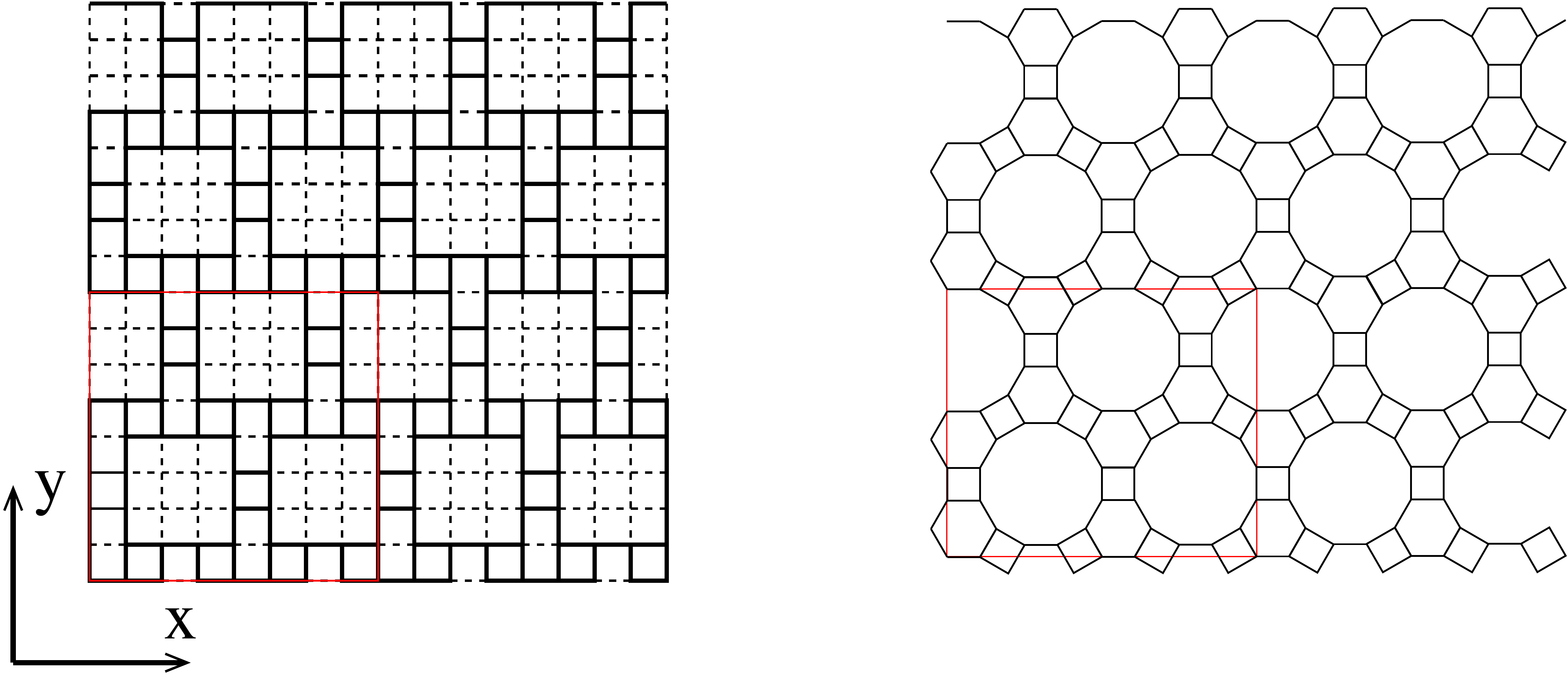}

  \hspace{1.8cm}    $8 \times 8$    \hspace{3.8cm}  $\left(6+2\sqrt{3}\right) \times \left(3+3\sqrt{3}\right)$
	\caption{Cross lattice. }
    \label{Fig:cross}
\end{center}
\end{figure*}

\begin{figure*}[ht]
\begin{center}
\includegraphics[scale=0.23]{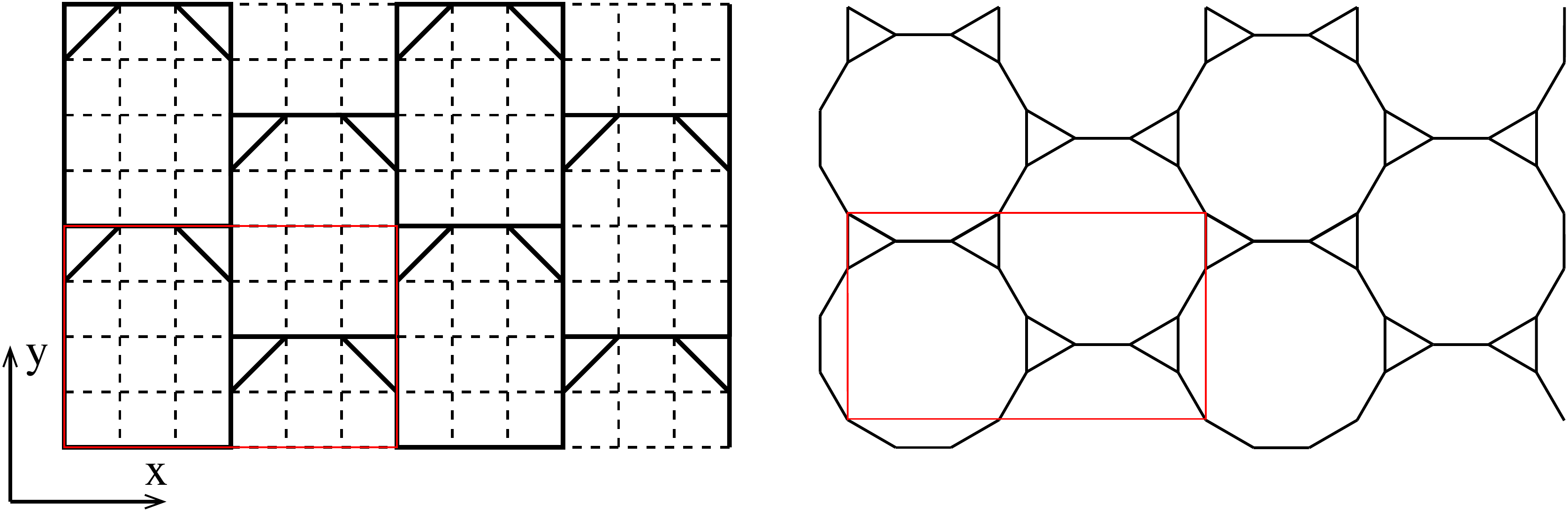}

	\hspace{1.6cm}    $6 \times 4$    \hspace{4.6cm}  $\left(3+2\sqrt{3}\right) \times \left(2+\sqrt{3}\right)$
	\caption{Three-twelve lattice. }
    \label{Fig:three-twelve}
\end{center}
\end{figure*}

\newpage

\vspace{5cm}

\setcounter{table}{0}
\renewcommand\thetable{S\arabic{table}}
\begin{table*}[hb]
\begin{center}
\caption{The number of vertices of Archimedean lattices with the linear size $L$.}
\label{Tab:lattice-size}
    \renewcommand\arraystretch{1.3}
    \setlength{\tabcolsep}{2.8mm}{
    \begin{tabular}[t]{llllll}
    \hline
    \hline
         Triangular & Frieze & Snub square & Snub hexagonal & Square & Ruby \\
         $L^2$ & $L^2/2$ & $L^2$ & $6L^2/7$ & $L^2$ & $6L^2/7$ \\  \cline{1-6}
         Kagome & Honeycomb & Four-eight & Cross & Three-twelve &  \\
         $3L^2/4$ & $L^2/2$ & $L^2/2$ & $3L^2/4$ & $L^2/3$ &  \\
    \hline
    \hline
    \end{tabular}}
\end{center}
\end{table*}

\begin{table*}[htbp]
\begin{center}
\caption{The number of independent samples taken at each patch size $\theta$ for one-patch disks on Archimedean lattices.
	For other models, the sample sizes are smaller but still being larger than or equal to $10^8$.}
\label{Tab:sample-size}
    \renewcommand\arraystretch{1.55}
    \setlength{\tabcolsep}{5.8mm}{
    \begin{tabular}[t]{lllll}
    \hline
    \hline
    \multirow{2}*{Triangular} & $L$ & $2-16$ & $32-128$ & $256$ \\  
         & Sample size & $10^{10}$ & $10^{9}$ & $10^{8}$ \\
    \hline 
      \multirow{2}*{Frieze} & $L$ & $4-16$ & $32-128$ & $256$ \\  
         & Sample size & $10^{10}$ & $10^{9}$ & $10^{8}$ \\
    \hline
      \multirow{2}*{Snub square} & $L$ & $4-16$ & $32-64$ & $128-256$ \\  
         & Sample size & $10^{10}$ & $10^{9}$ & $10^{8}$ \\
    \hline    
      \multirow{2}*{Snub hexagonal} & $L$ & $7-35$ & $70-140$ & $280$ \\  
         & Sample size & $10^{10}$ & $10^{9}$ & $10^{8}$ \\
    \hline
      \multirow{2}*{Square} & $L$ & $2-16$ & $32-128$ & $256$ \\  
         & Sample size & $10^{10}$ & $10^{9}$ & $10^{8}$ \\
    \hline    
      \multirow{2}*{Ruby} & $L$ & $7-14$ & $35-140$ & $280$ \\  
         & Sample size & $10^{10}$ & $10^{9}$ & $10^{8}$ \\
    \hline
      \multirow{2}*{Kagome} & $L$ & $2-16$ & $32-128$ & $256$ \\  
         & Sample size & $10^{10}$ & $10^{9}$ & $10^{8}$ \\
    \hline
    \multirow{2}*{Honeycomb} & $L$ & $4-32$ & $64-256$ & $512$ \\   
         & Sample size & $10^{10}$ & $10^{9}$ & $10^{8}$ \\
    \hline
      \multirow{2}*{Four-eight} & $L$ & $4-32$ & $64-256$ & $512$ \\  
         & Sample size & $10^{10}$ & $10^{9}$ & $10^{8}$ \\
    \hline    
      \multirow{2}*{Cross} & $L$ & $8-16$ & $32-128$ & $256$ \\  
         & Sample size & $10^{10}$ & $10^{9}$ & $10^{8}$ \\
    \hline
      \multirow{2}*{Three-twelve} & $L$ & $6-24$ & $48-192$ & $384$ \\  
         & Sample size & $10^{10}$ & $10^{9}$ & $10^{8}$ \\
    \hline    
    \hline    
    \end{tabular}}
\end{center}
\end{table*}

\begin{table*}[htbp]
\begin{center}
\caption{Simulation time for patchy particles on Archimedean lattices. The unit of time is one hour, 
	using one processor of an Intel Xeon Scalable Gold 6130 CPU under the hyper-threading mode.}
\label{Tab:time}
    \renewcommand\arraystretch{1.65}
    \setlength{\tabcolsep}{1.2mm}{
	    \begin{tabular}[t]{l|llllll|ll}
    \hline
    \hline
      \multirow{2}*{Lattice}  & \multicolumn{6}{c|}{Disk}   & \multicolumn{2}{c}{Sphere} \\  \cline{2-9}
        & One-patch & Two-patch & Three-patch & Four-patch & Five-patch & Six-patch & One-patch & Two-patch \\
    \hline
    Triangular & $28276$ & $7213$ & $333$ & $5906$ & $856$ & $2594$ & $7308$ & $1279$ \\
        Frieze & $45676$ & $3221$ & $2358$ & $2179$ & $1380$ & $2768$ & $1401$ & $1149$  \\
   Snub square & $26817$ & $2158$ & $1042$ & $459$ & $1235$ & $771$ & $4086$ & $579$  \\
Snub hexagonal & $44204$ & $2869$ & $2375$ & $2912$  & $1346$ & $1021$ & $6741$ & $466$ \\
      Square   & $38914$ & $3982$ & $818$ & $2012$ & $1025$ & $2491$ & $3186$ & $1518$  \\
          Ruby & $48448$ & $7251$ & $1392$ & $4147$ & $1184$ & $1311$ & $12140$ & $611$  \\
       Kagome  & $29382$ & $4825$ & $1171$ & $1785$ & $648$ & $1254$ & $5453$ &  $345$ \\
     Honeycomb & $60222$ & $1886$ & $3429$ & $14547$ & $1149$ & $1386$ & $7381$ & $345$  \\
    Four-eight & $52055$ & $3506$ & $531$ & $1068$ & $2056$ & $1451$ & $8403$ & $676$  \\
         Cross & $16923$ & $11950$ & $1738$ & $442$ & $1136$ & $1184$ & $5542$ & $278$  \\
  Three-twelve & $63629$ & $1567$ & $702$ & $644$ & $904$ & $1639$ & $17737$ & $627$  \\
    \hline
    \hline
    \end{tabular}}
\end{center}
\end{table*}

\begin{figure*}[ht]
\begin{center}
\includegraphics[scale=0.84]{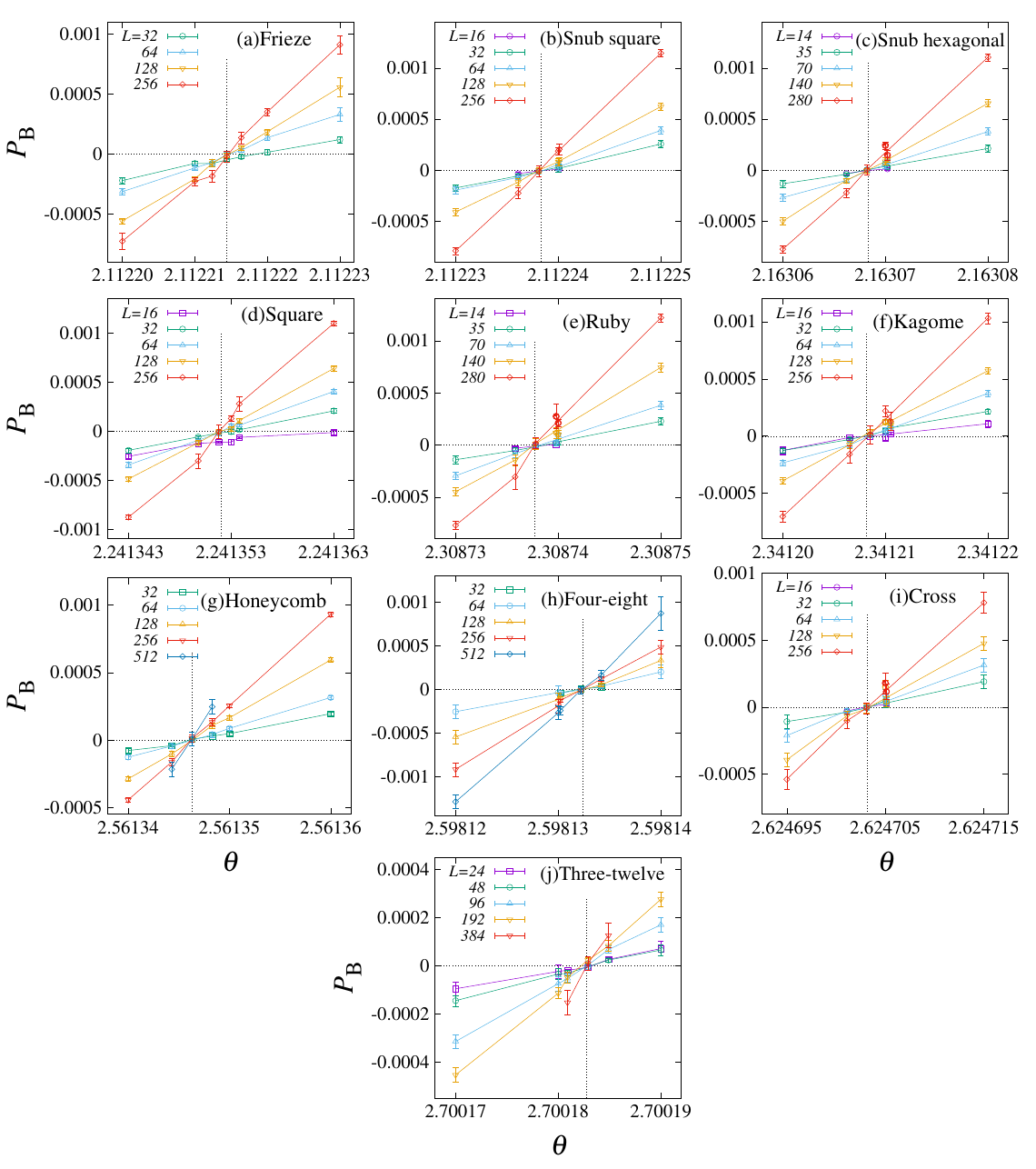}
	\caption{Intersection plots for one-patch disks on different lattices. 
	The vertical dashed lines show the estimated threshold values $\theta_c$.}
    \label{Fig:one-patch-disk}
\end{center}
\end{figure*}

\begin{figure*}[ht]
\begin{center}
\includegraphics[scale=0.8]{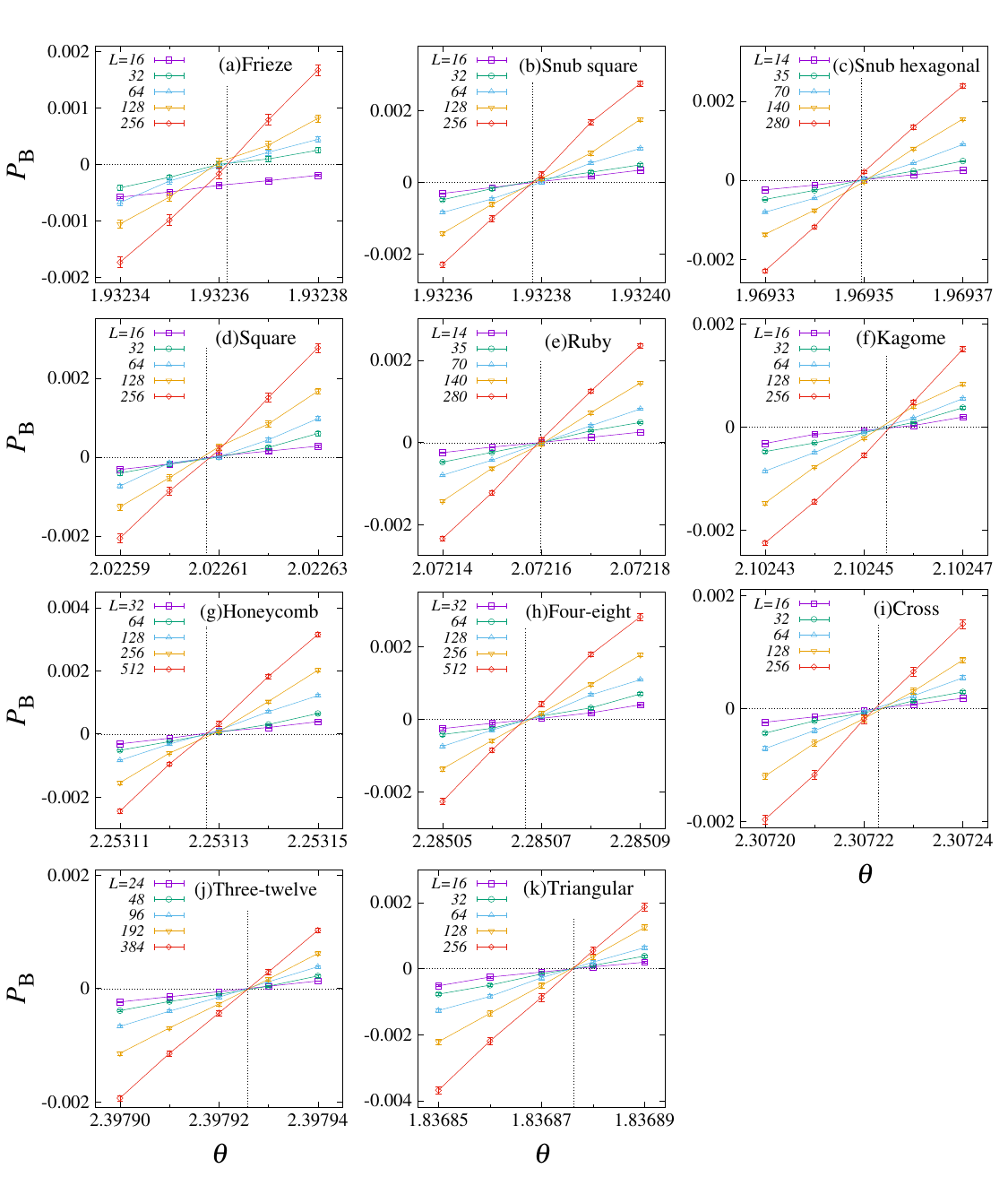}
	\caption{Intersection plots for one-patch spheres on different lattices.
	The vertical dashed lines show the estimated threshold values $\theta_c$.}
\label{Fig:one-patch-sphere}
\end{center}
\end{figure*}

\begin{figure*}[ht]
\begin{center}
\includegraphics[scale=0.8]{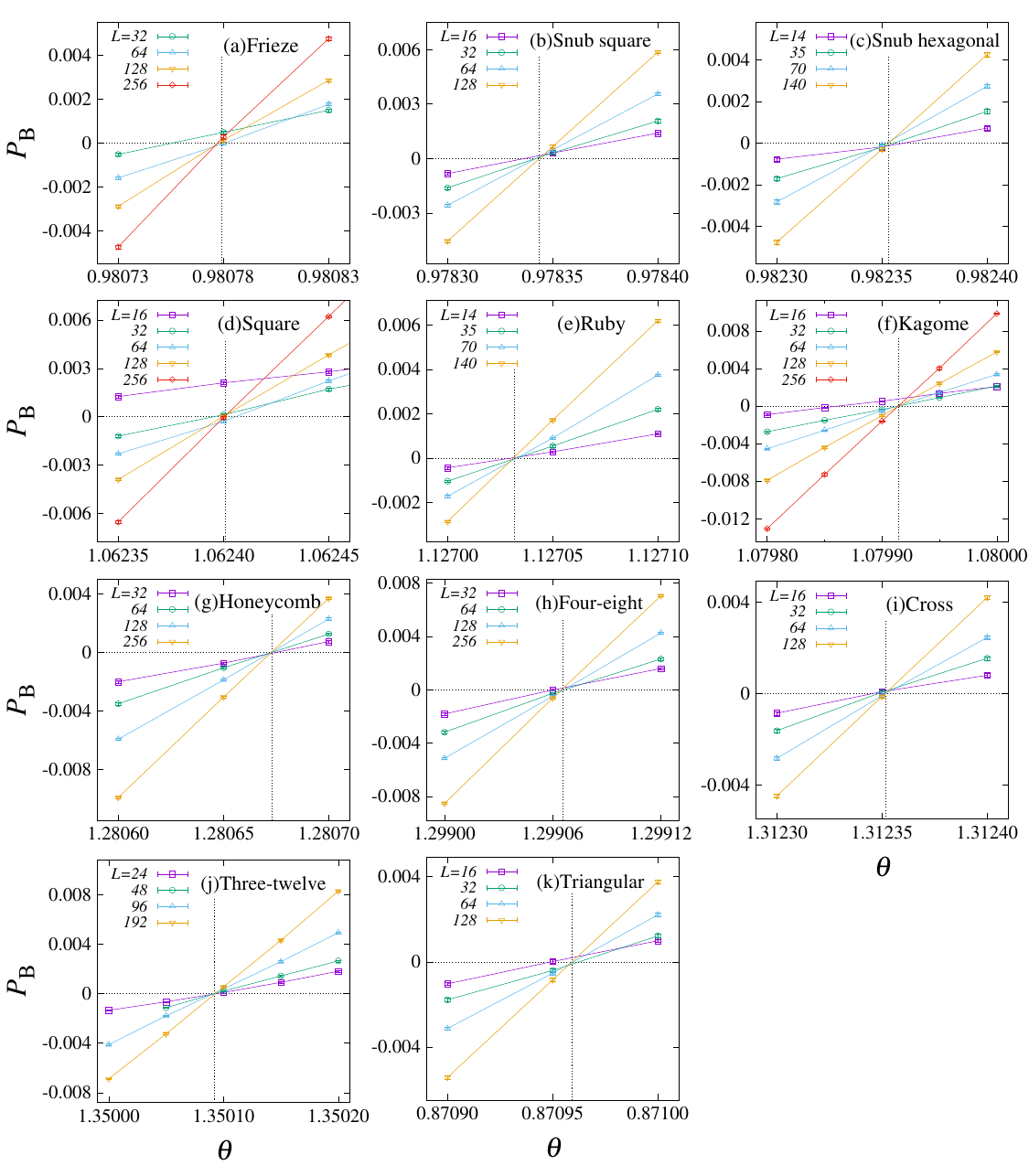}
	\caption{Intersection plots for two-patch disks on different lattices.
	The vertical dashed lines show the estimated threshold values $\theta_c$.}
\label{Fig:two-patch-disk}
\end{center}
\end{figure*}

\begin{figure*}[ht]
\begin{center}
\includegraphics[scale=0.8]{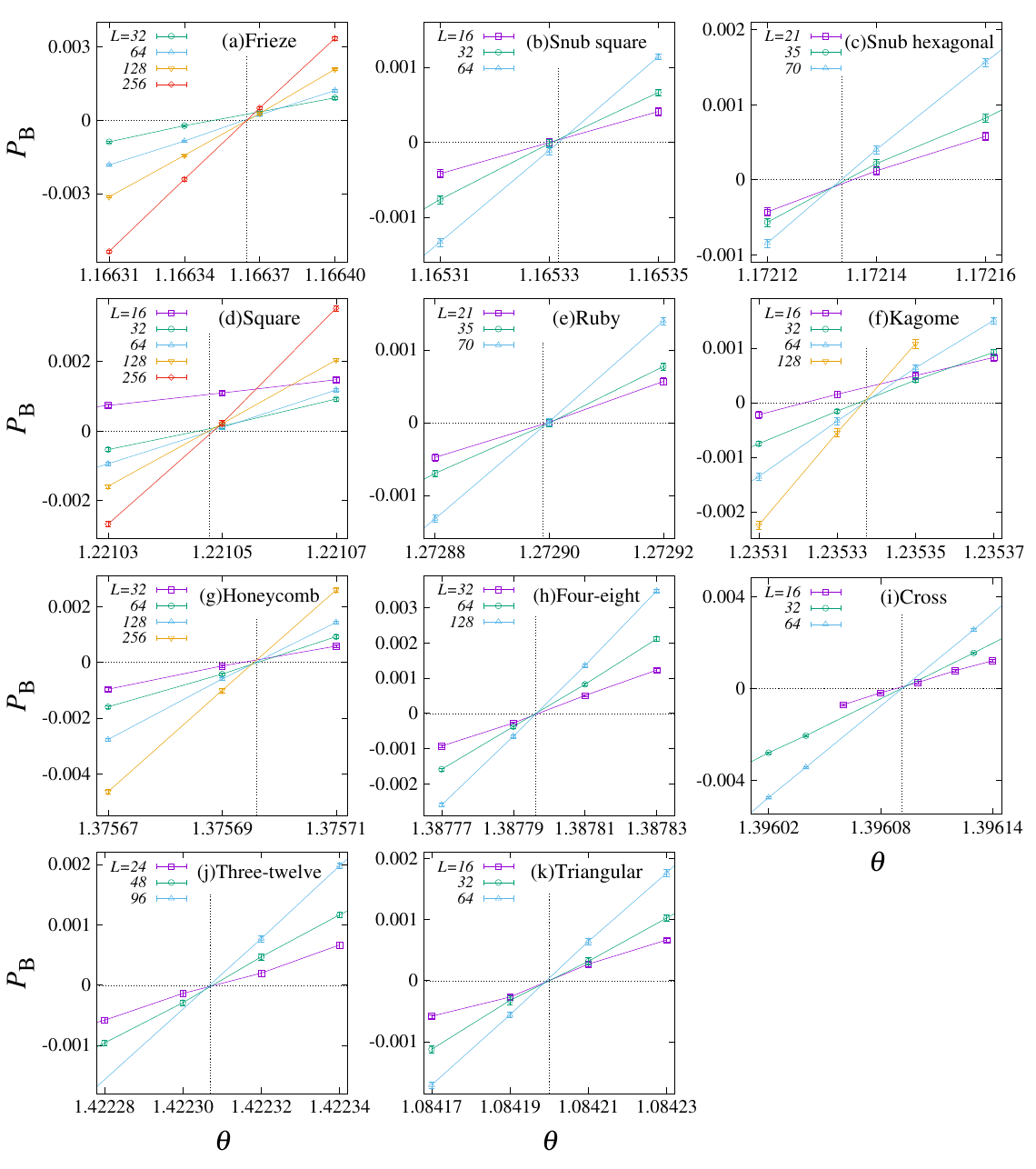}
	\caption{Intersection plots for two-patch spheres on different lattices.
	The vertical dashed lines show the estimated threshold values $\theta_c$.}
\label{Fig:two-patch-sphere}
\end{center}
\end{figure*}

\begin{figure*}[ht]
\begin{center}
\includegraphics[scale=0.8]{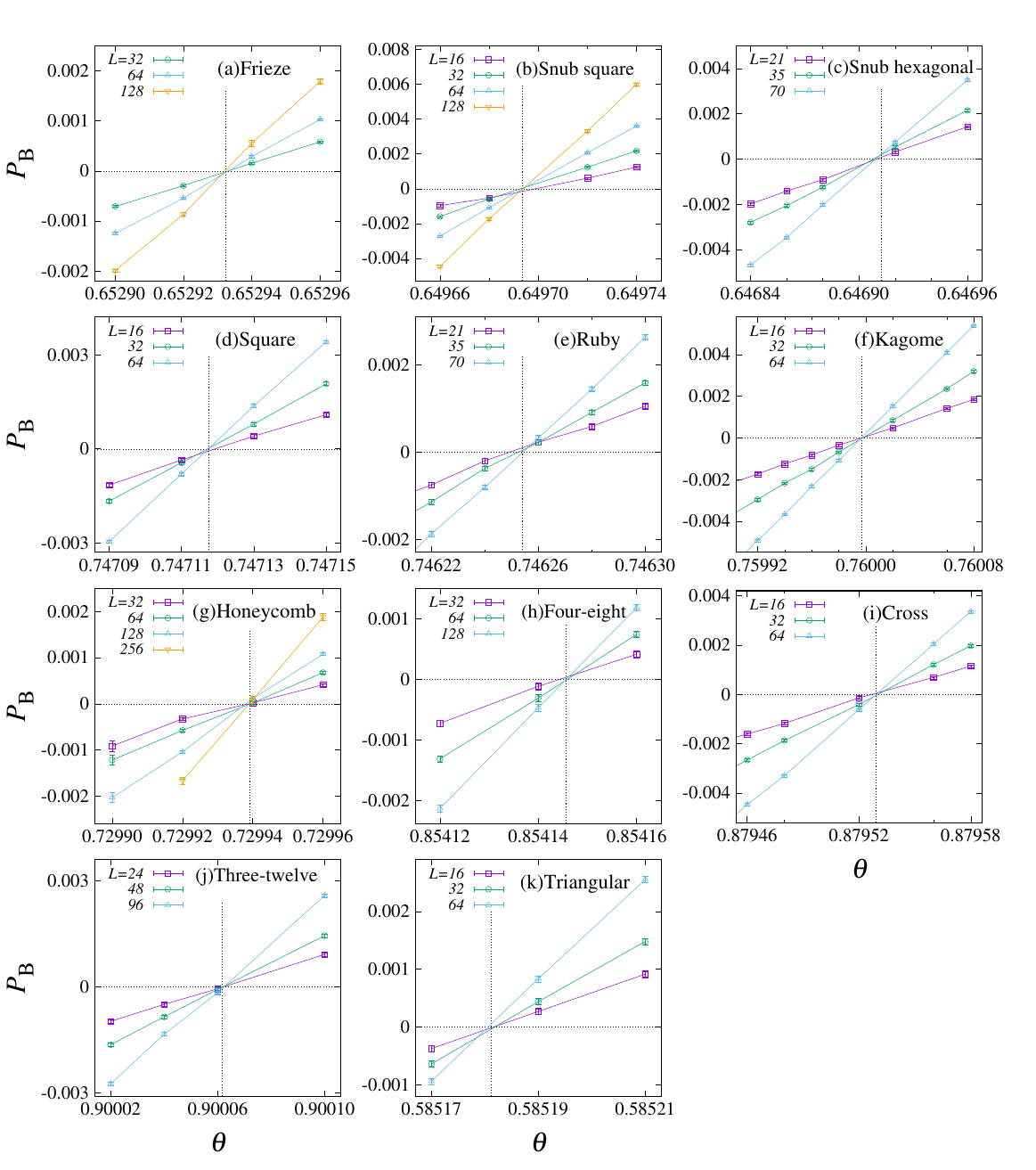}
	\caption{Intersection plots for three-patch disks on different lattices.
	The vertical dashed lines show the estimated threshold values $\theta_c$.}
    \label{Fig:three-patch-disk}
\end{center}
\end{figure*}

\begin{figure*}[ht]
\begin{center}
\includegraphics[scale=0.8]{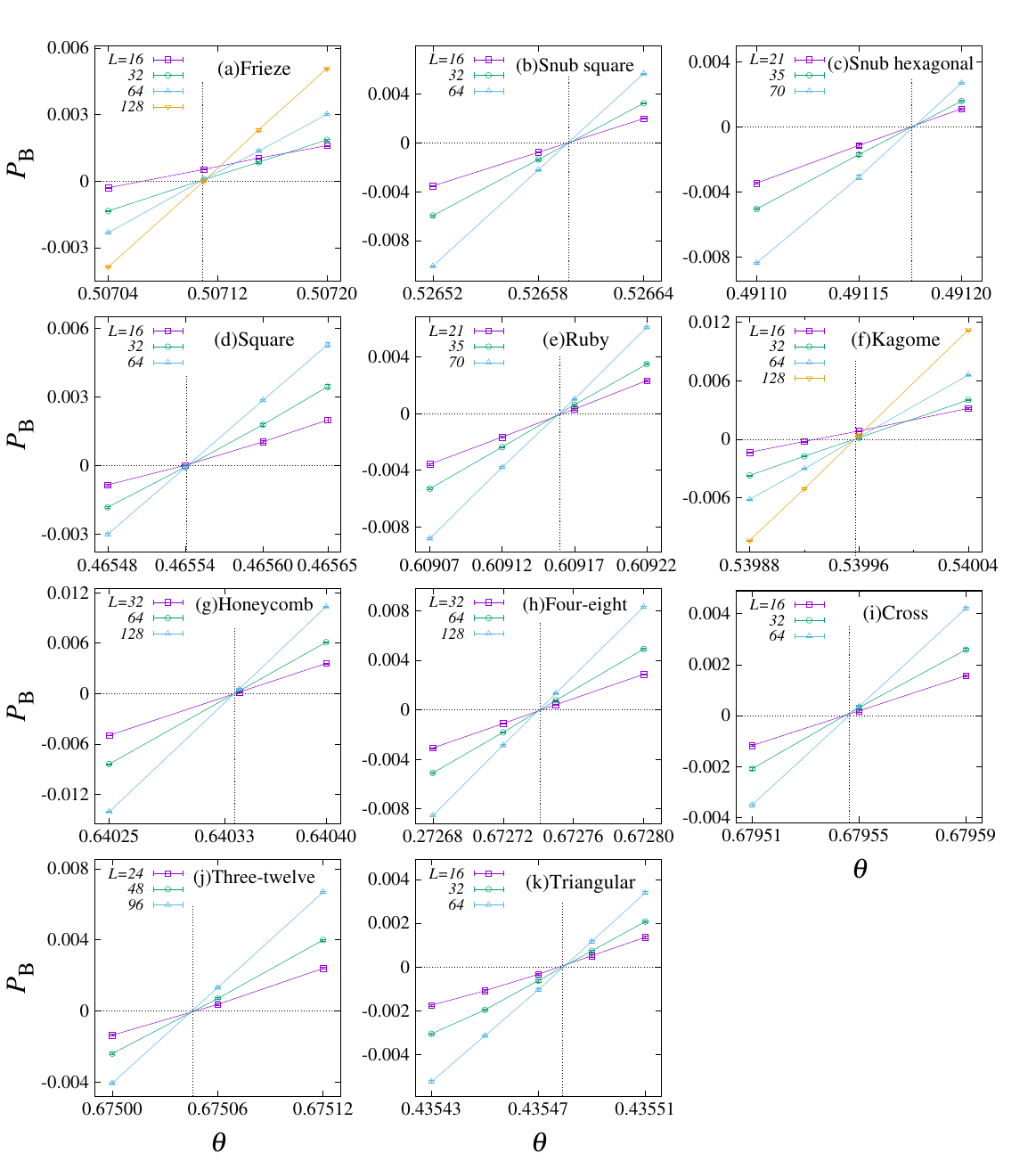}
	\caption{Intersection plots for four-patch disks on different lattices.
	The vertical dashed lines show the estimated threshold values $\theta_c$.}
\label{Fig:four-patch-disk}
\end{center}
\end{figure*}

\begin{figure*}[ht]
\begin{center}
\includegraphics[scale=0.8]{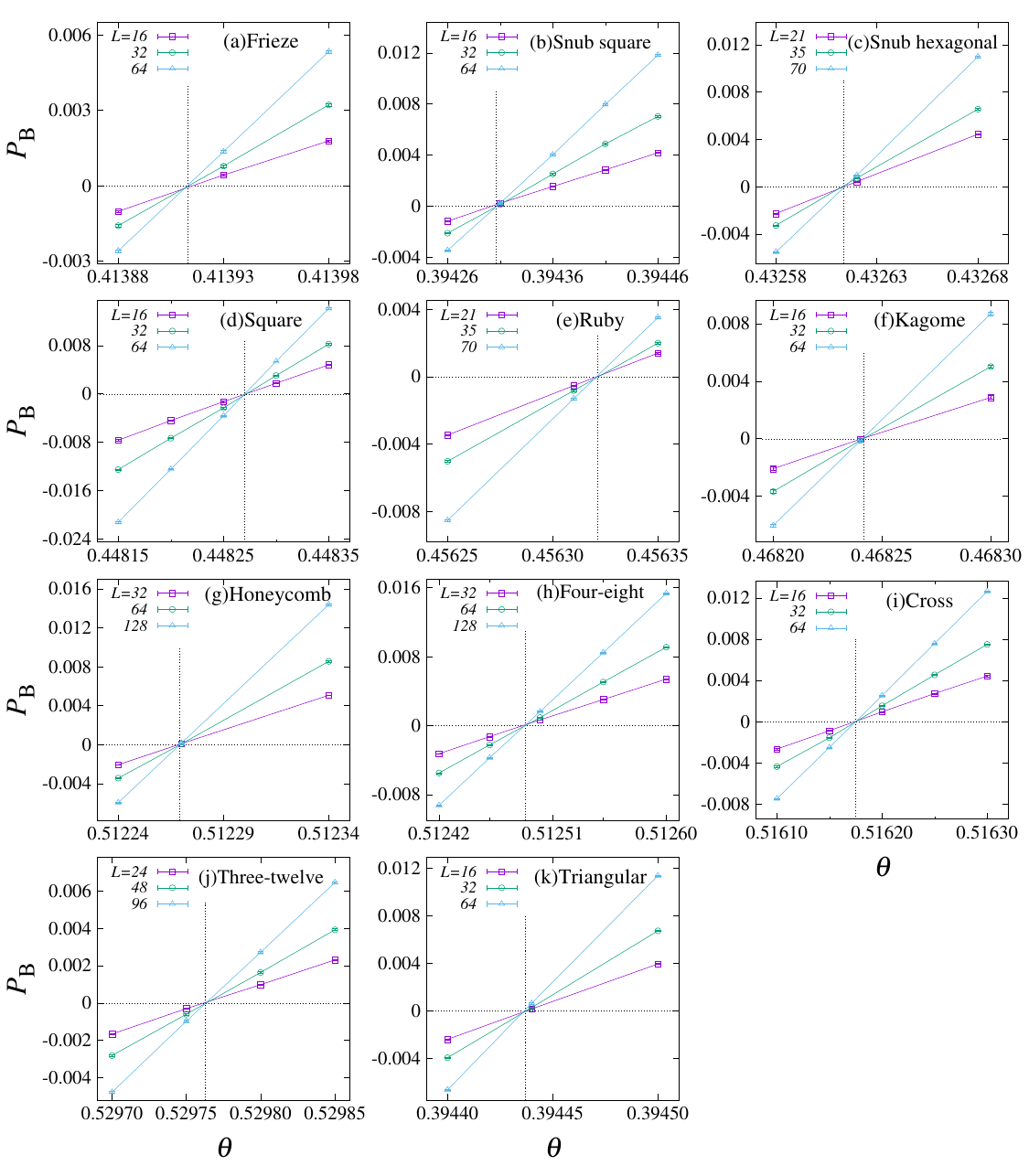}
	\caption{Intersection plots for five-patch disks on different lattices.
	The vertical dashed lines show the estimated threshold values $\theta_c$.}
    \label{Fig:five-patch-disk}
\end{center}
\end{figure*}

\begin{figure*}[ht]
\begin{center}
\includegraphics[scale=0.8]{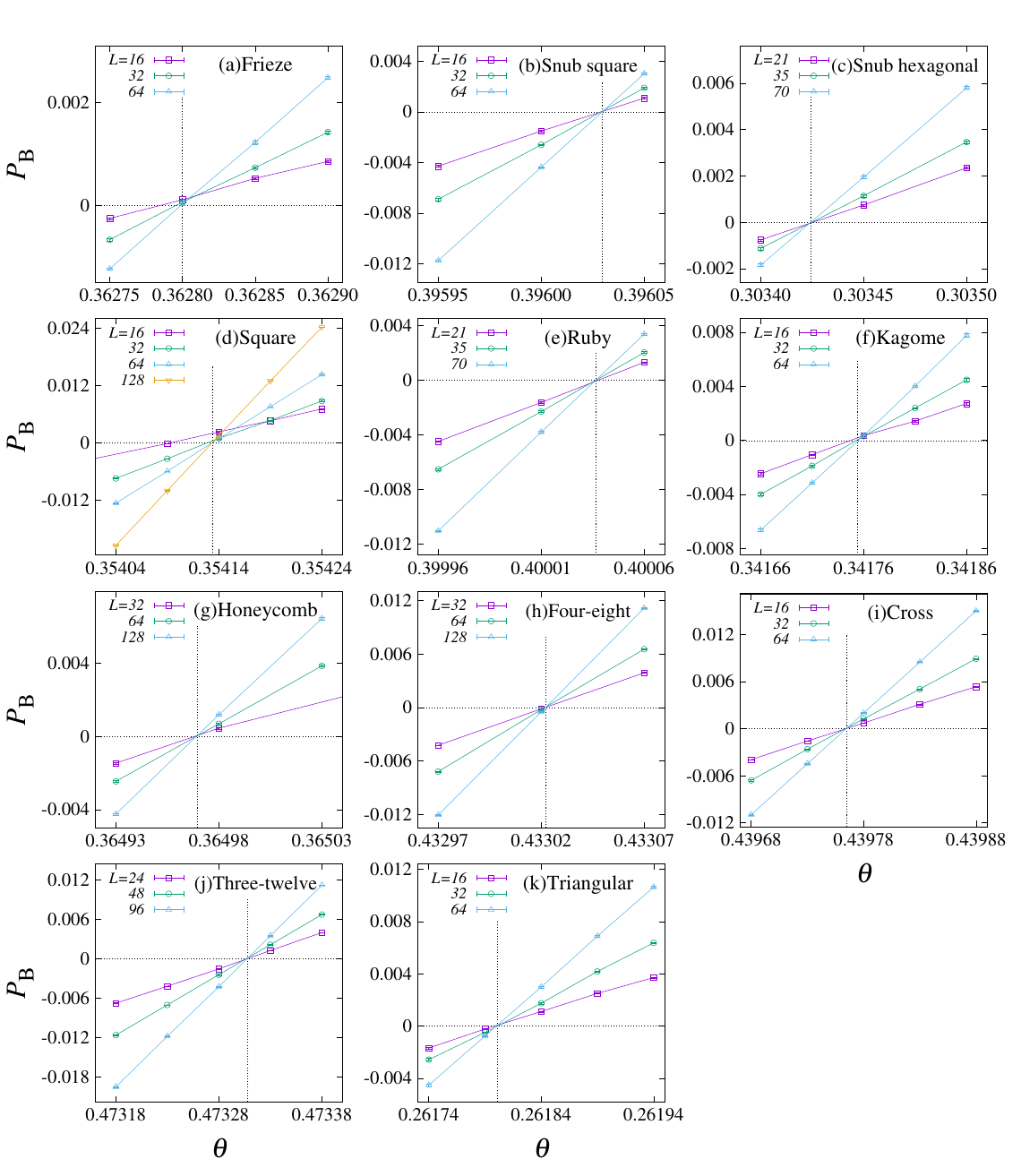}
	\caption{Intersection plots for six-patch disks on different lattices.
	The vertical dashed lines show the estimated threshold values $\theta_c$.}
\label{Fig:six-patch-disk}
\end{center}
\end{figure*}

\begin{figure*}[ht]
\begin{center}
\includegraphics[scale=0.8]{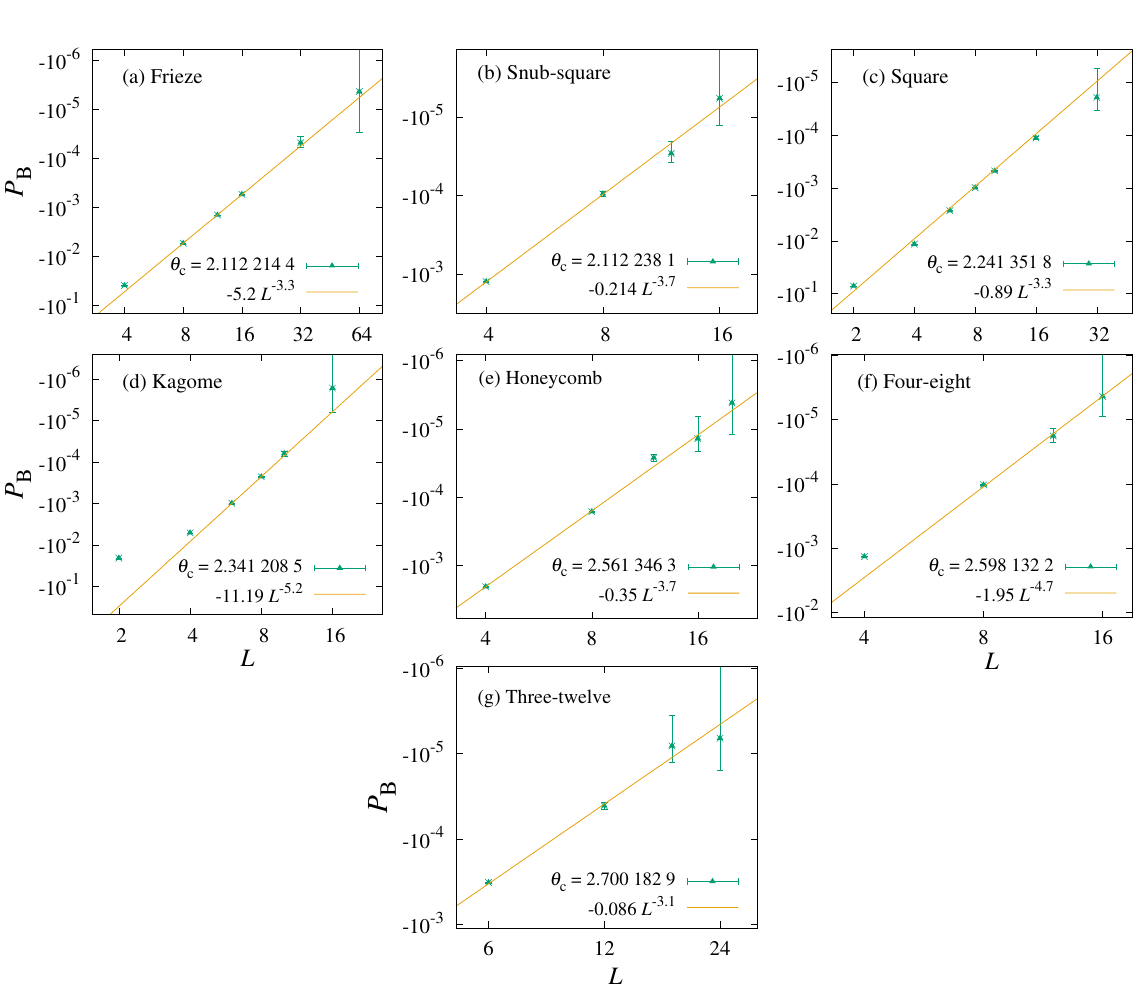}
	\caption{Plots of $\PB$ versus $L$ at the percolation threshold $\theta_c$ for one-patch disks on different lattices.}
\label{Fig:one-patch-disk-small-size}
\end{center}
\end{figure*}

\begin{figure*}[ht]
\begin{center}
\includegraphics[scale=0.8]{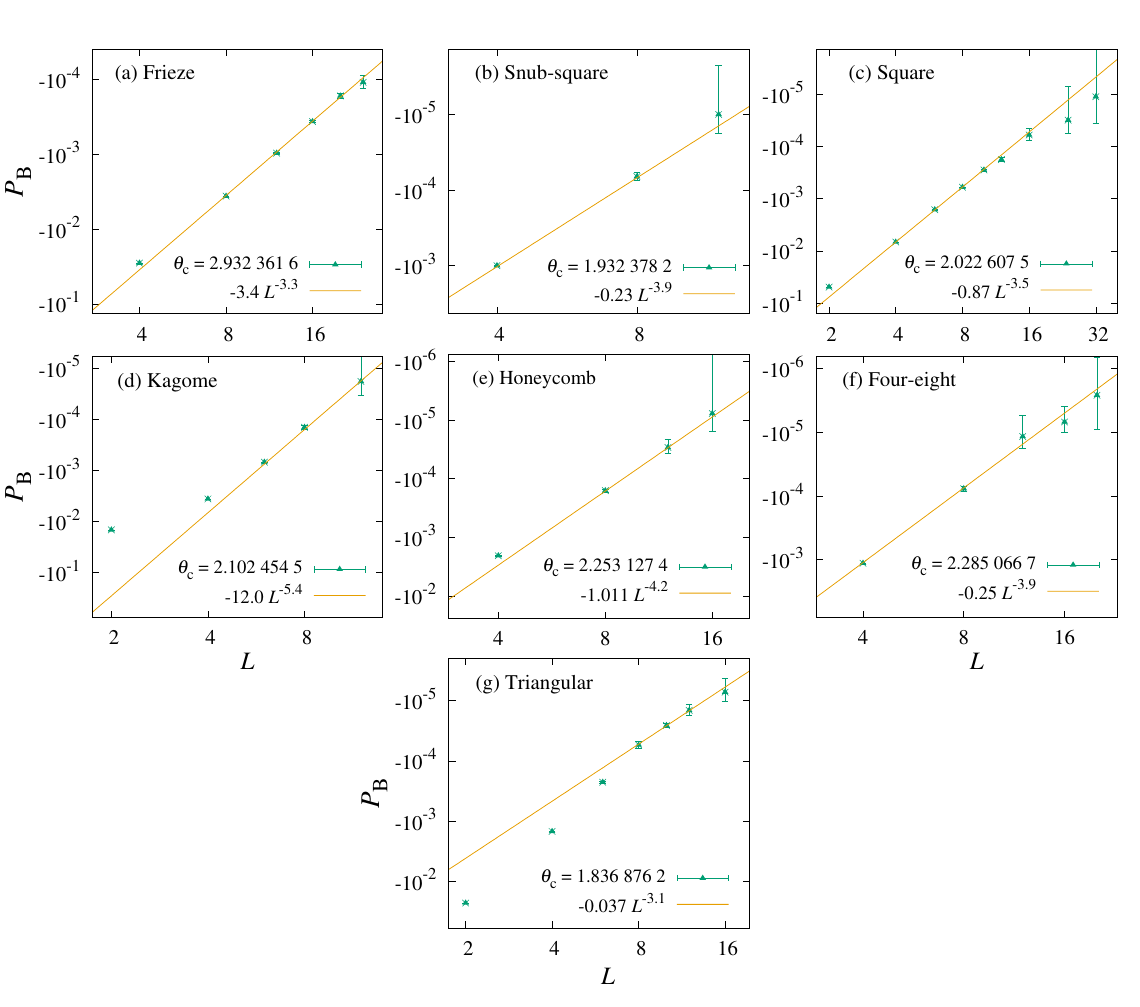}
	\caption{Plots of $\PB$ versus $L$ at the percolation threshold $\theta_c$ for one-patch spheres on different lattices.}
\label{Fig:one-patch-sphere-small-size}
\end{center}
\end{figure*}

\begin{figure*}[ht]
\begin{center}
\includegraphics[scale=0.6]{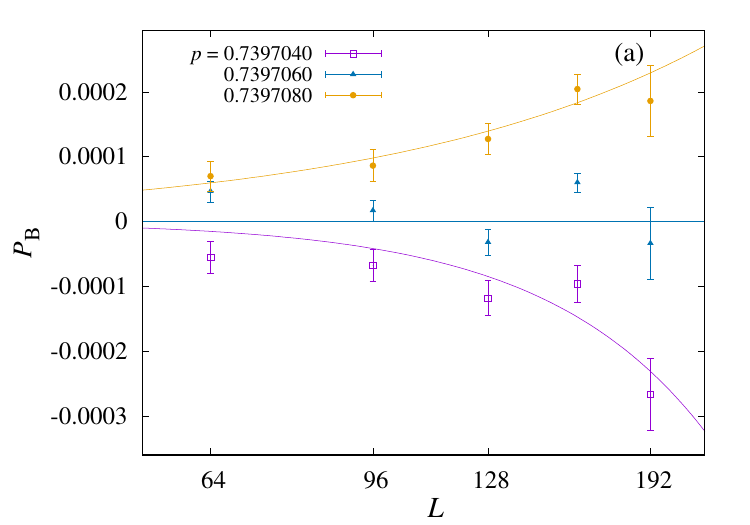} 
\includegraphics[scale=0.6]{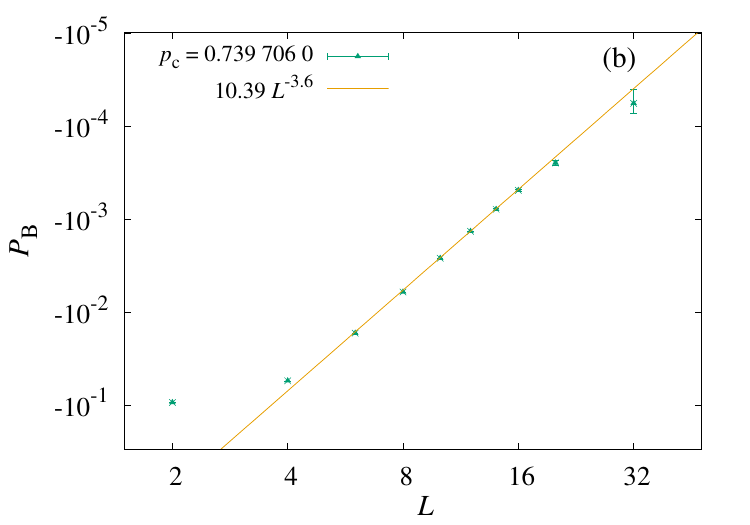} 
	\caption{The data of critical polynomial $\PB$ for site percolation on the Lieb lattice in 2D.}
\label{Fig:Lieb}
\end{center}
\end{figure*}

\begin{figure}[p]
\begin{adjustbox}{addcode={
\begin{minipage}{\width}}{
\caption{The plots show different patch-covering structures for a one-patch disk on the frieze lattice.
	For each plot, when the disk rotates clockwise within the given angle ($\theta_{i,j}=2\pi\chi_{i,j}$
	with $\chi_{i,j}$ values given in the text), the patch-covering states of edges do not change,
	Plots (a-e) are $3$-edge patch-covering structures, and plots (f-j) are four-edge patch-covering structures.}
\label{Fig:frieze-edge-covering}
\end{minipage}},rotate=90,center}
\includegraphics[width = 26cm]{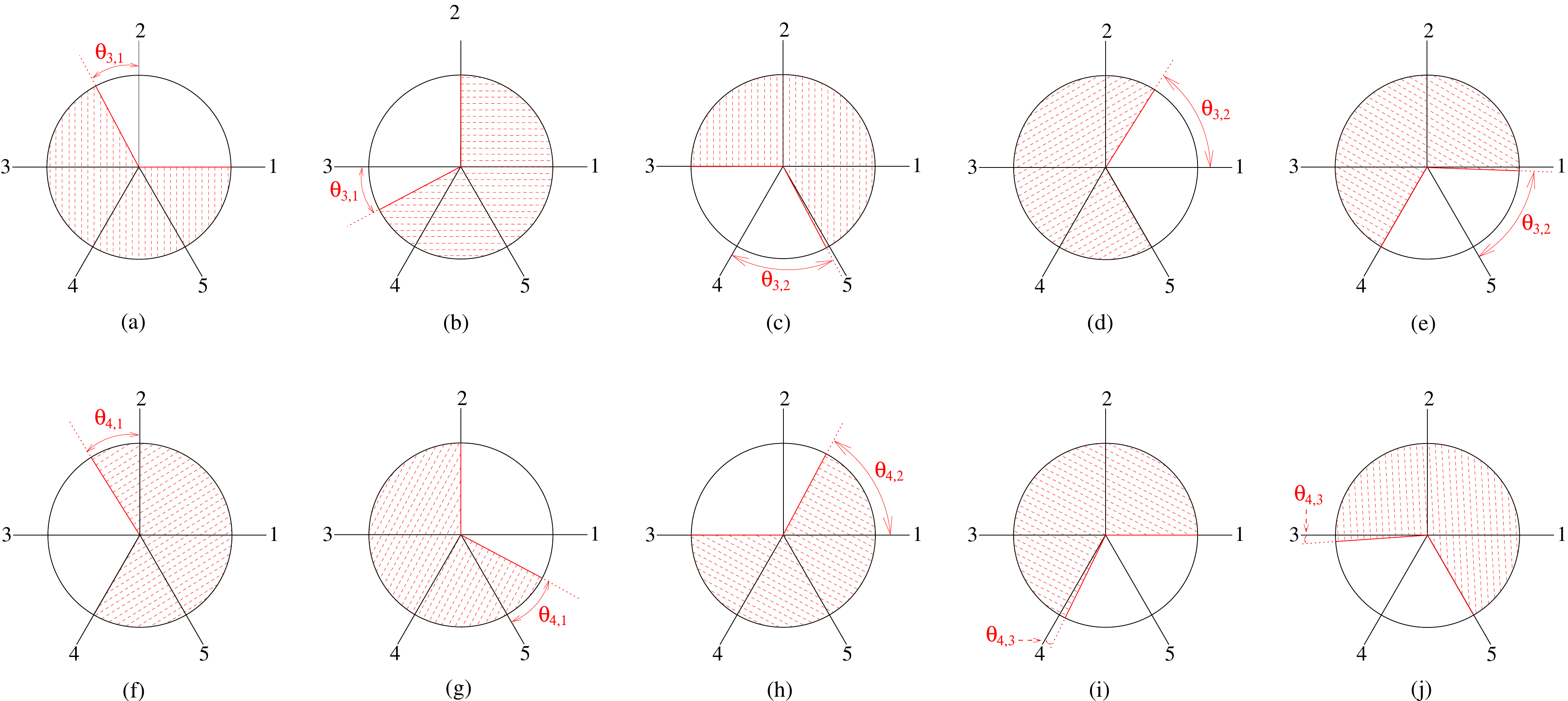}
\end{adjustbox}
\end{figure}

\begin{figure}[p]
\begin{adjustbox}{addcode={
\begin{minipage}{\width}}{
	\caption{A seven-patch disk rotates counterclockwise on a vertex of the triangular lattice. 
	Initially, patch $1$ is to the left of edge $1$ (the direction is defined assuming that the observer stands 
	at the disk center and faces outwards), with its right boundary touches the edge. 
	When the disk rotates by an angle $\theta_1$ ($\theta_1=2\pi\chi_1$, with $\chi_1$ given by Eq.~\ref{eq:tri-theta1} in the text), 
	the left boundary patch $5$ touches edge $5$ (the patch is to the right side of edge $5$). Within this angle $\theta_1$, 
	the patches always covers three edges, i.e. edge $2$, $3$ and $4$. And within the following angle $\theta_2$ 
	($\theta_2=2\pi\chi_2$, with $\chi_2$ given by Eq.~\ref{eq:tri-theta2} in the text), the patches always cover four edges, 
	i.e. edge $2$, $3$, $4$ and $5$. As the disk rotates counterclockwise further, $3$-edge patching-covering structures 
	and $4$-edge patch-covering structures appear in turn within an angle $\theta_1$ or $\theta_2$, respectively.
	At the bottom, the symbol \textcircled{k} means that edge $k$ is going to be uncovered by a patch to its left side,
	and \textcircled{k}* means that edge $k$ is going to be covered by a patch to its right side.
	The total angle between two \textcircled{1}s is $2\pi/7$.}
\label{Fig:seven-patch-rotate}
\end{minipage}},rotate=90,center}
\includegraphics[width = 26cm]{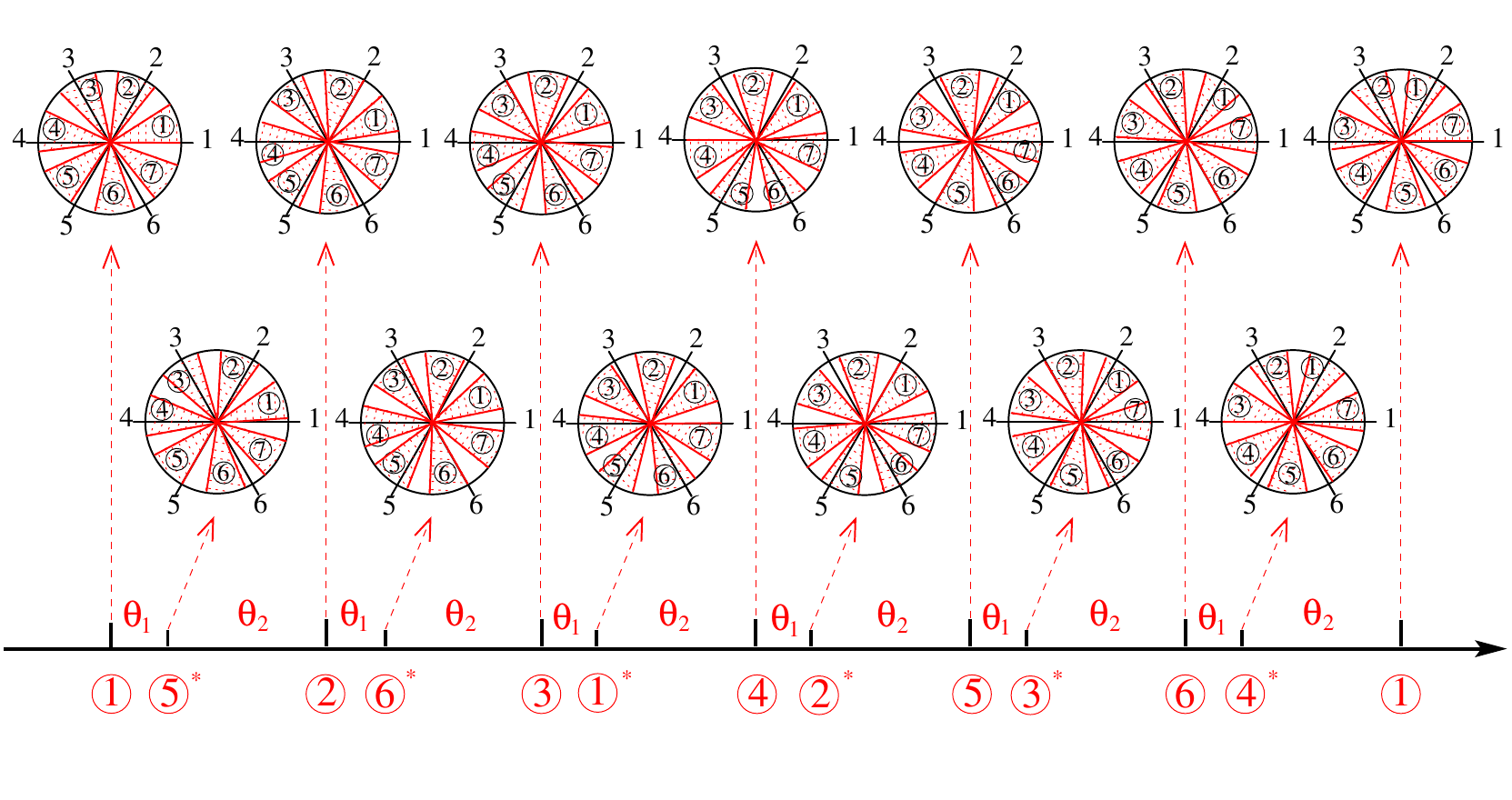}
\end{adjustbox}
\end{figure}

\begin{table*}[htbp]
\begin{center}
\caption{Probabilities of different patch-covering structures for patch sizes $\chi$ near ${\chi}_c$, 
	for $n$-patch disks on the triangular, square and honeycomb lattices. }
\label{Tab:contact-proba}
    \renewcommand\arraystretch{2.7}
    \setlength{\tabcolsep}{2.4mm}{
    \begin{tabular}[t]{llllllllll}
    \hline
    \hline
	    Lattice & \hspace{-4mm} $\mod(n,n_0)$ & Type & Probability \\
    \hline
              & \multirow{2}*{$1$} & {$3$}-edge &  $\left(\dfrac{1-{\chi}}{n}+\dfrac{n-1}{3} \cdot \dfrac{1}{n}-\dfrac{1}{3}\right) \cdot n \cdot 6 = 4-6{\chi}$  \\
              &              & {$4$}-edge &  $\left(\dfrac{{\chi}}{n}+\dfrac{n-1}{2} \cdot \dfrac{1}{n}-\dfrac{1}{2}\right) \cdot n \cdot 6 = 6{\chi}-3$  \\   \cline{2-4}
              & \multirow{2}*{$2$} & {$2$}-edge &  $\left(\dfrac{1-{\chi}}{n}+\dfrac{2n-1}{3} \cdot \dfrac{1}{n}-\dfrac{2}{3}\right) \cdot n \cdot 3 = 2-3{\chi}$  \\
              &              & {$4$}-edge &  $\left(\dfrac{{\chi}}{n}+\dfrac{2n-1}{3} \cdot \dfrac{1}{n}-\dfrac{2}{3}\right) \cdot n \cdot 3 = 3{\chi}-1$  \\   \cline{2-4}
  \multirow{2}*{\makecell[tl]{Triangular\\ $n_0=6$}}  & \multirow{2}*{$3$} & {$3$}-edge &  $\dfrac{{1-\chi}}{n} \cdot n \cdot 2 = 2-2{\chi}$  \\
              &              & {$6$}-edge &  $\left(\dfrac{1}{6}-\dfrac{n-3}{6} \cdot \dfrac{1}{n}-\dfrac{{1-\chi}}{n}\right) \cdot n \cdot 2 = 2{\chi}-1$  \\   \cline{2-4}
              & \multirow{2}*{$4$} & {$2$}-edge &  $\left(\dfrac{1-{\chi}}{n}+\dfrac{n-1}{3} \cdot \dfrac{1}{n}-\dfrac{1}{3}\right) \cdot n \cdot 3 = 2-3{\chi}$  \\
              &              & {$4$}-edge &  $\left(\dfrac{{\chi}}{n}+\dfrac{n-1}{3} \cdot \dfrac{1}{n}-\dfrac{1}{3}\right) \cdot n \cdot 3 = 3{\chi}-1$  \\   \cline{2-4}
              & \multirow{2}*{$5$} & {$3$}-edge &  $\left(\dfrac{1-{\chi}}{n}+\dfrac{2n-1}{3} \cdot \dfrac{1}{n}-\dfrac{2}{3}\right) \cdot n \cdot 6 = 4-6{\chi}$  \\
              &              & {$4$}-edge &  $\left(\dfrac{{\chi}}{n}+\dfrac{n-1}{2} \cdot \dfrac{1}{n}-\dfrac{1}{2}\right) \cdot n \cdot 6 = 6{\chi}-3$  \\
    \hline
              & \multirow{2}*{$1$} & {$2$}-edge &  $\left(\dfrac{{1-\chi}}{n}+\dfrac{n-1}{4} \cdot \dfrac{1}{n}-\dfrac{1}{4}\right) \cdot n \cdot 4 = 3-4{\chi}$   \\
              &              & {$3$}-edge &  $\left(\dfrac{{\chi}}{n}+\dfrac{n-1}{2} \cdot \dfrac{1}{n}-\dfrac{1}{2}\right) \cdot n \cdot 4 = 4{\chi}-2$   \\   \cline{2-4}
      \multirow{2}*{\makecell[tl]{Square \\ $n_0=4$}}       & \multirow{2}*{$2$} & {$2$}-edge &  $\dfrac{{1-\chi}}{n} \cdot n \cdot 2 = 2-2{\chi}$  \\
              &              & {$4$}-edge &  $\left(\dfrac{1}{4}-\dfrac{n-2}{4} \cdot \dfrac{1}{n}-\dfrac{{1-\chi}}{n}\right) \cdot n \cdot 2 = 2{\chi}-1$   \\    \cline{2-4}
              & \multirow{2}*{$3$} & {$2$}-edge &  $\left(\dfrac{1-{\chi}}{n}+\dfrac{3n-1}{4} \cdot \dfrac{1}{n}-\dfrac{3}{4}\right) \cdot n \cdot 4 = 3-4{\chi}$   \\
              &              & {$3$}-edge &  $\left(\dfrac{{\chi}}{n}+\dfrac{n-1}{2} \cdot \dfrac{1}{n}-\dfrac{1}{2}\right) \cdot n \cdot 4 = 4{\chi}-2$   \\
    \hline
              & \multirow{2}*{$1$} & {$2$}-edge &  $\dfrac{{1-\chi}}{n} \cdot n \cdot 3 = 3-3{\chi}$   \\
 \multirow{2}*{\makecell[tl]{Honeycomb \\ $n_0=3$}} &  & {$3$}-edge &   $\left(\dfrac{{\chi}}{n}+\dfrac{2n-2}{3} \cdot \dfrac{1}{n}-\dfrac{2}{3}\right) \cdot n \cdot 3 = 3{\chi}-2$  \\  \cline{2-4}
              & \multirow{2}*{$2$} & {$2$}-edge &  $\dfrac{{1-\chi}}{n} \cdot n \cdot 3 = 3-3{\chi}$   \\
              &              & {$3$}-edge &  $\left(\dfrac{{\chi}}{n}+\dfrac{n-2}{3} \cdot \dfrac{1}{n}-\dfrac{1}{3}\right) \cdot n \cdot 3 = 3{\chi}-2$   \\

    \hline
    \hline
    \end{tabular}}
\end{center}
\end{table*}

\newpage

\begin{table*}[htbp]
\begin{center}
\caption{Probabilities of different patch-covering structures for patch sizes $\chi$ near ${\chi}_c$,
        for $n$-patch disks on the snub hexagonal lattice. The period is $n_0=6$.}
\label{Tab:contact-proba-snub-hex}
    \renewcommand\arraystretch{2.7}
    \setlength{\tabcolsep}{2.0mm}{
    \begin{tabular}[t]{lllllllll}
    \hline
    \hline
	    \hspace{-4mm} $\mod(n,n_0)$ & Type & {On the triangular lattice} &  {On the snub hexagonal lattice}  \\
    \hline
                        \multirow{3}*{$1$} & {$3$}-edge &  & $\left(5-6{\chi} \right) \cdot \dfrac{4}{6} = \dfrac{10}{3}-4{\chi}$  &   \\
                                           & {$4$}-edge &  $\left(\dfrac{{1-\chi}}{n}+\dfrac{n-1}{6} \cdot \dfrac{1}{n}-\dfrac{1}{6}\right) \cdot n \cdot 6= 5-6{\chi}$ &  $\left(5-6{\chi}\right) \cdot \dfrac{2}{6}+\left(6{\chi}-4\right) \cdot \dfrac{5}{6} = 3{\chi}-\dfrac{5}{3}$  &    \\
                                           & {$5$}-edge &  $\left(\dfrac{1}{3}-\dfrac{n-1}{3} \cdot \dfrac{1}{n}-\dfrac{{1-\chi}}{n}\right) \cdot n \cdot 6= 6{\chi}-4$  &  $\left(6{\chi}-4 \right) \cdot \dfrac{1}{6} = {\chi}-\dfrac{2}{3}$  &   \\
    \hline

                        \multirow{4}*{$2$} & {$1$}-edge &    &  $\left(2-3{\chi}\right)\cdot \dfrac{2}{6} = \dfrac{2}{3}-{\chi}$  &  \\
                                           & {$2$}-edge &  $\left(\dfrac{1}{3}-\dfrac{n-2}{3} \cdot \dfrac{1}{n}-\dfrac{{\chi}}{n}\right) \cdot n \cdot 3 = 2-3{\chi}$   &  $\left(2-3{\chi}\right) \cdot \dfrac{4}{6} = \dfrac{4}{3}-2{\chi}$  &  \\
                                           & {$3$}-edge &   &  $\left(3{\chi}-1\right) \cdot \dfrac{4}{6} = 2{\chi}-\dfrac{2}{3}$  &   \\
                                           & {$4$}-edge &  $\left(\dfrac{{\chi}}{n}+\dfrac{n-2}{6} \cdot \dfrac{1}{n}-\dfrac{1}{6}\right) \cdot n \cdot 3 = 3{\chi}-1$  &  $\left(3{\chi}-1\right) \cdot \dfrac{2}{6} = {\chi}-\dfrac{1}{3}$  &   \\
    \hline

                        \multirow{4}*{$3$} & {$2$}-edge &    &  $ \left(2-2{\chi}\right) \cdot \dfrac{3}{6} = 1-{\chi}$ &   \\
                                           & {$3$}-edge &  $\dfrac{{1-\chi}}{n} \cdot n \cdot 2 = 2-2{\chi}$  & $\left(2-2{\chi}\right) \cdot \dfrac{3}{6} = 1-{\chi}$  &   \\
                                           & {$5$}-edge &    &  $\left(2{\chi}-1\right) \cdot \dfrac{6}{6} = 2{\chi}-1$  &   \\
                                           & {$6$}-edge &  $\left(\dfrac{{\chi}}{n}+\dfrac{n-3}{6} \cdot \dfrac{1}{n}-\dfrac{1}{6}\right) \cdot n \cdot 2 = 2{\chi}-1$  &    \\
    \hline

                        \multirow{4}*{$4$} & {$1$}-edge &   &  $\left(2-3{\chi}\right) \cdot \dfrac{2}{6} = \dfrac{2}{3}-{\chi}$  &   \\
                                           & {$2$}-edge & $\left(\dfrac{1}{6}-\dfrac{{n-4}}{6} \cdot \dfrac{1}{n}-\dfrac{{\chi}}{n}\right) \cdot n \cdot 3 = 2-3{\chi}$  & $\left(2-3{\chi}\right) \cdot \dfrac{4}{6} = \dfrac{4}{3}-2{\chi}$  &      \\
                                           & {$3$}-edge &   & $\left(3{\chi}-1\right) \cdot \dfrac{4}{6} = 2{\chi}-\dfrac{2}{3}$  &   \\
                                           & {$4$}-edge &  $\left(\dfrac{{\chi}}{n}+\dfrac{n-1}{3} \cdot \dfrac{1}{n}-\dfrac{1}{3}\right) \cdot n \cdot 3 = 3{\chi}-1$  & $\left(3{\chi}-1\right) \cdot \dfrac{2}{6} = {\chi}-\dfrac{1}{3}$  &  \\
    \hline

                        \multirow{3}*{$5$} & {$3$}-edge &    & $\left(5-6{\chi}\right) \cdot \dfrac{4}{6} = \dfrac{10}{3}-4{\chi}$ &  \\
                                           & {$4$}-edge &  $\left(\dfrac{1}{6}-\dfrac{{n-5}}{6} \cdot \dfrac{1}{n}-\dfrac{{\chi}}{n}\right) \cdot n \cdot 6 = 5-6{\chi}$  & $\left(5-6{\chi}\right) \cdot \dfrac{2}{6}+\left(6{\chi}-4\right) \cdot \dfrac{5}{6} = 3{\chi}-\dfrac{5}{3}$   &   \\
                                           & {$5$}-edge &  $\left(\dfrac{{\chi}}{n}+\dfrac{n-2}{3} \cdot \dfrac{1}{n}-\dfrac{1}{3}\right) \cdot n \cdot 6 = 6{\chi}-4$   &  $\left(6{\chi}-4\right) \cdot \dfrac{1}{6} = {\chi}-\dfrac{2}{3}$  &    \\
    \hline
    \hline
    \end{tabular}}
\end{center}
\end{table*}

\newpage

\begin{table*}[htbp]
\begin{center}
\caption{Probabilities of different patch-covering structures for patch sizes $\chi$ near ${\chi}_c$,
        for $n$-patch disks on the kagome lattice.  The period is $n_0=6$.
	The $2$-edge structure for $\mod(n,6)=3$ has the two
        edges at an angle of $2\pi/3$, while that for $\mod(n,6)=2$ or $4$ has the two edges in a straight line.}
\label{Tab:contact-proba-kagome}
    \renewcommand\arraystretch{2.7}
    \setlength{\tabcolsep}{2.4mm}{
    \begin{tabular}[t]{lllllllll}
    \hline
    \hline
	    \hspace{-4mm} $\mod(n,n_0)$ & Type & {On the triangular lattice} &  {On the kagome lattice}  \\
    \hline
                         \multirow{4}*{$1$} & {$2$}-edge &   &  $\left(5-6{\chi}\right) \cdot \dfrac{2}{6} = \dfrac{5}{3}-2{\chi}$   \\
                                            & {$3$}-edge &   &   $\left(5-6{\chi}\right) \cdot \dfrac{4}{6}+\left(6{\chi}-4\right) \cdot \dfrac{4}{6} = \dfrac{2}{3}$  \\
                                            & {$4$}-edge & $\left(\dfrac{{1-\chi}}{n}+\dfrac{n-1}{6} \cdot \dfrac{1}{n}-\dfrac{1}{6} \right) \cdot n \cdot 6 = 5-6{\chi}$  &  $\left(6{\chi}-4\right) \cdot \dfrac{2}{6} = 2{\chi}-\dfrac{4}{3}$  \\
                                            & {$5$}-edge & $\left(\dfrac{1}{3}-\dfrac{n-1}{3} \cdot \dfrac{1}{n}-\dfrac{{1-\chi}}{n}\right) \cdot n \cdot 6 = 6{\chi}-4$  &   \\
    \hline

                         \multirow{3}*{$2$} & {$2$}-edge &   &  $\left(3-3{\chi}\right) \cdot \dfrac{4}{6} = 2-2{\chi}$   \\
                                            & {$4$}-edge & $\left(\dfrac{{1-\chi}}{n}\right) \cdot n \cdot 3 = 3-3{\chi}$ &  $\left(3-3{\chi}\right) \cdot \dfrac{2}{6}+\left(3{\chi}-2\right) \cdot \dfrac{6}{6} = 2{\chi}-1$   \\
                                            & {$6$}-edge & $\left(\dfrac{1}{6}-\dfrac{n-2}{6} \cdot \dfrac{1}{n}-\dfrac{{1-\chi}}{n}\right) \cdot n \cdot 3 = 3{\chi}-2$ &  \\
    \hline

                         \multirow{4}*{$3$} & {$2$}-edge &   &  $\left(2-2{\chi}\right) \cdot \dfrac{6}{6}  = 2-2{\chi}$   \\
                                            & {$3$}-edge & $\left(\dfrac{{1-\chi}}{n}\right) \cdot n \cdot 2 = 2-2{\chi}$  &     \\
                                            & {$4$}-edge &   &  $\left(2{\chi}-1\right) \cdot \dfrac{6}{6} = 2{\chi}-1$   \\
                                            & {$6$}-edge & $\left(\dfrac{{\chi}}{n}+\dfrac{n-3}{6} \cdot \dfrac{1}{n}-\dfrac{1}{6}\right) \cdot n \cdot 2 = 2{\chi}-1$  &     \\
    \hline

                         \multirow{3}*{$4$} & {$2$}-edge &   &  $\left(3-3{\chi}\right) \cdot \dfrac{4}{6} = 2-2{\chi}$   \\
                                            & {$4$}-edge & $\dfrac{{1-\chi}}{n} \cdot n \cdot 3 = 3-3{\chi}$  &  $\left(3-3{\chi}\right) \cdot \dfrac{2}{6} + \left(3{\chi}-2\right) \cdot \dfrac{6}{6} = 2{\chi}-1$   \\
                                            & {$6$}-edge & $\left(\dfrac{{\chi}}{n}+\dfrac{n-4}{6} \cdot \dfrac{1}{n}-\dfrac{1}{6}\right) \cdot n \cdot 3 = 3{\chi}-2$  &    \\
    \hline

                         \multirow{4}*{$5$} & {$2$}-edge &   &  $\left(5-6{\chi}\right) \cdot \dfrac{2}{6} = \dfrac{5}{3}-2{\chi}$  \\
                                            & {$3$}-edge &   &  $\left(5-6{\chi}\right) \cdot \dfrac{4}{6}+\left(6{\chi}-4\right) \cdot \dfrac{4}{6} = \dfrac{2}{3}$   \\
                                            & {$4$}-edge & $\left(\dfrac{1}{6}-\dfrac{n-5}{6} \cdot \dfrac{1}{n}-\dfrac{{\chi}}{n}\right) \cdot 6 = 5-6{\chi}$  &  $\left(6{\chi}-4\right) \cdot \dfrac{2}{6} = 2{\chi}-\dfrac{4}{3}$   \\
                                            & {$5$}-edge & $\left(\dfrac{{\chi}}{n}+\dfrac{n-2}{3} \cdot \dfrac{1}{n}-\dfrac{1}{3}\right) \cdot 6 = 6{\chi}-4$  &   \\
    \hline
    \hline
    \end{tabular}}
\end{center}
\end{table*}

\newpage

\begin{table*}[htbp]
\begin{center}
\caption{Probabilities of different patch-covering structures for patch sizes $\chi$ near ${\chi}_c$,
        for $n$-patch disks on the four-eight lattice. 
	The period is $n_0=8$.
	For intermediate results on the regular octagon,
	the patch disk is placed at the center of the octagon and its patches cover edges connecting
	the center and vertices of the octagon. 
	This table should be read together with Table~\ref{Tab:contact-proba-four-eight-b}.}
\label{Tab:contact-proba-four-eight-a}
    \renewcommand\arraystretch{2.7}
    \setlength{\tabcolsep}{2.4mm}{
    \begin{tabular}[t]{lllllllll}
    \hline
    \hline
	    \hspace{-4mm} {$\mod(n,n_0)$} & Type & {On the regular octagon} &  {On the four-eight lattice}   \\
    \hline
                         \multirow{4}*{$1$} & {$2$}-edge &   &  $\left(7-8{\chi}\right) \cdot \dfrac{6}{8}+\left(8{\chi}-6\right) \cdot \dfrac{3}{8} = 3-3{\chi}$   \\
                                            & {$3$}-edge &   &   $\left(7-8{\chi}\right) \cdot \dfrac{2}{8}+\left(8{\chi}-6\right) \cdot \dfrac{5}{8} = 3{\chi}-2$  \\
                                            & {$6$}-edge & $\left(\dfrac{{1-\chi}}{n}+\dfrac{n-1}{8} \cdot \dfrac{1}{n}-\dfrac{1}{8} \right) \cdot n \cdot 8 = 7-8{\chi}$  &    \\
                                            & {$7$}-edge & $\left(\dfrac{1}{4}-\dfrac{n-1}{4} \cdot \dfrac{1}{n}-\dfrac{{1-\chi}}{n}\right) \cdot n \cdot 8 = 8{\chi}-6$  &   \\
    \hline

                         \multirow{4}*{$2$} & {$2$}-edge &   &  $\left(4-4{\chi}\right) \cdot \dfrac{6}{8} = 3-3{\chi}$   \\
                                            & {$3$}-edge &   &  $\left(4-4{\chi}\right) \cdot \dfrac{2}{8}+\left(4{\chi}-3\right) \cdot \dfrac{8}{8} = 3{\chi}-2$   \\
                                            & {$6$}-edge & $\left(\dfrac{{1-\chi}}{n}\right) \cdot n \cdot 4 = 4-4{\chi}$ &  \\
                                            & {$8$}-edge & $\left(\dfrac{1}{8}-\dfrac{n-2}{8} \cdot \dfrac{1}{n}-\dfrac{{1-\chi}}{n}\right) \cdot n \cdot 4 = 4{\chi}-3$  &  \\
    \hline

                         \multirow{5}*{$3$} & {$1$}-edge &   &  $\left(7-8{\chi}\right) \cdot \dfrac{2}{8} = \dfrac{7}{4}-2{\chi}$   \\
                                            & {$2$}-edge &   &  $\left(7-8{\chi}\right) \cdot \dfrac{2}{8}+\left(8{\chi}-6\right) \cdot \dfrac{3}{8} = {\chi}-\dfrac{1}{2}$   \\
                                            & {$3$}-edge &   &  $\left(7-8{\chi}\right) \cdot \dfrac{4}{8}+\left(8{\chi}-6\right) \cdot \dfrac{5}{8} = {\chi}-\dfrac{1}{4}$   \\                            
                                            & {$6$}-edge & $\left(\dfrac{{1-\chi}}{n}+\dfrac{3n-1}{8} \cdot \dfrac{1}{n}-\dfrac{3}{8} \right) \cdot n \cdot 8 = 7-8{\chi}$  &     \\
                                            & {$7$}-edge & $\left(\dfrac{{\chi}}{n}+\dfrac{n-3}{4} \cdot \dfrac{1}{n}-\dfrac{1}{4}\right) \cdot n \cdot 8 = 8{\chi}-6$  &   \\
    \hline

                                            & {$1$}-edge &   &  $\left(2-2{\chi}\right) \cdot \dfrac{4}{8} = 1-{\chi}$  \\
                         \multirow{3}*{$4$} & {$2$}-edge &   &  $\left(2-2{\chi}\right) \cdot \dfrac{4}{8} = 1-{\chi}$   \\
                                            & {$3$}-edge &   &  $\left(2{\chi}-1\right) \cdot \dfrac{8}{8} = 2{\chi}-1$  \\
                                            & {$4$}-edge & $\dfrac{{1-\chi}}{n} \cdot n \cdot 2 = 2-2{\chi}$  &     \\
                                            & {$8$}-edge & $\left(\dfrac{{\chi}}{n}+\dfrac{n-4}{8} \cdot \dfrac{1}{n}-\dfrac{1}{8}\right) \cdot n \cdot 2 = 2{\chi}-1$  &    \\

    \hline
    \hline
    \end{tabular}}
\end{center}
\end{table*}

\begin{table*}[htbp]
\begin{center}
\caption{Probabilities of different patch-covering structures for patch sizes $\chi$ near ${\chi}_c$,
        for $n$-patch disks on the four-eight lattice. 
	The period is $n_0=8$.
	For intermediate results on the regular octagon,
	the patch disk is placed at the center of the octagon and its patches cover edges connecting
	the center and vertices of the octagon. 
	This table should be read together with Table~\ref{Tab:contact-proba-four-eight-a}.}
\label{Tab:contact-proba-four-eight-b}
    \renewcommand\arraystretch{2.7}
    \setlength{\tabcolsep}{2.4mm}{
    \begin{tabular}[t]{lllllllll}
    \hline
    \hline
	    \hspace{-4mm} {$\mod(n,n_0)$} & Type & {On the regular octagon} &  {On the four-eight lattice}   \\
    \hline
    
                                            & {$1$}-edge &   &  $\left(7-8{\chi}\right) \cdot \dfrac{2}{8} = \dfrac{7}{4}-2{\chi}$   \\
                         \multirow{3}*{$5$} & {$2$}-edge &   &  $\left(7-8{\chi}\right) \cdot \dfrac{2}{8}+\left(8{\chi}-6\right) \cdot \dfrac{3}{8} = {\chi}-\dfrac{1}{2}$  \\
                                            & {$3$}-edge &   &  $\left(7-8{\chi}\right) \cdot \dfrac{4}{8}+\left(8{\chi}-6\right) \cdot \dfrac{5}{8} = {\chi}-\dfrac{1}{4}$   \\
                                            & {$6$}-edge & $\left(\dfrac{3}{8}-\dfrac{3n-7}{8} \cdot \dfrac{1}{n}-\dfrac{{\chi}}{n}\right) \cdot n \cdot 8 = 7-8{\chi}$  &    \\
                                            & {$7$}-edge & $\left(\dfrac{1}{4}-\dfrac{n-1}{4} \cdot \dfrac{1}{n}-\dfrac{{1-\chi}}{n}\right) \cdot n \cdot 8 = 8{\chi}-6$  &   \\    
    \hline

                         \multirow{4}*{$6$} & {$2$}-edge &   &  $\left(4-4{\chi}\right) \cdot \dfrac{6}{8} = 3-3{\chi}$   \\
                                            & {$3$}-edge &   &  $\left(4-4{\chi}\right) \cdot \dfrac{2}{8}+\left(4{\chi}-3\right) \cdot \dfrac{8}{8} = 3{\chi}-2$   \\
                                            & {$6$}-edge & $\left(\dfrac{{1-\chi}}{n}\right) \cdot n \cdot 4 = 4-4{\chi}$ &  \\
                                            & {$8$}-edge & $\left(\dfrac{{\chi}}{n}+\dfrac{n-6}{8} \cdot \dfrac{1}{n}-\dfrac{1}{8}\right) \cdot n \cdot 4 = 4{\chi}-3$  &  \\
    \hline

                         \multirow{4}*{$7$} & {$2$}-edge &   &  $\left(7-8{\chi}\right) \cdot \dfrac{6}{8}+\left(8{\chi}-6\right) \cdot \dfrac{3}{8} = 3-3{\chi}$   \\
                                            & {$3$}-edge &   &   $\left(7-8{\chi}\right) \cdot \dfrac{2}{8}+\left(8{\chi}-6\right) \cdot \dfrac{5}{8} = 3{\chi}-2$  \\
                                            & {$6$}-edge & $\left(\dfrac{1}{8}-\dfrac{n-7}{8} \cdot \dfrac{1}{n}-\dfrac{{\chi}}{n}\right) \cdot n \cdot 8 = 7-8{\chi}$  &    \\
                                            & {$7$}-edge & $\left(\dfrac{{\chi}}{n}+\dfrac{n-3}{4} \cdot \dfrac{1}{n}-\dfrac{1}{4}\right) \cdot n \cdot 8 = 8{\chi}-6$  &   \\

    \hline
    \hline
    \end{tabular}}
\end{center}
\end{table*}

\newpage

\begin{table*}[htbp]
\begin{center}
\caption{Probabilities of different patch-covering structures for patch sizes $\chi$ near ${\chi}_c$,
        for $n$-patch disks on the frieze lattice. 
	The period is $n_0=12$.
	For intermediate results on the regular dodecagon,
	the patch disk is placed at the center of the dodecagon and its patches cover edges connecting
	the center and vertices of the dodecagon. 
	This table should be read together with Tables~\ref{Tab:contact-proba-frieze-b}
	and \ref{Tab:contact-proba-frieze-c}.}
\label{Tab:contact-proba-frieze-a}
    \renewcommand\arraystretch{2.7}
    \setlength{\tabcolsep}{2.4mm}{
    \begin{tabular}[t]{lllllllll}
    \hline
    \hline
	    \hspace{-4mm} {$\mod(n,n_0)$} & Type & {On the regular dodecagon} &  {On the frieze lattice}   \\
    \hline
                         \multirow{4}*{$1$} & {$3$}-edge &   &  $\left(9-12{\chi}\right) \cdot \dfrac{8}{12}+\left(12{\chi}-8\right) \cdot \dfrac{3}{12} = 4-5{\chi}$   \\
                                            & {$4$}-edge &   &   $\left(9-12{\chi}\right) \cdot \dfrac{4}{12}+\left(12{\chi}-8\right) \cdot \dfrac{9}{12} = 5{\chi}-3$  \\
                                            & {$8$}-edge & $\left(\dfrac{{1-\chi}}{n}+\dfrac{n-1}{4} \cdot \dfrac{1}{n}-\dfrac{1}{4} \right) \cdot n \cdot 12 = 9-12{\chi}$  &    \\
                                            & {$9$}-edge & $\left(\dfrac{1}{3}-\dfrac{n-1}{3} \cdot \dfrac{1}{n}-\dfrac{{1-\chi}}{n}\right) \cdot n \cdot 12 = 12{\chi}-8$  &   \\
    \hline

                         \multirow{5}*{$2$} & {$2$}-edge &   &  $\left(4-6{\chi}\right) \cdot \dfrac{6}{12} = 2-3{\chi}$   \\
                                            & {$3$}-edge &   &  $\left(4-6{\chi}\right) \cdot \dfrac{6}{12}+\left(6{\chi}-3\right) \cdot \dfrac{8}{12} = {\chi}$   \\
                                            & {$4$}-edge &   &  $\left(6{\chi}-3\right) \cdot \dfrac{4}{12} = 2{\chi}-1$   \\
                                            & {$6$}-edge & $\left(\dfrac{{1-\chi}}{n}+\dfrac{n-2}{6} \cdot \dfrac{1}{n}-\dfrac{1}{6} \right) \cdot n \cdot 6 = 4-6{\chi}$ &  \\
                                            & {$8$}-edge & $\left(\dfrac{1}{4}-\dfrac{n-2}{4} \cdot \dfrac{1}{n}-\dfrac{{1-\chi}}{n}\right) \cdot n \cdot 6 = 6{\chi}-3$  &  \\
    \hline

                         \multirow{6}*{$3$} & {$2$}-edge &   &  $\left(3-4{\chi}\right) \cdot \dfrac{6}{12} = \dfrac{3}{2}-2{\chi}$   \\
                                            & {$3$}-edge &   &  $\left(3-4{\chi}\right) \cdot \dfrac{6}{12}+\left(4{\chi}-2\right) \cdot \dfrac{6}{12} = \dfrac{1}{2}$   \\
                                            & {$4$}-edge &   &  $\left(4{\chi}-2\right) \cdot \dfrac{3}{12} = {\chi}-\dfrac{1}{2}$   \\
                                            & {$5$}-edge &   &  $\left(4{\chi}-2\right) \cdot \dfrac{3}{12} = {\chi}-\dfrac{1}{2}$   \\
                                            & {$6$}-edge & $\left(\dfrac{{1-\chi}}{n}+\dfrac{n-3}{12} \cdot \dfrac{1}{n}-\dfrac{1}{12} \right) \cdot n \cdot 4 = 3-4{\chi}$  &     \\
                                            & {$9$}-edge & $\left(\dfrac{{\chi}}{n}+\dfrac{n-3}{6} \cdot \dfrac{1}{n}-\dfrac{1}{6}\right) \cdot n \cdot 4 = 4{\chi}-2$  &   \\
    \hline

                                            & {$1$}-edge &   &  $\left(2-3{\chi}\right) \cdot \dfrac{8}{12} = \dfrac{4}{3}-2{\chi}$  \\
                         \multirow{3}*{$4$} & {$2$}-edge &   &  $\left(3{\chi}-1\right) \cdot \dfrac{4}{12} = {\chi}-\dfrac{1}{3}$   \\
                                            & {$3$}-edge &   &  $\left(2-3{\chi}\right) \cdot \dfrac{4}{12} = \dfrac{2}{3}-{\chi}$  \\
                                            & {$4$}-edge &  $\left(\dfrac{1}{6}-\dfrac{n-4}{6} \cdot \dfrac{1}{n}-\dfrac{{\chi}}{n}\right) \cdot n \cdot 3 = 2-3{\chi}$  &  $\left(3{\chi}-1\right) \cdot \dfrac{8}{12} = 2{\chi}-\dfrac{2}{3}$  \\
                                            & {$8$}-edge & $\left(\dfrac{{\chi}}{n}+\dfrac{n-4}{12} \cdot \dfrac{1}{n}-\dfrac{1}{12}\right) \cdot n \cdot 3 = 3{\chi}-1$  &    \\

    \hline
    \hline
    \end{tabular}}
\end{center}
\end{table*}

\begin{table*}[htbp]
\begin{center}
\caption{Probabilities of different patch-covering structures for patch sizes $\chi$ near ${\chi}_c$,
        for $n$-patch disks on the frieze lattice. 
	The period is $n_0=12$.
	For intermediate results on the regular dodecagon,
	the patch disk is placed at the center of the dodecagon and its patches cover edges connecting
	the center and vertices of the dodecagon. 
	This table should be read together with Tables~\ref{Tab:contact-proba-frieze-a}
	and \ref{Tab:contact-proba-frieze-c}.}
\label{Tab:contact-proba-frieze-b}
    \renewcommand\arraystretch{2.7}
    \setlength{\tabcolsep}{2.4mm}{
    \begin{tabular}[t]{lllllllll}
    \hline
    \hline
	    \hspace{-4mm} {$\mod(n,n_0)$} & Type & {On the regular dodecagon} &  {On the frieze lattice}   \\
    \hline

                                            & {$1$}-edge &   &  $\left(8-12{\chi}\right) \cdot \dfrac{2}{12} = \dfrac{4}{3}-2{\chi}$   \\
                         \multirow{5}*{$5$} & {$2$}-edge &   &  $\left(8-12{\chi}\right) \cdot \dfrac{3}{12}+\left(12{\chi}-7\right) \cdot \dfrac{4}{12} = {\chi}-\dfrac{1}{3}$  \\
                                            & {$3$}-edge &   &  $\left(8-12{\chi}\right) \cdot \dfrac{2}{12}+\left(12{\chi}-7\right) \cdot \dfrac{2}{12} = \dfrac{1}{6}$   \\
                                            & {$4$}-edge &   &  $\left(8-12{\chi}\right) \cdot \dfrac{4}{12}+\left(12{\chi}-7\right) \cdot \dfrac{4}{12} = \dfrac{1}{3}$   \\
                                            & {$5$}-edge &   &  $\left(8-12{\chi}\right) \cdot \dfrac{1}{12}+\left(12{\chi}-7\right) \cdot \dfrac{2}{12} = {\chi}-\dfrac{1}{2}$   \\
                                            & {$7$}-edge & $\left(\dfrac{1}{3}-\dfrac{n-2}{3} \cdot \dfrac{1}{n}-\dfrac{{\chi}}{n}\right) \cdot n \cdot 12 = 8-12{\chi}$  &    \\
                                            & {$8$}-edge & $\left(\dfrac{1}{12}-\dfrac{n-5}{12} \cdot \dfrac{1}{n}-\dfrac{{1-\chi}}{n}\right) \cdot n \cdot 12 = 12{\chi}-7$  &   \\
    \hline

                         \multirow{5}*{$6$} & {$1$}-edge &   &  $\left(2-2{\chi}\right) \cdot \dfrac{6}{12} = 1-{\chi}$   \\
                                            & {$4$}-edge &   &  $\left(2-2{\chi}\right) \cdot \dfrac{6}{12} = 1-{\chi}$   \\
                                            & {$5$}-edge &   &  $\left(2{\chi}-1\right) \cdot \dfrac{12}{12} = 2{\chi}-1$   \\
                                            & {$6$}-edge & $\left(\dfrac{{1-\chi}}{n}\right) \cdot n \cdot 2 = 2-2{\chi}$ &  \\
                                            & {$12$}-edge & $\left(\dfrac{{\chi}}{n}+\dfrac{n-6}{12} \cdot \dfrac{1}{n}-\dfrac{1}{12}\right) \cdot n \cdot 2 = 2{\chi}-1$  &  \\               
    \hline

                                            & {$1$}-edge &   &  $\left(8-12{\chi}\right) \cdot \dfrac{2}{12} = \dfrac{4}{3}-2{\chi}$   \\
                                            & {$2$}-edge &   &   $\left(8-12{\chi}\right) \cdot \dfrac{3}{12}+\left(12{\chi}-7\right) \cdot \dfrac{4}{12} = {\chi}-\dfrac{1}{3}$  \\
                                            & {$3$}-edge &   &   $\left(8-12{\chi}\right) \cdot \dfrac{2}{12}+\left(12{\chi}-7\right) \cdot \dfrac{2}{12} = \dfrac{1}{6}$  \\
                         \multirow{1}*{$7$} & {$4$}-edge &   &   $\left(8-12{\chi}\right) \cdot \dfrac{4}{12}+\left(12{\chi}-7\right) \cdot \dfrac{4}{12} = \dfrac{1}{3}$  \\
                                            & {$5$}-edge &   &   $\left(8-12{\chi}\right) \cdot \dfrac{1}{12}+\left(12{\chi}-7\right) \cdot \dfrac{2}{12} = {\chi}-\dfrac{1}{2}$  \\
                                            & {$7$}-edge & $\left(\dfrac{{1-\chi}}{n}+\dfrac{n-1}{3} \cdot \dfrac{1}{n}-\dfrac{1}{3}\right) \cdot n \cdot 12 = 8-12{\chi}$  &    \\
                                            & {$8$}-edge & $\left(\dfrac{{\chi}}{n}+\dfrac{n-7}{12} \cdot \dfrac{1}{n}-\dfrac{1}{12}\right) \cdot n \cdot 12 = 12{\chi}-7$  &   \\             

    \hline
    \hline
    \end{tabular}}
\end{center}
\end{table*}

\begin{table*}[htbp]
\begin{center}
\caption{Probabilities of different patch-covering structures for patch sizes $\chi$ near ${\chi}_c$,
        for $n$-patch disks on the frieze lattice. 
	The period is $n_0=12$.
	For intermediate results on the regular dodecagon,
	the patch disk is placed at the center of the dodecagon and its patches cover edges connecting
	the center and vertices of the dodecagon. 
	This table should be read together with Tables~\ref{Tab:contact-proba-frieze-a}
	and \ref{Tab:contact-proba-frieze-b}.}
\label{Tab:contact-proba-frieze-c}
    \renewcommand\arraystretch{2.7}
    \setlength{\tabcolsep}{2.4mm}{
    \begin{tabular}[t]{lllllllll}
    \hline
    \hline
	    \hspace{-4mm} {$\mod(n,n_0)$} & Type & {On the regular dodecagon} &  {On the frieze lattice}   \\
    \hline

                                            & {$1$}-edge &   &  $\left(2-3{\chi}\right) \cdot \dfrac{8}{12} = \dfrac{4}{3}-2{\chi}$  \\
                         \multirow{3}*{$8$} & {$2$}-edge &   &  $\left(3{\chi}-1\right) \cdot \dfrac{4}{12} = {\chi}-\dfrac{1}{3}$   \\
                                            & {$3$}-edge &   &  $\left(2-3{\chi}\right) \cdot \dfrac{4}{12} = \dfrac{2}{3}-{\chi}$  \\
                                            & {$4$}-edge & $\left(\dfrac{1}{12}-\dfrac{n-8}{12} \cdot \dfrac{1}{n}-\dfrac{{\chi}}{n}\right) \cdot n \cdot 3 = 2-3{\chi}$  &  $\left(3{\chi}-1\right) \cdot \dfrac{8}{12} = 2{\chi}-\dfrac{2}{3}$  \\
                                            & {$8$}-edge & $\left(\dfrac{{\chi}}{n}+\dfrac{n-2}{6} \cdot \dfrac{1}{n}-\dfrac{1}{6}\right) \cdot n \cdot 3 = 3{\chi}-1$  &    \\
    \hline

                         \multirow{6}*{$9$} & {$2$}-edge &   &  $\left(3-4{\chi}\right) \cdot \dfrac{6}{12} = \dfrac{3}{2}-2{\chi}$   \\
                                            & {$3$}-edge &   &  $\left(3-4{\chi}\right) \cdot \dfrac{6}{12}+\left(4{\chi}-2\right) \cdot \dfrac{6}{12} = \dfrac{1}{2}$   \\
                                            & {$4$}-edge &   &  $\left(4{\chi}-2\right) \cdot \dfrac{3}{12} = {\chi}-\dfrac{1}{2}$   \\
                                            & {$5$}-edge &   &  $\left(4{\chi}-2\right) \cdot \dfrac{3}{12} = {\chi}-\dfrac{1}{2}$   \\
                                            & {$6$}-edge & $\left(\dfrac{1}{12}-\dfrac{n-9}{12} \cdot \dfrac{1}{n}-\dfrac{{\chi}}{n}\right) \cdot n \cdot 4 = 3-4{\chi}$  &     \\
                                            & {$9$}-edge & $\left(\dfrac{{\chi}}{n}+\dfrac{n-3}{6} \cdot \dfrac{1}{n}-\dfrac{1}{6}\right) \cdot n \cdot 4 = 4{\chi}-2$  &   \\
    \hline

                        \multirow{5}*{$10$} & {$2$}-edge &   &  $\left(4-6{\chi}\right) \cdot \dfrac{6}{12} = 2-3{\chi}$   \\
                                            & {$3$}-edge &   &  $\left(4-6{\chi}\right) \cdot \dfrac{6}{12}+\left(6{\chi}-3\right) \cdot \dfrac{8}{12} = {\chi}$   \\
                                            & {$4$}-edge &   &  $\left(6{\chi}-3\right) \cdot \dfrac{4}{12} = 2{\chi}-1$   \\
                                            & {$6$}-edge & $\left(\dfrac{1}{6}-\dfrac{n-4}{6} \cdot \dfrac{1}{n}-\dfrac{{\chi}}{n}\right) \cdot n \cdot 6 = 4-6{\chi}$ &  \\
                                            & {$8$}-edge & $\left(\dfrac{{\chi}}{n}+\dfrac{n-2}{4} \cdot \dfrac{1}{n}-\dfrac{1}{4}\right) \cdot n \cdot 6 = 6{\chi}-3$  &  \\
    \hline

                        \multirow{4}*{$11$} & {$3$}-edge &   &  $\left(9-12{\chi}\right) \cdot \dfrac{8}{12}+\left(12{\chi}-8\right) \cdot \dfrac{3}{12} = 4-5{\chi}$   \\
                                            & {$4$}-edge &   &   $\left(9-12{\chi}\right) \cdot \dfrac{4}{12}+\left(12{\chi}-8\right) \cdot \dfrac{9}{12} = 5{\chi}-3$  \\
                                            & {$8$}-edge & $\left(\dfrac{1}{4}-\dfrac{n-3}{4} \cdot \dfrac{1}{n}-\dfrac{{\chi}}{n}\right) \cdot n \cdot 12 = 9-12{\chi}$  &    \\
                                            & {$9$}-edge & $\left(\dfrac{{\chi}}{n}+\dfrac{n-2}{3} \cdot \dfrac{1}{n}-\dfrac{1}{3}\right) \cdot n \cdot 12 = 12{\chi}-8$  &   \\

    \hline
    \hline
    \end{tabular}}
\end{center}
\end{table*}

\newpage

\begin{table*}[htbp]
\begin{center}
\caption{Probabilities of different patch-covering structures for patch sizes $\chi$ near ${\chi}_c$,
        for $n$-patch disks on the snub square lattice. 
	The period is $n_0=12$.
	For intermediate results on the regular dodecagon,
	the patch disk is placed at the center of the dodecagon and its patches cover edges connecting
	the center and vertices of the dodecagon. 
	This table should be read together with Tables~\ref{Tab:contact-proba-snub-square-b}
	and \ref{Tab:contact-proba-snub-square-c}.}
\label{Tab:contact-proba-snub-square-a}
    \renewcommand\arraystretch{2.7}
    \setlength{\tabcolsep}{2.4mm}{
    \begin{tabular}[t]{lllllllll}
    \hline
    \hline
	    \hspace{-4mm}{$\mod(n,n_0)$} & Type & {On the regular dodecagon} &  {On the snub square lattice}   \\
    \hline
                         \multirow{4}*{$1$} & {$3$}-edge &   &  $\left(9-12{\chi}\right) \cdot \dfrac{8}{12}+\left(12{\chi}-8\right) \cdot \dfrac{3}{12} = 4-5{\chi}$   \\
                                            & {$4$}-edge &   &   $\left(9-12{\chi}\right) \cdot \dfrac{4}{12}+\left(12{\chi}-8\right) \cdot \dfrac{9}{12} = 5{\chi}-3$  \\
                                            & {$8$}-edge & $\left(\dfrac{{1-\chi}}{n}+\dfrac{n-1}{4} \cdot \dfrac{1}{n}-\dfrac{1}{4} \right) \cdot n \cdot 12 = 9-12{\chi}$  &    \\
                                            & {$9$}-edge & $\left(\dfrac{1}{3}-\dfrac{n-1}{3} \cdot \dfrac{1}{n}-\dfrac{{1-\chi}}{n}\right) \cdot n \cdot 12 = 12{\chi}-8$  &   \\
    \hline

                         \multirow{5}*{$2$} & {$2$}-edge &   &  $\left(4-6{\chi}\right) \cdot \dfrac{6}{12} = 2-3{\chi}$   \\
                                            & {$3$}-edge &   &  $\left(4-6{\chi}\right) \cdot \dfrac{6}{12}+\left(6{\chi}-3\right) \cdot \dfrac{8}{12} = {\chi}$   \\
                                            & {$4$}-edge &   &  $\left(6{\chi}-3\right) \cdot \dfrac{4}{12} = 2{\chi}-1$   \\
                                            & {$6$}-edge & $\left(\dfrac{1}{3}-\dfrac{n-2}{3} \cdot \dfrac{1}{n}-\dfrac{{\chi}}{n}\right) \cdot n \cdot 6 = 4-6{\chi}$ &  \\
                                            & {$8$}-edge & $\left(\dfrac{{\chi}}{n}+\dfrac{n-2}{4} \cdot \dfrac{1}{n}-\dfrac{1}{4}\right) \cdot n \cdot 6 = 6{\chi}-3$  &  \\
    \hline

                         \multirow{5}*{$3$} & {$2$}-edge &   &  $\left(3-4{\chi}\right) \cdot \dfrac{6}{12} = \dfrac{3}{2}-2{\chi}$   \\
                                            & {$3$}-edge &   &  $\left(3-4{\chi}\right) \cdot \dfrac{6}{12}+\left(4{\chi}-2\right) \cdot \dfrac{3}{12} = 1-{\chi}$   \\
                                            & {$4$}-edge &   &  $\left(4{\chi}-2\right) \cdot \dfrac{9}{12} = {3\chi}-\dfrac{3}{2}$   \\
                                            & {$6$}-edge & $\left(\dfrac{1}{4}-\dfrac{n-3}{4} \cdot \dfrac{1}{n}-\dfrac{{\chi}}{n}\right) \cdot n \cdot 4 = 3-4{\chi}$  &     \\
                                            & {$9$}-edge & $\left(\dfrac{{\chi}}{n}+\dfrac{n-3}{6} \cdot \dfrac{1}{n}-\dfrac{1}{6}\right) \cdot n \cdot 4 = 4{\chi}-2$  &   \\
    \hline

                                            & {$3$}-edge &   &  $\left(3-3{\chi}\right) \cdot \dfrac{8}{12} = 2-2{\chi}$  \\
                         \multirow{5}*{$4$} & {$4$}-edge &   &  $\left(3-3{\chi}\right) \cdot \dfrac{4}{12}+\left(3{\chi}-2\right) \cdot \dfrac{4}{12} = \dfrac{1}{3}$   \\
                                            & {$5$}-edge &   &  $\left(3{\chi}-2\right) \cdot \dfrac{8}{12} = 2{\chi}-\dfrac{4}{3}$  \\
                                            & {$8$}-edge &  $\left(\dfrac{1-{\chi}}{n}\right) \cdot n \cdot 3 = 3-3{\chi}$  &    \\
                                            & {$12$}-edge & $\left(\dfrac{{\chi}}{n}+\dfrac{n-4}{6} \cdot \dfrac{1}{n}-\dfrac{1}{6}\right) \cdot n \cdot 3 = 3{\chi}-2$  &    \\

    \hline
    \hline
    \end{tabular}}
\end{center}
\end{table*}

\begin{table*}[htbp]
\begin{center}
\caption{Probabilities of different patch-covering structures for patch sizes $\chi$ near ${\chi}_c$,
        for $n$-patch disks on the snub square lattice. 
	The period is $n_0=12$.
	For intermediate results on the regular dodecagon,
	the patch disk is placed at the center of the dodecagon and its patches cover edges connecting
	the center and vertices of the dodecagon. 
	This table should be read together with Tables~\ref{Tab:contact-proba-snub-square-a}
	and \ref{Tab:contact-proba-snub-square-c}.}
\label{Tab:contact-proba-snub-square-b}
    \renewcommand\arraystretch{2.7}
    \setlength{\tabcolsep}{2.4mm}{
    \begin{tabular}[t]{lllllllll}
    \hline
    \hline
	    \hspace{-4mm}{$\mod(n,n_0)$} & Type & {On the regular dodecagon} &  {On the snub square lattice}   \\
    \hline

                                            & {$0$}-edge &   &  $\left(8-12{\chi}\right) \cdot \dfrac{1}{12} = \dfrac{2}{3}-{\chi}$ \\
                                            & {$1$}-edge &   &  $\left(8-12{\chi}\right) \cdot \dfrac{2}{12}+\left(12{\chi}-7\right) \cdot \dfrac{2}{12} = \dfrac{1}{6}$   \\
                         \multirow{4}*{$5$} & {$2$}-edge &   &  $\left(8-12{\chi}\right) \cdot \dfrac{2}{12}+\left(12{\chi}-7\right) \cdot \dfrac{2}{12} = \dfrac{1}{6}$  \\
                                            & {$3$}-edge &   &  $\left(8-12{\chi}\right) \cdot \dfrac{2}{12}+\left(12{\chi}-7\right) \cdot \dfrac{2}{12} = \dfrac{1}{6}$   \\
                                            & {$4$}-edge &   &  $\left(8-12{\chi}\right) \cdot \dfrac{2}{12}+\left(12{\chi}-7\right) \cdot \dfrac{2}{12} = \dfrac{1}{6}$   \\
                                            & {$5$}-edge &   &  $\left(8-12{\chi}\right) \cdot \dfrac{3}{12}+\left(12{\chi}-7\right) \cdot \dfrac{4}{12} = {\chi}-\dfrac{1}{3}$   \\
                                            & {$7$}-edge & $\left(\dfrac{1}{3}-\dfrac{n-2}{3} \cdot \dfrac{1}{n}-\dfrac{{\chi}}{n}\right) \cdot n \cdot 12 = 8-12{\chi}$  &    \\
                                            & {$8$}-edge & $\left(\dfrac{1}{12}-\dfrac{n-5}{12} \cdot \dfrac{1}{n}-\dfrac{{1-\chi}}{n}\right) \cdot n \cdot 12 = 12{\chi}-7$  &   \\
    \hline

                         \multirow{5}*{$6$} & {$2$}-edge &   &  $\left(2-2{\chi}\right) \cdot \dfrac{6}{12} = 1-{\chi}$   \\
                                            & {$3$}-edge &   &  $\left(2-2{\chi}\right) \cdot \dfrac{6}{12} = 1-{\chi}$   \\
                                            & {$5$}-edge &   &  $\left(2{\chi}-1\right) \cdot \dfrac{12}{12} = 2{\chi}-1$   \\
                                            & {$6$}-edge & $\left(\dfrac{{1-\chi}}{n}\right) \cdot n \cdot 2 = 2-2{\chi}$ &  \\
                                            & {$12$}-edge & $\left(\dfrac{{\chi}}{n}+\dfrac{n-6}{12} \cdot \dfrac{1}{n}-\dfrac{1}{12}\right) \cdot n \cdot 2 = 2{\chi}-1$  &  \\
    \hline

                                            & {$0$}-edge &   &   $\left(8-12{\chi}\right) \cdot \dfrac{1}{12} = \dfrac{2}{3}-{\chi}$   \\
                                            & {$1$}-edge &   &   $\left(8-12{\chi}\right) \cdot \dfrac{2}{12}+\left(12{\chi}-7\right) \cdot \dfrac{2}{12} = \dfrac{1}{6}$   \\
                                            & {$2$}-edge &   &   $\left(8-12{\chi}\right) \cdot \dfrac{2}{12}+\left(12{\chi}-7\right) \cdot \dfrac{2}{12} = \dfrac{1}{6}$  \\
                         \multirow{2}*{$7$} & {$3$}-edge &   &   $\left(8-12{\chi}\right) \cdot \dfrac{2}{12}+\left(12{\chi}-7\right) \cdot \dfrac{2}{12} = \dfrac{1}{6}$  \\
                                            & {$4$}-edge &   &  $\left(8-12{\chi}\right) \cdot \dfrac{2}{12}+\left(12{\chi}-7\right) \cdot \dfrac{2}{12} = \dfrac{1}{6}$  \\
                                            & {$5$}-edge &   &   $\left(8-12{\chi}\right) \cdot \dfrac{3}{12}+\left(12{\chi}-7\right) \cdot \dfrac{4}{12} = {\chi}-\dfrac{1}{3}$  \\
                                            & {$7$}-edge & $\left(\dfrac{{1-\chi}}{n}+\dfrac{n-1}{3} \cdot \dfrac{1}{n}-\dfrac{1}{3}\right) \cdot n \cdot 12 = 8-12{\chi}$  &    \\
                                            & {$8$}-edge & $\left(\dfrac{{\chi}}{n}+\dfrac{n-7}{12} \cdot \dfrac{1}{n}-\dfrac{1}{12}\right) \cdot n \cdot 12 = 12{\chi}-7$  &   \\

    \hline
    \hline
    \end{tabular}}
\end{center}
\end{table*}

\begin{table*}[htbp]
\begin{center}
\caption{Probabilities of different patch-covering structures for patch sizes $\chi$ near ${\chi}_c$,
        for $n$-patch disks on the snub square lattice. 
	The period is $n_0=12$.
	For intermediate results on the regular dodecagon,
	the patch disk is placed at the center of the dodecagon and its patches cover edges connecting
	the center and vertices of the dodecagon. 
	This table should be read together with Tables~\ref{Tab:contact-proba-snub-square-a}
	and \ref{Tab:contact-proba-snub-square-b}.}
\label{Tab:contact-proba-snub-square-c}
    \renewcommand\arraystretch{2.7}
    \setlength{\tabcolsep}{2.4mm}{
    \begin{tabular}[t]{lllllllll}
    \hline
    \hline
	    \hspace{-4mm}{$\mod(n,n_0)$} & Type & {On the regular dodecagon} &  {On the snub square lattice}   \\
    \hline

                                            & {$3$}-edge &   &  $\left(3-3{\chi}\right) \cdot \dfrac{8}{12} = 2-2{\chi}$  \\
                         \multirow{3}*{$8$} & {$4$}-edge &   &  $\left(3-3{\chi}\right) \cdot \dfrac{4}{12}+\left(3{\chi}-2\right) \cdot \dfrac{4}{12} = \dfrac{1}{3}$   \\
                                            & {$5$}-edge &   &  $\left(3{\chi}-2\right) \cdot \dfrac{8}{12} = 2{\chi}-\dfrac{4}{3}$  \\
                                            & {$8$}-edge & $\left(\dfrac{1-{\chi}}{n}\right) \cdot n \cdot 3 = 3-3{\chi}$  &    \\
                                            & {$12$}-edge & $\left(\dfrac{{\chi}}{n}+\dfrac{n-8}{12} \cdot \dfrac{1}{n}-\dfrac{1}{12}\right) \cdot n \cdot 3 = 3{\chi}-2$  &    \\                  
    \hline

                         \multirow{5}*{$9$} & {$2$}-edge &   &  $\left(3-4{\chi}\right) \cdot \dfrac{6}{12} = \dfrac{3}{2}-2{\chi}$   \\
                                            & {$3$}-edge &   &  $\left(3-4{\chi}\right) \cdot \dfrac{6}{12}+\left(4{\chi}-2\right) \cdot \dfrac{3}{12} = 1-{\chi}$   \\
                                            & {$4$}-edge &   &  $\left(4{\chi}-2\right) \cdot \dfrac{9}{12} = 3{\chi}-\dfrac{3}{2}$   \\
                                            & {$6$}-edge & $\left(\dfrac{1}{12}-\dfrac{n-9}{12} \cdot \dfrac{1}{n}-\dfrac{{\chi}}{n}\right) \cdot n \cdot 4 = 3-4{\chi}$  &     \\
                                            & {$9$}-edge & $\left(\dfrac{{\chi}}{n}+\dfrac{n-3}{6} \cdot \dfrac{1}{n}-\dfrac{1}{6}\right) \cdot n \cdot 4 = 4{\chi}-2$  &   \\                      
    \hline

                        \multirow{5}*{$10$} & {$2$}-edge &   &  $\left(4-6{\chi}\right) \cdot \dfrac{6}{12} = 2-3{\chi}$   \\
                                            & {$3$}-edge &   &  $\left(4-6{\chi}\right) \cdot \dfrac{6}{12}+\left(6{\chi}-3\right) \cdot \dfrac{8}{12} = {\chi}$   \\
                                            & {$4$}-edge &   &  $\left(6{\chi}-3\right) \cdot \dfrac{4}{12} = 2{\chi}-1$   \\
                                            & {$6$}-edge & $\left(\dfrac{1}{6}-\dfrac{n-4}{6} \cdot \dfrac{1}{n}-\dfrac{{\chi}}{n}\right) \cdot n \cdot 6 = 4-6{\chi}$ &  \\                        
                                            & {$8$}-edge & $\left(\dfrac{{\chi}}{n}+\dfrac{n-2}{4} \cdot \dfrac{1}{n}-\dfrac{1}{4}\right) \cdot n \cdot 6 = 6{\chi}-3$  &  \\
    \hline

                        \multirow{4}*{$11$} & {$3$}-edge &   &  $\left(9-12{\chi}\right) \cdot \dfrac{8}{12}+\left(12{\chi}-8\right) \cdot \dfrac{3}{12} = 4-5{\chi}$   \\
                                            & {$4$}-edge &   &   $\left(9-12{\chi}\right) \cdot \dfrac{4}{12}+\left(12{\chi}-8\right) \cdot \dfrac{9}{12} = 5{\chi}-3$  \\
                                            & {$8$}-edge & $\left(\dfrac{1}{4}-\dfrac{n-3}{4} \cdot \dfrac{1}{n}-\dfrac{{\chi}}{n}\right) \cdot n \cdot 12 = 9-12{\chi}$  &    \\
                                            & {$9$}-edge & $\left(\dfrac{{\chi}}{n}+\dfrac{n-2}{3} \cdot \dfrac{1}{n}-\dfrac{1}{3}\right) \cdot n \cdot 12 = 12{\chi}-8$  &   \\                    

    \hline
    \hline
    \end{tabular}}
\end{center}
\end{table*}

\newpage

\begin{table*}[htbp]
\begin{center}
\caption{Probabilities of different patch-covering structures for patch sizes $\chi$ near ${\chi}_c$,
        for $n$-patch disks on the ruby lattice. 
	The period is $n_0=12$.
	For intermediate results on the regular dodecagon,
	the patch disk is placed at the center of the dodecagon and its patches cover edges connecting
	the center and vertices of the dodecagon. 
	This table should be read together with Tables~\ref{Tab:contact-proba-ruby-b}
	and \ref{Tab:contact-proba-ruby-c}.}
\label{Tab:contact-proba-ruby-a}
    \renewcommand\arraystretch{2.7}
    \setlength{\tabcolsep}{2.4mm}{
    \begin{tabular}[t]{lllllllll}
    \hline
    \hline
	    \hspace{-4mm}{$\mod(n,n_0)$} & Type & {On the regular dodecagon} &  {On the ruby lattice}   \\
    \hline
                         \multirow{5}*{$1$} & {$2$}-edge &   &  $\left(9-12{\chi}\right) \cdot \dfrac{4}{12}+\left(12{\chi}-8\right) \cdot \dfrac{1}{12} = \dfrac{7}{3}-3{\chi}$   \\
                                            & {$3$}-edge &   &   $\left(9-12{\chi}\right) \cdot \dfrac{8}{12}+\left(12{\chi}-8\right) \cdot \dfrac{10}{12} = 2{\chi}-\dfrac{2}{3}$  \\
                                            & {$4$}-edge &   &   $\left(12{\chi}-8\right) \cdot \dfrac{1}{12} = {\chi}-\dfrac{2}{3}$  \\
                                            & {$8$}-edge & $\left(\dfrac{{1-\chi}}{n}+\dfrac{n-1}{4} \cdot \dfrac{1}{n}-\dfrac{1}{4} \right) \cdot n \cdot 12 = 9-12{\chi}$  &    \\
                                            & {$9$}-edge & $\left(\dfrac{1}{3}-\dfrac{n-1}{3} \cdot \dfrac{1}{n}-\dfrac{{1-\chi}}{n}\right) \cdot n \cdot 12 = 12{\chi}-8$  &   \\
    \hline

                         \multirow{5}*{$2$} & {$2$}-edge &   &  $\left(5-6{\chi}\right) \cdot \dfrac{4}{12} = \dfrac{5}{3}-2{\chi}$   \\
                                            & {$3$}-edge &   &  $\left(5-6{\chi}\right) \cdot \dfrac{8}{12}+\left(6{\chi}-4\right) \cdot \dfrac{4}{12} = \dfrac{2}{3}$   \\
                                            & {$4$}-edge &   &  $\left(6{\chi}-4\right) \cdot \dfrac{8}{12} = 2{\chi}-\dfrac{4}{3}$   \\
                                            & {$8$}-edge & $\left(\dfrac{1-{\chi}}{n}+\dfrac{n-2}{12} \cdot \dfrac{1}{n}-\dfrac{1}{12}\right) \cdot n \cdot 6 = 5-6{\chi}$ &  \\                    
                                            & {$10$}-edge & $\left(\dfrac{{\chi}}{n}+\dfrac{n-2}{3} \cdot \dfrac{1}{n}-\dfrac{1}{3}\right) \cdot n \cdot 6 = 6{\chi}-4$  &  \\
    \hline

                                            & {$1$}-edge &   &  $\left(3-4{\chi}\right) \cdot \dfrac{6}{12} = \dfrac{3}{2}-2{\chi}$   \\
                         \multirow{4}*{$3$} & {$2$}-edge &   &  $\left(4{\chi}-2\right) \cdot \dfrac{3}{12} = {\chi}-\dfrac{1}{2}$   \\
                                            & {$3$}-edge &   &  $\left(3-4{\chi}\right) \cdot \dfrac{6}{12} + \left(4{\chi}-2\right) \cdot \dfrac{6}{12} = \dfrac{1}{2}$   \\                       
                                            & {$4$}-edge &   &  $\left(4{\chi}-2\right) \cdot \dfrac{3}{12} = {\chi}-\dfrac{1}{2}$   \\                                                             
                                            & {$6$}-edge & $\left(\dfrac{1}{4}-\dfrac{n-3}{4} \cdot \dfrac{1}{n}-\dfrac{{\chi}}{n}\right) \cdot n \cdot 4 = 3-4{\chi}$  &     \\
                                            & {$9$}-edge & $\left(\dfrac{{\chi}}{n}+\dfrac{n-3}{6} \cdot \dfrac{1}{n}-\dfrac{1}{6}\right) \cdot n \cdot 4 = 4{\chi}-2$  &   \\
    \hline

                         \multirow{4}*{$4$} & {$2$}-edge &   &  $\left(3-3{\chi}\right) \cdot \dfrac{8}{12} = 2-2{\chi}$   \\
                                            & {$4$}-edge &   &  $\left(3-3{\chi}\right) \cdot \dfrac{4}{12}+\left(3{\chi}-2\right) \cdot \dfrac{12}{12} = 2{\chi}-1$  \\
                                            & {$8$}-edge &  $\left(\dfrac{1-{\chi}}{n}\right) \cdot n \cdot 3 = 3-3{\chi}$  &    \\
                                            & {$12$}-edge & $\left(\dfrac{{\chi}}{n}+\dfrac{n-4}{6} \cdot \dfrac{1}{n}-\dfrac{1}{6}\right) \cdot n \cdot 3 = 3{\chi}-2$  &    \\

    \hline
    \hline
    \end{tabular}}
\end{center}
\end{table*}

\begin{table*}[htbp]
\begin{center}
\caption{Probabilities of different patch-covering structures for patch sizes $\chi$ near ${\chi}_c$,
        for $n$-patch disks on the ruby lattice. 
	The period is $n_0=12$.
	For intermediate results on the regular dodecagon,
	the patch disk is placed at the center of the dodecagon and its patches cover edges connecting
	the center and vertices of the dodecagon. 
	This table should be read together with Tables~\ref{Tab:contact-proba-ruby-a}
	and \ref{Tab:contact-proba-ruby-c}.}
\label{Tab:contact-proba-ruby-b}
    \renewcommand\arraystretch{2.7}
    \setlength{\tabcolsep}{2.4mm}{
    \begin{tabular}[t]{lllllllll}
    \hline
    \hline
	    \hspace{-4mm}{$\mod(n,n_0)$} & Type & {On the regular dodecagon} &  {On the ruby lattice}   \\
    \hline

                         \multirow{6}*{$5$} & {$1$}-edge &   &  $\left(9-12{\chi}\right) \cdot \dfrac{2}{12} = \dfrac{3}{2}-2{\chi}$   \\
                                            & {$2$}-edge &   &  $\left(9-12{\chi}\right) \cdot \dfrac{4}{12}+\left(12{\chi}-8\right) \cdot \dfrac{5}{12} = {\chi}-\dfrac{1}{3}$   \\
                                            & {$3$}-edge &   &  $\left(9-12{\chi}\right) \cdot \dfrac{2}{12}+\left(12{\chi}-8\right) \cdot \dfrac{2}{12} = \dfrac{1}{6}$   \\
                                            & {$4$}-edge &   &  $\left(9-12{\chi}\right) \cdot \dfrac{4}{12}+\left(12{\chi}-8\right) \cdot \dfrac{5}{12} = {\chi}-\dfrac{1}{3}$   \\
                                            & {$8$}-edge & $\left(\dfrac{{1-\chi}}{n}+\dfrac{n-1}{4} \cdot \dfrac{1}{n}-\dfrac{1}{4}\right) \cdot n \cdot 12 = 9-12{\chi}$  &    \\
                                            & {$9$}-edge & $\left(\dfrac{{\chi}}{n}+\dfrac{n-2}{3} \cdot \dfrac{1}{n}-\dfrac{1}{3}\right) \cdot n \cdot 12 = 12{\chi}-8$  &   \\
    \hline

                         \multirow{4}*{$6$} & {$2$}-edge &   &  $\left(2-2{\chi}\right) \cdot \dfrac{12}{12} = 2-2{\chi}$   \\
                                            & {$4$}-edge &   &  $\left(2{\chi}-1\right) \cdot \dfrac{12}{12} = 2{\chi}-1$   \\
                                            & {$6$}-edge & $\left(\dfrac{{1-\chi}}{n}\right) \cdot n \cdot 2 = 2-2{\chi}$  &  \\
                                            & {$12$}-edge & $\left(\dfrac{{\chi}}{n}+\dfrac{n-6}{12} \cdot \dfrac{1}{n}-\dfrac{1}{12}\right) \cdot n \cdot 2 = 2{\chi}-1$  &  \\
    \hline

                                            & {$1$}-edge &   &  $\left(9-12{\chi}\right) \cdot \dfrac{2}{12} = \dfrac{3}{2}-2{\chi}$ \\
                         \multirow{4}*{$7$} & {$2$}-edge &   &  $\left(9-12{\chi}\right) \cdot \dfrac{4}{12}+\left(12{\chi}-8\right) \cdot \dfrac{5}{12} = {\chi}-\dfrac{1}{3}$ \\
                                            & {$3$}-edge &   &  $\left(9-12{\chi}\right) \cdot \dfrac{2}{12}+\left(12{\chi}-8\right) \cdot \dfrac{2}{12} = \dfrac{1}{6}$  \\
                                            & {$4$}-edge &   &  $\left(9-12{\chi}\right) \cdot \dfrac{4}{12}+\left(12{\chi}-8\right) \cdot \dfrac{5}{12} = {\chi}-\dfrac{1}{3}$   \\
                                            & {$8$}-edge & $\left(\dfrac{1}{4}-\dfrac{n-3}{4} \cdot \dfrac{1}{n}-\dfrac{{\chi}}{n}\right) \cdot n \cdot 12 = 9-12{\chi}$  &    \\
                                            & {$9$}-edge & $\left(\dfrac{1}{3}-\dfrac{n-1}{3} \cdot \dfrac{1}{n}-\dfrac{1-{\chi}}{n}\right) \cdot n \cdot 12 = 12{\chi}-8$  &   \\
    \hline

                         \multirow{4}*{$8$} & {$2$}-edge &   &  $\left(3-3{\chi}\right) \cdot \dfrac{12}{12} = 2-2{\chi}$   \\
                                            & {$4$}-edge &   &  $\left(3-3{\chi}\right) \cdot \dfrac{6}{12}+\left(3{\chi}-2\right) \cdot \dfrac{12}{12} = 2{\chi}-1$  \\
                                            & {$8$}-edge & $\left(\dfrac{1-{\chi}}{n}\right) \cdot n \cdot 3 = 3-3{\chi}$  &    \\
                                            & {$12$}-edge & $\left(\dfrac{{\chi}}{n}+\dfrac{n-8}{12} \cdot \dfrac{1}{n}-\dfrac{1}{12}\right) \cdot n \cdot 3 = 3{\chi}-2$  &    \\

    \hline
    \hline
    \end{tabular}}
\end{center}
\end{table*}

\begin{table*}[htbp]
\begin{center}
\caption{Probabilities of different patch-covering structures for patch sizes $\chi$ near ${\chi}_c$,
        for $n$-patch disks on the ruby lattice. 
	The period is $n_0=12$.
	For intermediate results on the regular dodecagon,
	the patch disk is placed at the center of the dodecagon and its patches cover edges connecting
	the center and vertices of the dodecagon. 
	This table should be read together with Tables~\ref{Tab:contact-proba-ruby-a}
	and \ref{Tab:contact-proba-ruby-b}.}
\label{Tab:contact-proba-ruby-c}
    \renewcommand\arraystretch{2.7}
    \setlength{\tabcolsep}{2.4mm}{
    \begin{tabular}[t]{lllllllll}
    \hline
    \hline
	    \hspace{-4mm}{$\mod(n,n_0)$} & Type & {On the regular dodecagon} &  {On the ruby lattice}   \\
    \hline

                                            & {$1$}-edge &   &  $\left(3-4{\chi}\right) \cdot \dfrac{6}{12}= \dfrac{3}{2}-2{\chi}$   \\
                         \multirow{4}*{$9$} & {$2$}-edge &   &  $\left(4{\chi}-2\right) \cdot \dfrac{3}{12} = {\chi}-\dfrac{1}{2}$   \\
                                            & {$3$}-edge &   &  $\left(3-4{\chi}\right) \cdot \dfrac{6}{12} + \left(4{\chi}-2\right) \cdot \dfrac{6}{12} = \dfrac{1}{2}$   \\
                                            & {$4$}-edge &   &  $\left(4{\chi}-2\right) \cdot \dfrac{3}{12} = {\chi}-\dfrac{1}{2}$   \\
                                            & {$6$}-edge & $\left(\dfrac{1}{12}-\dfrac{n-9}{12} \cdot \dfrac{1}{n}-\dfrac{{\chi}}{n}\right) \cdot n \cdot 4 = 3-4{\chi}$  &     \\
                                            & {$9$}-edge & $\left(\dfrac{{\chi}}{n}+\dfrac{n-3}{6} \cdot \dfrac{1}{n}-\dfrac{1}{6}\right) \cdot n \cdot 4 = 4{\chi}-2$  &   \\
    \hline

                        \multirow{5}*{$10$} & {$2$}-edge &   &  $\left(5-6{\chi}\right) \cdot \dfrac{4}{12} = \dfrac{5}{3}-2{\chi}$   \\
                                            & {$3$}-edge &   &  $\left(5-6{\chi}\right) \cdot \dfrac{8}{12}+\left(6{\chi}-4\right) \cdot \dfrac{4}{12} = \dfrac{2}{3}$   \\
                                            & {$4$}-edge &   &  $\left(6{\chi}-4\right) \cdot \dfrac{8}{12} = 2{\chi}-\dfrac{4}{3}$   \\
                                            & {$8$}-edge & $\left(\dfrac{1}{12}-\dfrac{n-10}{12} \cdot \dfrac{1}{n}-\dfrac{{\chi}}{n}\right) \cdot n \cdot 6 = 5-6{\chi}$ &  \\
                                            & {$10$}-edge & $\left(\dfrac{{\chi}}{n}+\dfrac{n-4}{6} \cdot \dfrac{1}{n}-\dfrac{1}{6}\right) \cdot n \cdot 6 = 6{\chi}-4$  &  \\
    \hline

                        \multirow{5}*{$11$} & {$2$}-edge &   &  $\left(9-12{\chi}\right) \cdot \dfrac{4}{12}+\left(12{\chi}-8\right) \cdot \dfrac{1}{12} = \dfrac{7}{3}-3{\chi}$   \\
                                            & {$3$}-edge &   &   $\left(9-12{\chi}\right) \cdot \dfrac{8}{12}+\left(12{\chi}-8\right) \cdot \dfrac{10}{12} = 2{\chi}-\dfrac{2}{3}$  \\
                                            & {$4$}-edge &   &   $\left(12{\chi}-8\right) \cdot \dfrac{1}{12} = {\chi}-\dfrac{2}{3}$  \\
                                            & {$8$}-edge & $\left(\dfrac{1}{4}-\dfrac{n-3}{4} \cdot \dfrac{1}{n}-\dfrac{{\chi}}{n}\right) \cdot n \cdot 12 = 9-12{\chi}$  &    \\
                                            & {$9$}-edge & $\left(\dfrac{{\chi}}{n}+\dfrac{n-2}{3} \cdot \dfrac{1}{n}-\dfrac{1}{3}\right) \cdot n \cdot 12 = 12{\chi}-8$  &   \\                    

    \hline
    \hline
    \end{tabular}}
\end{center}
\end{table*}

\newpage

\begin{table*}[htbp]
\begin{center}
\caption{Probabilities of different patch-covering structures for patch sizes $\chi$ near ${\chi}_c$,
        for $n$-patch disks on the cross lattice. 
	The period is $n_0=12$.
	For intermediate results on the regular dodecagon,
	the patch disk is placed at the center of the dodecagon and its patches cover edges connecting
	the center and vertices of the dodecagon. 
	This table should be read together with Tables~\ref{Tab:contact-proba-cross-b}
	and \ref{Tab:contact-proba-cross-c}.}
\label{Tab:contact-proba-cross-a}
    \renewcommand\arraystretch{2.7}
    \setlength{\tabcolsep}{2.4mm}{
    \begin{tabular}[t]{lllllllll}
    \hline
    \hline
	    \hspace{-4mm}{$\mod(n,n_0)$} & Type & {On the regular dodecagon} &  {On the cross lattice}   \\
    \hline
                         \multirow{4}*{$1$} & {$2$}-edge &   &  $\left(11-12{\chi}\right) \cdot \dfrac{6}{12}+\left(12{\chi}-10\right) \cdot \dfrac{3}{12} = 3-3{\chi}$   \\
                                            & {$3$}-edge &   &   $\left(11-12{\chi}\right) \cdot \dfrac{6}{12}+\left(12{\chi}-10\right) \cdot \dfrac{9}{12} = 3{\chi}-2$  \\
                                            & {$10$}-edge & $\left(\dfrac{{1-\chi}}{n}+\dfrac{n-1}{12} \cdot \dfrac{1}{n}-\dfrac{1}{12} \right) \cdot n \cdot 12 = 11-12{\chi}$  &    \\
                                            & {$11$}-edge & $\left(\dfrac{1}{6}-\dfrac{n-1}{6} \cdot \dfrac{1}{n}-\dfrac{{1-\chi}}{n}\right) \cdot n \cdot 12 = 12{\chi}-10$  &   \\
    \hline

                         \multirow{4}*{$2$} & {$2$}-edge &   &  $\left(6-6{\chi}\right) \cdot \dfrac{6}{12} = 3-3{\chi}$   \\
                                            & {$3$}-edge &   &  $\left(6-6{\chi}\right) \cdot \dfrac{6}{12}+\left(6{\chi}-5\right) \cdot \dfrac{12}{12} = 3{\chi}-2$   \\
                                            & {$10$}-edge & $\left(\dfrac{1-{\chi}}{n}\right) \cdot n \cdot 6 = 6-6{\chi}$ &  \\
                                            & {$12$}-edge & $\left(\dfrac{1}{12}-\dfrac{n-2}{12} \cdot \dfrac{1}{n}-\dfrac{1-{\chi}}{n}\right) \cdot n \cdot 6 = 6{\chi}-5$  &  \\
    \hline

                         \multirow{5}*{$3$} & {$1$}-edge &   &  $\left(4-4{\chi}\right) \cdot \dfrac{3}{12} = 1-{\chi}$   \\
                                            & {$2$}-edge &   &  $\left(4-4{\chi}\right) \cdot \dfrac{3}{12} = 1-{\chi}$   \\
                                            & {$3$}-edge &   &  $\left(4-4{\chi}\right) \cdot \dfrac{6}{12} + \left(4{\chi}-3\right) \cdot \dfrac{12}{12} = 2{\chi}-1$   \\                         
                                            & {$9$}-edge & $\left(\dfrac{1-{\chi}}{n}\right) \cdot n \cdot 4 = 4-4{\chi}$  &     \\
                                            & {$12$}-edge & $\left(\dfrac{{\chi}}{n}+\dfrac{n-3}{4} \cdot \dfrac{1}{n}-\dfrac{1}{4}\right) \cdot n \cdot 4 = 4{\chi}-3$  &   \\
    \hline

                                            & {$1$}-edge &   &  $\left(3-3{\chi}\right) \cdot \dfrac{4}{12} = 1-{\chi}$  \\
                         \multirow{3}*{$4$} & {$2$}-edge &   &  $\left(3-3{\chi}\right) \cdot \dfrac{4}{12} = 1-{\chi}$   \\
                                            & {$3$}-edge &   &  $\left(3-3{\chi}\right) \cdot \dfrac{4}{12}+\left(3{\chi}-2\right) \cdot \dfrac{12}{12} = 2{\chi}-1$  \\
                                            & {$8$}-edge &  $\left(\dfrac{1-{\chi}}{n}\right) \cdot n \cdot 3 = 3-3{\chi}$  &    \\
                                            & {$12$}-edge & $\left(\dfrac{{\chi}}{n}+\dfrac{n-4}{6} \cdot \dfrac{1}{n}-\dfrac{1}{6}\right) \cdot n \cdot 3 = 3{\chi}-2$  &    \\

    \hline
    \hline
    \end{tabular}}
\end{center}
\end{table*}

\begin{table*}[htbp]
\begin{center}
\caption{Probabilities of different patch-covering structures for patch sizes $\chi$ near ${\chi}_c$,
        for $n$-patch disks on the cross lattice. 
	The period is $n_0=12$.
	For intermediate results on the regular dodecagon,
	the patch disk is placed at the center of the dodecagon and its patches cover edges connecting
	the center and vertices of the dodecagon. 
	This table should be read together with Tables~\ref{Tab:contact-proba-cross-a}
	and \ref{Tab:contact-proba-cross-c}.}
\label{Tab:contact-proba-cross-b}
    \renewcommand\arraystretch{2.7}
    \setlength{\tabcolsep}{2.4mm}{
    \begin{tabular}[t]{lllllllll}
    \hline
    \hline
	    \hspace{-4mm}{$\mod(n,n_0)$} & Type & {On the regular dodecagon} &  {On the cross lattice}   \\
    \hline
                         \multirow{5}*{$5$} & {$1$}-edge &   &  $\left(10-12{\chi}\right) \cdot \dfrac{2}{12}+\left(12{\chi}-9\right) \cdot \dfrac{1}{12} = \dfrac{11}{12}-{\chi}$   \\
                                            & {$2$}-edge &   &  $\left(10-12{\chi}\right) \cdot \dfrac{5}{12}+\left(12{\chi}-9\right) \cdot \dfrac{4}{12} = \dfrac{7}{6}-{\chi}$   \\
                                            & {$3$}-edge &   &  $\left(10-12{\chi}\right) \cdot \dfrac{5}{12}+\left(12{\chi}-9\right) \cdot \dfrac{7}{12} = 2{\chi}-\dfrac{13}{12}$   \\
                                            & {$9$}-edge & $\left(\dfrac{1}{6}-\dfrac{n-5}{6} \cdot \dfrac{1}{n}-\dfrac{{\chi}}{n}\right) \cdot n \cdot 12 = 10-12{\chi}$  &    \\
                                            & {$10$}-edge & $\left(\dfrac{1}{4}-\dfrac{n-1}{4} \cdot \dfrac{1}{n}-\dfrac{{1-\chi}}{n}\right) \cdot n \cdot 12 = 12{\chi}-9$  &   \\
    \hline

                         \multirow{5}*{$6$} & {$1$}-edge &   &  $\left(2-2{\chi}\right) \cdot \dfrac{6}{12} = 1-{\chi}$   \\
                                            & {$2$}-edge &   &  $\left(2-2{\chi}\right) \cdot \dfrac{6}{12} = 1-{\chi}$   \\
                                            & {$3$}-edge &   &  $\left(2{\chi}-1\right) \cdot \dfrac{12}{12} = 2{\chi}-1$   \\
                                            & {$6$}-edge & $\left(\dfrac{{1-\chi}}{n}\right) \cdot n \cdot 2 = 2-2{\chi}$ &  \\
                                            & {$12$}-edge & $\left(\dfrac{{\chi}}{n}+\dfrac{n-6}{12} \cdot \dfrac{1}{n}-\dfrac{1}{12}\right) \cdot n \cdot 2 = 2{\chi}-1$  &  \\                    
    \hline

                                            & {$1$}-edge &   &  $\left(10-12{\chi}\right) \cdot \dfrac{2}{12}+\left(12{\chi}-9\right) \cdot \dfrac{1}{12} = \dfrac{11}{12}-{\chi}$  \\
                         \multirow{3}*{$7$} & {$2$}-edge &   &  $\left(10-12{\chi}\right) \cdot \dfrac{5}{12}+\left(12{\chi}-9\right) \cdot \dfrac{4}{12} = \dfrac{7}{6}-{\chi}$  \\
                                            & {$3$}-edge &   &  $\left(10-12{\chi}\right) \cdot \dfrac{5}{12}+\left(12{\chi}-9\right) \cdot \dfrac{7}{12} = 2{\chi}-\dfrac{13}{12}$  \\
                                            & {$9$}-edge & $\left(\dfrac{{1-\chi}}{n}+\dfrac{n-1}{6} \cdot \dfrac{1}{n}-\dfrac{1}{6}\right) \cdot n \cdot 12 = 10-12{\chi}$  &    \\
                                            & {$10$}-edge & $\left(\dfrac{{\chi}}{n}+\dfrac{n-3}{4} \cdot \dfrac{1}{n}-\dfrac{1}{4}\right) \cdot n \cdot 12 = 12{\chi}-9$  &   \\                   
    \hline

                                            & {$1$}-edge &   &  $\left(3-3{\chi}\right) \cdot \dfrac{4}{12} = 1-{\chi}$  \\
                         \multirow{3}*{$8$} & {$2$}-edge &   &  $\left(3-3{\chi}\right) \cdot \dfrac{4}{12} = 1-{\chi}$   \\
                                            & {$3$}-edge &   &  $\left(3-3{\chi}\right) \cdot \dfrac{4}{12}+\left(3{\chi}-2\right) \cdot \dfrac{12}{12} = 2{\chi}-1$  \\
                                            & {$8$}-edge & $\left(\dfrac{1-{\chi}}{n}\right) \cdot n \cdot 3 = 3-3{\chi}$  &    \\
                                            & {$12$}-edge & $\left(\dfrac{{\chi}}{n}+\dfrac{n-8}{12} \cdot \dfrac{1}{n}-\dfrac{1}{12}\right) \cdot n \cdot 3 = 3{\chi}-2$  &    \\                  
    \hline
    \hline
    \end{tabular}}
\end{center}
\end{table*}

\begin{table*}[htbp]
\begin{center}
\caption{Probabilities of different patch-covering structures for patch sizes $\chi$ near ${\chi}_c$,
        for $n$-patch disks on the cross lattice. 
	The period is $n_0=12$.
	For intermediate results on the regular dodecagon,
	the patch disk is placed at the center of the dodecagon and its patches cover edges connecting
	the center and vertices of the dodecagon. 
	This table should be read together with Tables~\ref{Tab:contact-proba-cross-a}
	and \ref{Tab:contact-proba-cross-b}.}
\label{Tab:contact-proba-cross-c}
    \renewcommand\arraystretch{2.7}
    \setlength{\tabcolsep}{2.4mm}{
    \begin{tabular}[t]{lllllllll}
    \hline
    \hline
	    \hspace{-4mm}{$\mod(n,n_0)$} & Type & {On the regular dodecagon} &  {On the cross lattice}   \\
    \hline

                         \multirow{5}*{$9$} & {$1$}-edge &   &  $\left(4-4{\chi}\right) \cdot \dfrac{3}{12} = 1-{\chi}$   \\
                                            & {$2$}-edge &   &  $\left(4-4{\chi}\right) \cdot \dfrac{3}{12} = 1-{\chi}$   \\
                                            & {$3$}-edge &   &  $\left(4-4{\chi}\right) \cdot \dfrac{6}{12}+\left(4{\chi}-3\right) \cdot \dfrac{12}{12} = 2{\chi}-1$   \\
                                            & {$9$}-edge & $\left(\dfrac{1-{\chi}}{n}\right) \cdot n \cdot 4 = 4-4{\chi}$  &     \\
                                            & {$12$}-edge & $\left(\dfrac{{\chi}}{n}+\dfrac{n-9}{12} \cdot \dfrac{1}{n}-\dfrac{1}{12}\right) \cdot n \cdot 4 = 4{\chi}-3$  &   \\
    \hline

                        \multirow{4}*{$10$} & {$2$}-edge &   &  $\left(6-6{\chi}\right) \cdot \dfrac{6}{12} = 3-3{\chi}$   \\
                                            & {$3$}-edge &   &  $\left(6-6{\chi}\right) \cdot \dfrac{6}{12}+\left(6{\chi}-5\right) \cdot \dfrac{12}{12} = 3{\chi}-2$   \\
                                            & {$10$}-edge & $\left(\dfrac{1-{\chi}}{n}\right) \cdot n \cdot 6 = 6-6{\chi}$ &  \\
                                            & {$12$}-edge & $\left(\dfrac{{\chi}}{n}+\dfrac{n-10}{12} \cdot \dfrac{1}{n}-\dfrac{1}{12}\right) \cdot n \cdot 6 = 6{\chi}-5$  &  \\
    \hline

                        \multirow{4}*{$11$} & {$2$}-edge &   &  $\left(11-12{\chi}\right) \cdot \dfrac{6}{12}+\left(12{\chi}-10\right) \cdot \dfrac{3}{12} = 3-3{\chi}$   \\
                                            & {$3$}-edge &   &   $\left(11-12{\chi}\right) \cdot \dfrac{6}{12}+\left(12{\chi}-10\right) \cdot \dfrac{9}{12} = 3{\chi}-2$  \\
                                            & {$9$}-edge & $\left(\dfrac{1}{12}-\dfrac{n-11}{12} \cdot \dfrac{1}{n}-\dfrac{{\chi}}{n}\right) \cdot n \cdot 12 = 11-12{\chi}$  &    \\
                                            & {$10$}-edge & $\left(\dfrac{{\chi}}{n}+\dfrac{n-5}{6} \cdot \dfrac{1}{n}-\dfrac{1}{6}\right) \cdot n \cdot 12 = 12{\chi}-10$  &   \\

    \hline
    \hline
    \end{tabular}}
\end{center}
\end{table*}

\newpage

\begin{table*}[htbp]
\begin{center}
\caption{Probabilities of different patch-covering structures for patch sizes $\chi$ near ${\chi}_c$,
        for $n$-patch disks on the three-twelve lattice. 
	The period is $n_0=12$.
	For intermediate results on the regular dodecagon,
	the patch disk is placed at the center of the dodecagon and its patches cover edges connecting
	the center and vertices of the dodecagon. 
	This table should be read together with Tables~\ref{Tab:contact-proba-three-twelve-b}
	and \ref{Tab:contact-proba-three-twelve-c}.}
\label{Tab:contact-proba-three-twelve-a}
    \renewcommand\arraystretch{2.5}
    \setlength{\tabcolsep}{2.0mm}{
    \begin{tabular}[t]{lllllllll}
    \hline
    \hline
	    \hspace{-4mm}{$\mod(n,n_0)$} & Type & {On the regular dodecagon} &  {On the three-twelve lattice}   \\
    \hline
                         \multirow{4}*{$1$} & {$2$}-edge &   &  $\left(11-12{\chi}\right) \cdot \dfrac{6}{12}+\left(12{\chi}-10\right) \cdot \dfrac{3}{12} = 3-3{\chi}$   \\
                                            & {$3$}-edge &   &   $\left(11-12{\chi}\right) \cdot \dfrac{6}{12}+\left(12{\chi}-10\right) \cdot \dfrac{9}{12} = 3{\chi}-2$  \\
                                            & {$10$}-edge & $\left(\dfrac{{1-\chi}}{n}+\dfrac{n-1}{12} \cdot \dfrac{1}{n}-\dfrac{1}{12} \right) \cdot n \cdot 12 = 11-12{\chi}$  &    \\
                                            & {$11$}-edge & $\left(\dfrac{1}{6}-\dfrac{n-1}{6} \cdot \dfrac{1}{n}-\dfrac{{1-\chi}}{n}\right) \cdot n \cdot 12 = 12{\chi}-10$  &   \\
    \hline

                         \multirow{4}*{$2$} & {$2$}-edge &   &  $\left(6-6{\chi}\right) \cdot \dfrac{6}{12} = 3-3{\chi}$   \\
                                            & {$3$}-edge &   &  $\left(6-6{\chi}\right) \cdot \dfrac{6}{12}+\left(6{\chi}-5\right) \cdot \dfrac{12}{12} = 3{\chi}-2$   \\
                                            & {$10$}-edge & $\left(\dfrac{1-{\chi}}{n}\right) \cdot n \cdot 6 = 6-6{\chi}$ &  \\
                                            & {$12$}-edge & $\left(\dfrac{1}{12}-\dfrac{n-2}{12} \cdot \dfrac{1}{n}-\dfrac{1-{\chi}}{n}\right) \cdot n \cdot 6 = 6{\chi}-5$  &  \\
    \hline

                         \multirow{4}*{$3$} & {$2$}-edge &   &  $\left(4-4{\chi}\right) \cdot \dfrac{9}{12} = 3-3{\chi}$   \\
                                            & {$3$}-edge &   &  $\left(4-4{\chi}\right) \cdot \dfrac{3}{12} + \left(4{\chi}-3\right) \cdot \dfrac{12}{12} = 3{\chi}-2$   \\
                                            & {$9$}-edge & $\left(\dfrac{1-{\chi}}{n}\right) \cdot n \cdot 4 = 4-4{\chi}$  &     \\
                                            & {$12$}-edge & $\left(\dfrac{{\chi}}{n}+\dfrac{n-3}{4} \cdot \dfrac{1}{n}-\dfrac{1}{4}\right) \cdot n \cdot 4 = 4{\chi}-3$  &   \\
    \hline

                         \multirow{4}*{$4$} & {$2$}-edge &   &  $\left(3-3{\chi}\right) \cdot \dfrac{12}{12} = 3-3{\chi}$   \\
                                            & {$3$}-edge &   &  $\left(3{\chi}-2\right) \cdot \dfrac{12}{12} = 3{\chi}-2$  \\
                                            & {$8$}-edge &  $\left(\dfrac{1-{\chi}}{n}\right) \cdot n \cdot 3 = 3-3{\chi}$  &    \\
                                            & {$12$}-edge & $\left(\dfrac{{\chi}}{n}+\dfrac{n-4}{6} \cdot \dfrac{1}{n}-\dfrac{1}{6}\right) \cdot n \cdot 3 = 3{\chi}-2$  &    \\
    \hline

                         \multirow{5}*{$5$} & {$1$}-edge &   &  $\left(11-12{\chi}\right) \cdot \dfrac{2}{12} = \dfrac{11}{6}-2{\chi}$   \\
                                            & {$2$}-edge &   &  $\left(11-12{\chi}\right) \cdot \dfrac{2}{12}+\left(12{\chi}-10\right) \cdot \dfrac{2}{12} = \dfrac{1}{6}$   \\
                                            & {$3$}-edge &   &  $\left(11-12{\chi}\right) \cdot \dfrac{8}{12}+\left(12{\chi}-10\right) \cdot \dfrac{10}{12} = 2{\chi}-1$   \\
                                            & {$10$}-edge & $\left(\dfrac{{1-\chi}}{n}+\dfrac{5n-1}{12} \cdot \dfrac{1}{n}-\dfrac{5}{12}\right) \cdot n \cdot 12 = 11-12{\chi}$  &    \\
                                            & {$11$}-edge & $\left(\dfrac{{\chi}}{n}+\dfrac{n-5}{6} \cdot \dfrac{1}{n}-\dfrac{1}{6}\right) \cdot n \cdot 12 = 12{\chi}-10$  &   \\

    \hline
    \hline
    \end{tabular}}
\end{center}
\end{table*}

\begin{table*}[htbp]
\begin{center}
\caption{Probabilities of different patch-covering structures for patch sizes $\chi$ near ${\chi}_c$,
        for $n$-patch disks on the three-twelve lattice. 
	The period is $n_0=12$.
	For intermediate results on the regular dodecagon,
	the patch disk is placed at the center of the dodecagon and its patches cover edges connecting
	the center and vertices of the dodecagon. 
	This table should be read together with Tables~\ref{Tab:contact-proba-three-twelve-a}
	and \ref{Tab:contact-proba-three-twelve-c}.}
\label{Tab:contact-proba-three-twelve-b}
    \renewcommand\arraystretch{2.7}
    \setlength{\tabcolsep}{2.0mm}{
    \begin{tabular}[t]{lllllllll}
    \hline
    \hline
	    \hspace{-4mm}{$\mod(n,n_0)$} & Type & {On the regular dodecagon} &  {On the three-twelve lattice}   \\
    \hline

                                            & {$1$}-edge &   &  $\left(2-2{\chi}\right) \cdot \dfrac{6}{12} = 1-{\chi}$   \\
                         \multirow{3}*{$6$} & {$2$}-edge &   &  $\left(2-2{\chi}\right) \cdot \dfrac{6}{12} = 1-{\chi}$   \\
                                            & {$3$}-edge &   &  $\left(2{\chi}-1\right) \cdot \dfrac{12}{12} = 2{\chi}-1$   \\
                                            & {$6$}-edge & $\left(\dfrac{{1-\chi}}{n}\right) \cdot n \cdot 2 = 2-2{\chi}$  &  \\
                                            & {$12$}-edge & $\left(\dfrac{{\chi}}{n}+\dfrac{n-6}{12} \cdot \dfrac{1}{n}-\dfrac{1}{12}\right) \cdot n \cdot 2 = 2{\chi}-1$  &  \\
    \hline

                                            & {$1$}-edge &   &  $\left(11-12{\chi}\right) \cdot \dfrac{2}{12} = \dfrac{11}{6}-2{\chi}$  \\
                         \multirow{3}*{$7$} & {$2$}-edge &   &  $\left(11-12{\chi}\right) \cdot \dfrac{2}{12}+\left(12{\chi}-10\right) \cdot \dfrac{2}{12} = \dfrac{1}{6}$ \\
                                            & {$3$}-edge &   &  $\left(11-12{\chi}\right) \cdot \dfrac{8}{12}+\left(12{\chi}-10\right) \cdot \dfrac{10}{12} = 2{\chi}-1$  \\
                                            & {$10$}-edge & $\left(\dfrac{5}{12}-\dfrac{5n-11}{12} \cdot \dfrac{1}{n}-\dfrac{{\chi}}{n}\right) \cdot n \cdot 12 = 11-12{\chi}$  &    \\
                                            & {$11$}-edge & $\left(\dfrac{1}{6}-\dfrac{n-1}{6} \cdot \dfrac{1}{n}-\dfrac{1-{\chi}}{n}\right) \cdot n \cdot 12 = 12{\chi}-10$  &   \\
    \hline

                         \multirow{4}*{$8$} & {$2$}-edge &   &  $\left(3-3{\chi}\right) \cdot \dfrac{12}{12} = 3-3{\chi}$   \\
                                            & {$4$}-edge &   &  $\left(3{\chi}-2\right) \cdot \dfrac{12}{12} = 3{\chi}-2$  \\
                                            & {$8$}-edge & $\left(\dfrac{1-{\chi}}{n}\right) \cdot n \cdot 3 = 3-3{\chi}$  &    \\
                                            & {$12$}-edge & $\left(\dfrac{{\chi}}{n}+\dfrac{n-8}{12} \cdot \dfrac{1}{n}-\dfrac{1}{12}\right) \cdot n \cdot 3 = 3{\chi}-2$  &    \\
    \hline
					                  
 	                     \multirow{4}*{$9$} & {$2$}-edge &   &  $\left(4-4{\chi}\right) \cdot \dfrac{9}{12} = 3-3{\chi}$   \\
                                            & {$3$}-edge &   &  $\left(4-4{\chi}\right) \cdot \dfrac{3}{12}+\left(4{\chi}-3\right) \cdot \dfrac{12}{12} = 3{\chi}-2$   \\
                                            & {$9$}-edge & $\left(\dfrac{1-{\chi}}{n}\right) \cdot n \cdot 4 = 4-4{\chi}$  &     \\
                                            & {$12$}-edge & $\left(\dfrac{{\chi}}{n}+\dfrac{n-9}{12} \cdot \dfrac{1}{n}-\dfrac{1}{12}\right) \cdot n \cdot 4 = 4{\chi}-3$  &   \\

    \hline
    \hline
    \end{tabular}}
\end{center}
\end{table*}

\begin{table*}[htbp]
\begin{center}
\caption{Probabilities of different patch-covering structures for patch sizes $\chi$ near ${\chi}_c$,
        for $n$-patch disks on the three-twelve lattice. 
	The period is $n_0=12$.
	For intermediate results on the regular dodecagon,
	the patch disk is placed at the center of the dodecagon and its patches cover edges connecting
	the center and vertices of the dodecagon. 
	This table should be read together with Tables~\ref{Tab:contact-proba-three-twelve-a}
	and \ref{Tab:contact-proba-three-twelve-b}.}
\label{Tab:contact-proba-three-twelve-c}
    \renewcommand\arraystretch{2.7}
    \setlength{\tabcolsep}{2.0mm}{
    \begin{tabular}[t]{lllllllll}
    \hline
    \hline
	    \hspace{-4mm}{$\mod(n,n_0)$} & Type & {On the regular dodecagon} &  {On the three-twelve lattice}   \\
    \hline

                        \multirow{4}*{$10$} & {$2$}-edge &   &  $\left(6-6{\chi}\right) \cdot \dfrac{6}{12} = 3-3{\chi}$   \\
                                            & {$3$}-edge &   &  $\left(6-6{\chi}\right) \cdot \dfrac{6}{12}+\left(6{\chi}-5\right) \cdot \dfrac{12}{12} = 3{\chi}-2$   \\
                                            & {$10$}-edge & $\left(\dfrac{1-{\chi}}{n}\right) \cdot n \cdot 6 = 6-6{\chi}$ &  \\
                                            & {$12$}-edge & $\left(\dfrac{{\chi}}{n}+\dfrac{n-10}{12} \cdot \dfrac{1}{n}-\dfrac{1}{12}\right) \cdot n \cdot 6 = 6{\chi}-5$  &  \\    
    \hline

                        \multirow{4}*{$11$} & {$2$}-edge &   &  $\left(11-12{\chi}\right) \cdot \dfrac{6}{12}+\left(12{\chi}-10\right) \cdot \dfrac{3}{12} = 3-3{\chi}$   \\
                                            & {$3$}-edge &   &   $\left(11-12{\chi}\right) \cdot \dfrac{6}{12}+\left(12{\chi}-10\right) \cdot \dfrac{9}{12} = 3{\chi}-2$  \\
                                            & {$9$}-edge & $\left(\dfrac{1}{12}-\dfrac{n-11}{12} \cdot \dfrac{1}{n}-\dfrac{{\chi}}{n}\right) \cdot n \cdot 12 = 11-12{\chi}$  &    \\
                                            & {$10$}-edge & $\left(\dfrac{{\chi}}{n}+\dfrac{n-5}{6} \cdot \dfrac{1}{n}-\dfrac{1}{6}\right) \cdot n \cdot 12 = 12{\chi}-10$  &   \\

    \hline
    \hline
    \end{tabular}}
\end{center}
\end{table*}

\newpage

\begin{figure*}[ht]
\begin{center}
\hspace{-15.0cm}    $(a)$ Honeycomb

\includegraphics[scale=0.3]{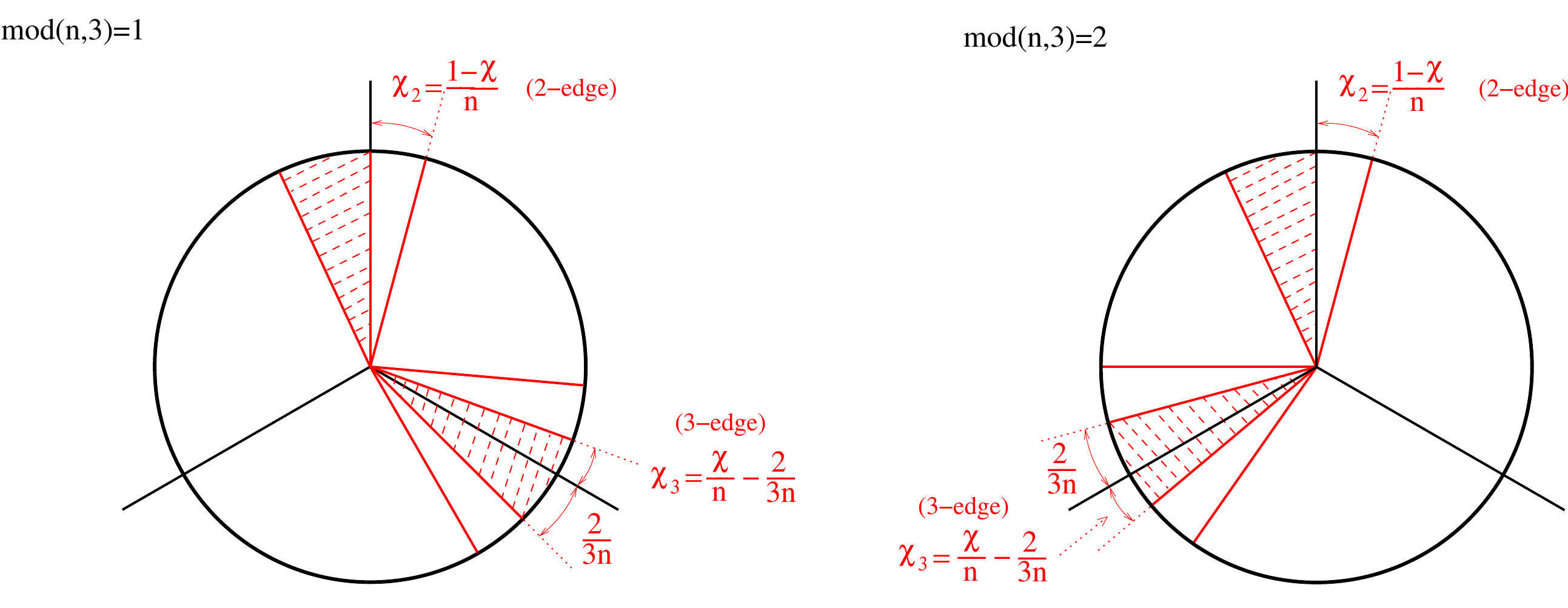} \\
\hspace{-15.7cm}    $(b)$ Square

\includegraphics[scale=0.3]{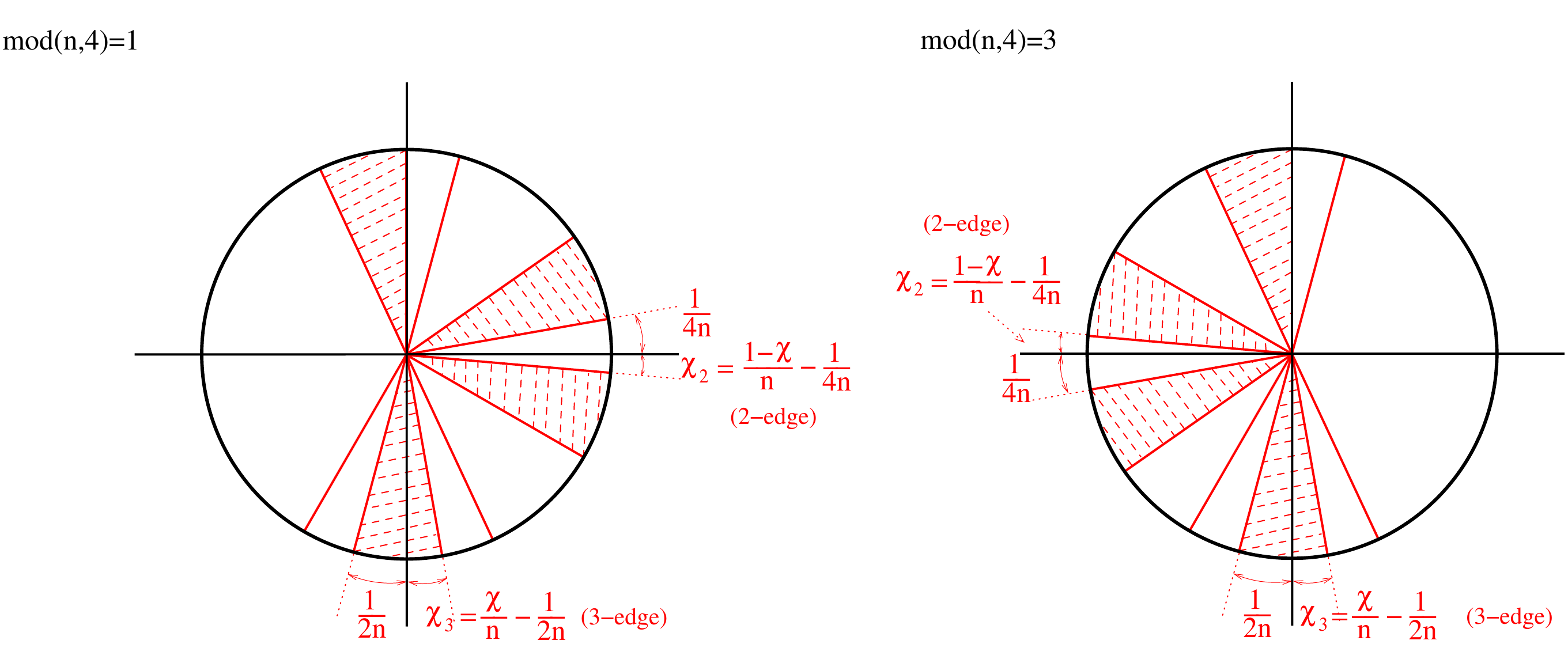} \\
\hspace{-15.0cm}    $(c)$ Triangular

\includegraphics[scale=0.3]{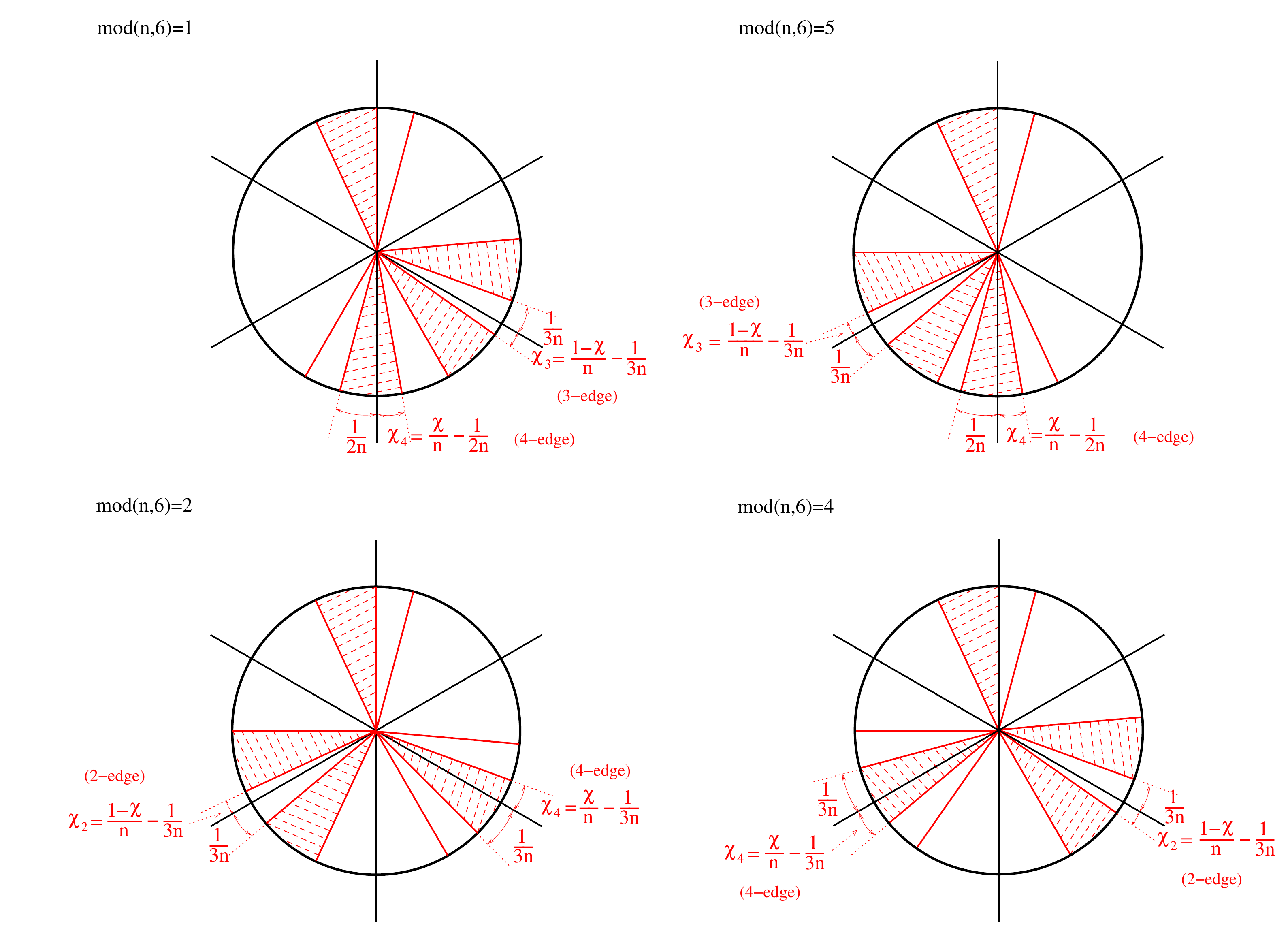} \\

	\caption{Plots demonstrating symmetries between $n$-patch disks with $\mod(n,n_0)=m$ and $n_0-m$.
		 The angles $\chi_i$ for $i$-edge patch-covering structures are from expressions in Table~\ref{Tab:contact-proba}.}
    \label{Fig:symmetry-m}
\end{center}
\end{figure*}